\documentclass[aps,preprint,superscriptaddress,
showpacs,showkeys,
nofootinbib,eqsecnum,floats,altaffilletter, floatfix]{revtex4}
\usepackage{graphicx,color}
\usepackage{float}
\usepackage{amsmath}
\usepackage{graphicx}
\usepackage{bm}
\usepackage[normalem]{ulem}
\usepackage{xcolor}
\usepackage{color,colordvi}
\usepackage{epsfig}
\usepackage{natbib}
\usepackage{mciteplus}
\usepackage{mathcomp}
\usepackage{axodraw}
\usepackage{slashbox}
\usepackage[colorlinks=true,urlcolor=blue]{hyperref}

\usepackage{subfig} 
\usepackage{multirow}

\begin{document}
\begin{flushright}
\end{flushright}
\newcommand  {\ba} {\begin{eqnarray}}
\newcommand  {\ea} {\end{eqnarray}}
\def\cM{{\cal M}}
\def\cO{{\cal O}}
\def\cK{{\cal K}}
\def\cS{{\cal S}}
\def\tRP{\,$\tilde{t}$-{RPV}\,}
\def\cRP{\,$\chi^+$\!-RPV\,}
\def\RPl{\,RPC-like\;}
\def\bftRP{\,$\mathbf{\tilde{t}}$-$\mathbf{RPV}$\,}
\def\bfcRP{\,$\boldsymbol{\chi^{~}}$\!-$\mathbf{RPV}$\,}
\def\bfRPl{\,$\mathbf{RPC}${\bf-like}\,}
\def\RPlRPl{2t4b2j}
\def\RPlcRP{1t5b2j}
\def\RPltRP{1t3b2j}
\def\cRPcRP{6b2j}
\def\cRPtRP{4b2j}
\def\tRPtRP{2b2j}
\newcommand{\sara}[1]{\textcolor{blue}{#1}}
\newcommand{\sarazero}[1]{#1}

\newcommand\sst{\bgroup\markoverwith{\textcolor{blue}{\rule[0.5ex]{2pt}{0.4pt}}}\ULon}
\newcommand{\saracom}[1]{\textcolor{blue}{\bf{[#1]}}}

\newcommand{\lorenzo}[1]{\textcolor{red}{#1}}
\newcommand{\lorenzozero}[1]{#1}

\newcommand\lst{\bgroup\markoverwith{\textcolor{red}{\rule[0.5ex]{2pt}{0.4pt}}}\ULon}
\newcommand{\lorenzocom}[1]{\textcolor{red}{\bf{[#1]}}}

\newcommand{\gilbert}[1]{\textcolor{black}{ \bf{#1}}}
\newcommand{\gilbertzero}[1]{#1}

\newcommand\gst{\bgroup\markoverwith{\textcolor{black}{\rule[0.5ex]{2pt}{0.6pt}}}\ULon}
\newcommand{\gilbertcom}[1]{\textcolor{black}{\bf{[#1]}}}

\newcommand{\ignore}[1]{}
\newcommand{\gilbertcomhide}[1]{}
\newcommand{\lorenzocomhide}[1]{}
\newcommand{\saracomhide}[1]{}
\newcommand{\gsthide}[1]{}
\newcommand{\lsthide}[1]{}
\newcommand{\ssthide}[1]{}

\newcommand{\mh}{m_{h^0}}
\newcommand{\mw}{m_W}
\newcommand{\mz}{m_Z}
\newcommand{\mt}{m_t}
\newcommand{\mb}{m_b}
\newcommand{\be}{\beta}\newcommand{\al}{\alpha}
\newcommand{\lam}{\lambda}
\newcommand{\no}{\nonumber}
\makeatletter
\makeatother
\def\ga{\mathrel{\raise.3ex\hbox{$>$\kern-.75em\lower1ex\hbox{$\sim$}}}}
\def\la{\mathrel{\raise.3ex\hbox{$<$\kern-.75em\lower1ex\hbox{$\sim$}}}}
\newcommand*{\TeV}{\ifmmode {\mathrm{\ Te\kern -0.1em V}}\else
                   \textrm{Te\kern -0.1em V}\fi}%
\newcommand*{\GeV}{\ifmmode {\mathrm{\ Ge\kern -0.1em V}}\else
                   \textrm{Ge\kern -0.1em V}\fi}%
\newcommand*{\MeV}{\ifmmode {\mathrm{\ Me\kern -0.1em V}}\else
                   \textrm{Me\kern -0.1em V}\fi}%
\newcommand*{\keV}{\ifmmode {\mathrm{\ ke\kern -0.1em V}}\else
                   \textrm{ke\kern -0.1em V}\fi}%
\newcommand*{\eV}{\ifmmode  {\mathrm{\ e\kern -0.1em V}}\else
                   \textrm{e\kern -0.1em V}\fi}%

\title{Stashing the stops in multijet events at the LHC}
\author{Sara Diglio}
\altaffiliation[Current address: ]{SUBATECH, Ecole des Mines de Nantes, CNRS/IN2P3, Universit\'e de Nantes, Nantes, France}
\email[electronic address: ]{diglio@subatech.in2p3.fr}
\affiliation{Centre de Physique des Particules de Marseille (CPPM), UMR 7346
IN2P3-Univ. Aix-Marseille, Marseille, F-France}
 \author{Lorenzo Feligioni}
 \email{lorenzo@cppm.in2p3.fr}
\affiliation{Centre de Physique des Particules de Marseille (CPPM), UMR 7346 IN2P3-Univ. Aix-Marseille, Marseille, F-France}
 \author{Gilbert Moultaka}
 \email{gilbert.moultaka@umontpellier.fr}
 \thanks{corresponding author}
\affiliation{Laboratoire Charles Coulomb (L2C), UMR 5221 CNRS-Universit\'e de Montpellier, Montpellier, F-France}

\date{\today}

\begin{abstract}

While the presence of a light stop is increasingly disfavored by the experimental 
limits set on R-parity conserving scenarios, 
the naturalness of supersymmetry could still be safely concealed in the more challenging
final states predicted by the existence of non-null R-parity violating couplings. 
Although R-parity violating signatures are extensively looked for at the Large Hadron 
Collider, these searches always assume 100\% branching ratios for the direct decays of 
supersymmetric particles into Standard Model ones. In this paper we scrutinize the 
implications of relaxing this assumption by focusing on one motivated scenario where 
the lightest stop is heavier than a chargino and a neutralino. Considering a class of 
R-parity baryon number violating couplings, we show on general grounds that while the 
direct decay of the stop into Standard Model particles is dominant for large values of 
these couplings,  smaller values give rise, instead, to the 
dominance of a plethora of longer decay chains and richer final states that have been so
far barely analyzed at the LHC, thus weakening the impact of the present experimental stop mass limits. 
We characterize the case for R-parity baryon number violating couplings in the 
$10^{-7} - 10^{-1}$ range, in two different benchmark points scenarios within the 
model-independent setting of the low-energy phenomenological Minimal Supersymmetric 
Standard Model. We identify the different relevant experimental signatures from stop pair
production and decays, estimate 
the corresponding proton--proton cross sections at $\sqrt{s}=14$~\TeV\ 
and discuss signal versus background issues.
    
\end{abstract}


\maketitle
\section{Introduction}
\noindent

The discovery at the Large Hadron Collider (LHC) of a weakly-coupled, spin-0 particle compatible with the Higgs boson~\cite{Englert:1964aa,Higgs:1964aa,Higgs:1964ab,Guralnik:1964aa}, by both 
ATLAS~\cite{Aad:2012tfa} and CMS~\cite{Chatrchyan:2012xdj} collaborations, with a mass of approximately 125 GeV~\cite{Aad:2015zhl},  constrains 
all theoretical extensions to the Standard Model (SM)
that aim at a mechanism for spontaneous Electroweak Symmetry Breaking (EWSB) relieved of the \textit{naturalness problem}. In the coming years the measurements of the properties of this new particle will shed further light on the possibility of new physics at the \TeV~scale.
 While the presence of a new class 
of phenomena at the \TeV~scale is predicted by a large variety of models
 which address the various theoretical shortcomings 
of the SM, the LHC Run 1 and first Run 2 data sets analysed so far gave 
no evidence for new physics Beyond the Standard Model (BSM). Indirect
manifestations might be hiding in heavy flavor rare decays anomalies reported by LHCb
\cite{Aaij:2014ora,Aaij:2015oid} with moderate to sizable statistical significance \cite{Descotes-Genon:2015uva}, 
and the well established neutrino oscillation phenomena \cite{Agashe:2014kda} can be viewed as clear indications 
for the need for BSM physics \cite{Deppisch:2015qwa}. 

Supersymmetry (SUSY)
~\cite{Golfand:1971iw,Ramond:1971gb,Neveu:1971iv,Gervais:1971wu,Volkov:1973ix,Wess:1973kz} has long been considered to be an elegant way of
triggering the EWSB, relating it radiatively through perturbative quantum effects to 
possible new physics at much higher
scales, such as Grand Unification, while stabilizing the various scales without 
unnatural fine-tunings.  
It can also provide in its R-parity conserving (RPC) version several dark matter 
candidates, the most popular being a neutralino when
the lightest supersymmetric particle (LSP). Nonetheless, the naturalness of the Higgs 
potential favors light third-generation squarks whose RPC striking signatures have yet 
to be observed at hadron colliders, pushing the limits on the mass of such particles at 
the boundary of what is accepted to be natural. This could be a hint that the role of SUSY as a panacea for all SM standing problems should be 
revised. In 
particular if R-parity violating (RPV) operators~\cite{Fayet:1974pd, Salam:1974xa} 
in the superpotential are not artificially suppressed to allow for instance for a neutralino dark matter, RPV SUSY could be welcome for a natural EWSB since most of LHC constraints based on searches for missing energy signatures would not be valid anymore.  
One thus expects the interest in RPV SUSY searches at the LHC to build up significantly in the coming years \cite{Evans:2012bf}.

From the theoretical point of view, it is attractive to view R-parity breaking as a dynamical 
issue. The magnitude of the RPV couplings could then be related to residual low-energy effects 
of some ultraviolet completions of the minimal SUSY extension of the Standard Model, 
see e.g. \cite{Csaki:2015fea,Mohapatra:2015fua} for recent reviews. 
On a more fundamental level,  
whether R-parity is conserved or not as a residual discrete symmetry of  
continuous R-symmetries, could also depend on the breaking mechanisms of the latter, 
which is an open question intimately related to the origin of SUSY breaking itself 
\cite{Nelson:1993nf}. 
The presence of RPV operators with small couplings, but still sufficiently large to trigger 
prompt decays within the detector, is thus not unlikely. It can also preserve 
some of the appealing features of the RPC scenarios; e.g. a very light metastable 
gravitino can provide a viable dark matter candidate, and the stability of the proton can be 
protected by other discrete symmetries \cite{Ibanez:1991pr}.

Limits on RPV scenarios have been given by 
ATLAS~\cite{ATLAS-CONF-2016-037,Aad:2012ypy,ATLAS-CONF-2016-075,ATLAS-CONF-2016-057,ATLAS-CONF-2016-094,Aad:2016kww,ATLAS-CONF-2016-022}  and CMS~\cite{Chatrchyan:2013xsw, Chatrchyan:2013fea, CMS:yut, Chatrchyan:2013gia, CMS:2013qda, Khachatryan:2016iqn,CMS-PAS-SUS-16-013,CMS-PAS-SUS-14-020,Khachatryan:2014lpa}. These limits rely on simplifying model assumptions. 
In particular, 
the mass limits on the lighter stop
assume  in the case of hadronic stop RPV decays
100\% branching ratio into two body final states \cite{Khachatryan:2014lpa,Aad:2016kww,ATLAS-CONF-2016-022}. 
It follows that, apart from the qualitative requirement of prompt decays, the derived 
limits are independent of the size of the RPV couplings themselves, and thus 
insensitive to the experimental limits on the latter \cite{Barbier:2004ez}.
While this assumption is clearly valid if 
the stop were the LSP, it calls for more model-dependence in the opposite case.

It has been pointed out in Ref.~\cite{Evans:2014gfa} that busier final states with high 
$b$-quark multiplicities, not looked for by the LHC experiments so far, can become the 
dominant stop decay channels in regions of the parameter space where part of the 
neutralino/chargino sector is lighter than the lightest stop, thus mitigating the 
present LHC limits. Furthermore, as shown in Refs.~\cite{Liu:2015bma,Csaki:2015uza,Monteux:2016gag}, existing experimental searches performed at the LHC, such as di-jet resonant production, top-quark pairs, four-tops and displaced decays, can be re-interpreted as limits for a class of RPV couplings involving the stop and SM quarks.
  
In the present paper we go a step further by considering extensively the sensitivity to the magnitudes of the RPV 
  couplings for stop-pairs production. 
  This pinpoints the critical role of the size of these couplings in unveiling  the final states 
  that are dominant among all the different combinations of stop decay chains. It also unfolds the experimental strategy to be sensitive to 
  stop pair production and decays, spanning several orders of magnitude for the 
value of the RPV couplings.

The rest of the paper is organized as follows: in Section II we recall the main 
theoretical ingredients of the RPV sector of the Minimal Supersymmetric Standard 
Model (RPV-MSSM), as well as the present LHC limits on RPV stop searches, discussing
possible new search channels, and give the simplifying model assumptions we make. 
In Section III we describe the general features of the stop pair production
and decays, classify all the possible decay channels triggered by R-parity 
violation and motivate the allowed range of the corresponding couplings. 
In Section IV we give an analytical discussion
of the sensitivity to the considered RPV coupling, while in Section V we identify two 
classes of benchmark points of the model.
The stop pair production total cross-section and decay channels for these two benchmark 
points are evaluated in Section VI 
illustrating quantitatively the phenomenological sensitivities to the RPV 
coupling and to the stop-chargino mass splitting.
Section VII is devoted to a discussion of the signal and background issues
for each of the promising final states. Finally, the conclusion and prospectives are given in Section VIII.

\section{Model assumptions}

\subsection{RPV-MSSM}

The superpotential of the RPV-MSSM (see for instance \cite{Barbier:2004ez}) has three distinct parts: 
\begin{equation}
W_{\text{RPV}} = W_{\text{RPC}} + W_{\not{L}} + W_{\not{B}} \ .
\end{equation}
The R-parity conserving part, 
\begin{equation}
W_{\text{RPC}}= (Y^L)_{i j} \hat{L}_i \cdot \hat{H}_1 \, \hat{E}^c_j + (Y^D)_{i j} \hat{Q}_i \cdot \hat{H}_1 \, \hat{D}^c_j 
   + (Y^U)_{i j} \hat{Q}_i \cdot \hat{H}_2 \, \hat{U}^c_j + \mu \hat{H}_2.\hat{H}_1
   \label{eq:WRPC} \ ,
\end{equation}
involves the Yukawa coupling matrices $Y^L, Y^D, Y^U$ and the Higgs mixing
parameter $\mu$.
The R-parity violating part, $W_{\not{L}} + W_{\not{B}}$, 
splits into a lepton number violating sector
involving bilinear and trilinear couplings,
\begin{equation}
W_{\not{L}} = \frac{1}{2} \lambda_{ijk} \hat{L}_i \cdot \hat{L}_j \, \hat{E}^c_k + 
                \lambda'_{ijk} \hat{L}_i \cdot \hat{Q}_j \, \hat{D}^c_k 
                          + \mu_i \hat{L}_i \cdot \hat{H}_2  \label{eq:WLNV},
\end{equation}
 and a baryon number violating sector involving trilinear couplings,
\begin{equation}
W_{\not{B}} = \frac{1}{2} \lambda''_{ijk} {\hat{U}^{\alpha c}_i} 
{\hat{D}^{\beta c}_j} {\hat{D}^{\gamma c}_k} \epsilon_{\alpha \beta \gamma} \label{eq:WBNV}.
\end{equation}

\noindent
The chiral superfields $\hat{L}$ and $\hat{Q}$ denote respectively the lepton and
quark $SU(2)$ doublets, $\hat{E}, \hat{D}$ and $\hat{U}$ the corresponding
singlets, and $\hat{H}_1$ and $\hat{H}_2$ are the two Higgs doublets, together 
with their conventional $U(1)_Y$ hypercharges. Summation over 
repeated indices is understood in all the above expressions where
$\alpha,\beta,\gamma=1,2,3$  denote the $SU(3)$ color indices, the dots 
($A \cdot B \equiv \epsilon_{a b} A^a B^b$) define SU(2) invariants, 
the $i,j,k=1,2,3$  are generation indices, and $c$ indicates charge conjugation.
Also the trilinear RPV couplings should satisfy the relation:
\begin{equation}
\lambda_{ijk} = - \lambda_{jik} \ \mbox{and} \
\lambda''_{ijk} = - \lambda''_{ikj} \ ,
\end{equation}
as an immediate consequence of the antisymmetry of the 
$\epsilon_{a b}$ and $\epsilon_{\alpha \beta \gamma}$ symbols respectively.

Recall that to account for SUSY breaking, assumed to be soft in the visible
sector, the low energy MSSM  is expected to have additional RPC and RPV 
terms in the Lagrangian density with the following general structure, 
\begin{equation}
{\cal L}^{\text{soft}}_{\text{RPC}}= -V^{\text{soft}}_{\text{RPC}} -\frac12 (M_1 \tilde{B}\tilde{B} + M_2 \tilde{W}\tilde{W} + M_3 \tilde{g}\tilde{g}), 
\label{eq:Lsoft-RPC}
\end{equation}
where
\begin{eqnarray}
V^{\text{soft}}_{\text{RPC}}&=& (m_{\tilde{Q}}^2)_{i j} \tilde{Q}_i^\dag \tilde{Q}_j + (m_{\tilde{U}}^2)_{i j} \tilde{U}_i^\dag \tilde{U}_j + (m_{\tilde{D}}^2)_{i j} \tilde{D}_i^\dag \tilde{D}_j +
   (m_{\tilde{L}}^2)_{i j} \tilde{L}_i^\dag \tilde{L}_j + (m_{\tilde{E}}^2)_{i j} \tilde{E}_i^\dag \tilde{E}_j + m_{H_1}^2 |\tilde{H}_1|^2  \nonumber 
\\
 && + m_{H_2}^2 |\tilde{H}_2|^2 + \big((T^l)_{i j} \tilde{L}_i \cdot \tilde{H}_1 \, \tilde{E}^c_j + (T^d)_{i j} \tilde{Q}_i \cdot \tilde{H}_1 \, \tilde{D}^c_j 
   + (T^u)_{i j} \tilde{Q}_i \cdot \tilde{H}_2 \, \tilde{U}^c_j + B_{\mu} \tilde{H}_2.\tilde{H}_1 + \text{h.c.}\big) \nonumber \\
   \label{eq:Vsoft-RPC}
\end{eqnarray}
involves the RPC soft SUSY breaking scalar masses, trilinear couplings and Higgs mixing, 
and
\begin{equation}
{\cal L}^{\text{soft}}_{\text{RPV}}= - V^{\text{soft}}_{\not{L}} - V^{\text{soft}}_{\not{B}},
\label{eq:Lsoft-RPV}
\end{equation}
where
\begin{equation}
V^{\text{soft}}_{\not{L}} = \frac{1}{2} T_{ijk} \tilde{L}_i \cdot \tilde{L}_j \, \tilde{E}^c_k + 
                T'_{ijk} \tilde{L}_i \cdot \tilde{Q}_j \, \tilde{D}^c_k 
                          + B_i \tilde{L}_i \cdot \tilde{H}_2 + \tilde{m}^2_{1 i} \tilde{H}_1^\dag \tilde{L}_i +  \text{h.c.},
                          \label{eq:Vsoft-LNV}
\end{equation}
and
\begin{equation}
V^{\text{soft}}_{\not{B}} = \frac{1}{2} T''_{ijk} {\tilde{U}^{\alpha c}_i} 
{\tilde{D}^{\beta c}_j} {\tilde{D}^{\gamma c}_k} \epsilon_{\alpha \beta \gamma} + \text{h.c.}
\label{eq:Vsoft-BNV}
\end{equation}
involve respectively the lepton and baryon number violating soft SUSY breaking bilinear
and trilinear couplings. 
In Eq.~(\ref{eq:Lsoft-RPC}) the twiddled fields denote the $U(1)_Y$, $SU(2)_L$, and 
$SU(3)$ gauginos in the Weyl representation where we have suppressed the gauge 
indices, and $M_1, M_2, M_3$ denote their soft masses. The fields in   
Eqs.~(\ref{eq:Vsoft-RPC}, \ref{eq:Vsoft-LNV}, \ref{eq:Vsoft-BNV}) are the scalar 
components of the chiral superfields entering the superpotentials (\ref{eq:WRPC} 
--\ref{eq:WBNV}) and the $m^2$'s, $B_\mu, B_i$, $T^{l,d,u}, T,T'T''$ are the bilinear 
and trilinear soft-susy breaking parameters. We define also $\tan \beta \equiv \frac{v_2}{v_1}$, the ratio of the vacuum expectation values developed 
by $H_2$ and $H_1$ after EWSB.

In the sequel we do not rely on specific high scale model assumptions which can trigger the EWSB and correlate the various low-energy 
SUSY preserving and soft breaking parameters, or possibly provide a dynamical origin to the RPV couplings 
\cite{Csaki:2015fea,Mohapatra:2015fua}. 
Given the low-energy phenomenological assumptions we rely on,  
the process of stop production and decays under consideration depends 
only on a reduced 
set of MSSM parameters insensitive to such correlations. Furthermore, we assume conservatively 
minimal flavor violation (MFV) \cite{D'Ambrosio:2002ex}, since the heavy versus light quark content of the final states
is instrumental to our study.

\subsection{LHC searches and new channels}
The likeliness of a relatively light stop, motivated by natural SUSY and a large 
mass splitting between 
the two stop states that could account for the observed Higgs boson mass (at least within the
MSSM), together with the more general expectation that  the third (s)quark generation plays 
a central role in triggering the electroweak symmetry breaking, makes the search for light stops particularly compelling. This is true both in RPC and RPV scenarios.
The present LHC mass limits from direct production in the RPC scenarios are of order 800 GeV \cite{CMS-PAS-SUS-16-029,CMS-PAS-SUS-16-028,ATLAS-CONF-2016-050,ATLAS-CONF-2016-077} and the exploitable range is expected to cross the \TeV~scale towards the end of Run 2. 
Moreover, some of the all-leptonic RPV searches have already increased this limit in some cases 
up to 1020 GeV \cite{Chatrchyan:2013xsw}. 
Lighter stops could however still be hiding in the all-hadronic channels final states with very low missing energy, as would be 
typically the case in RPV scenarios if dominated by baryon number violating couplings $\lambda_{33i}''$, cf. Eq.~(\ref{eq:WBNV}).
Searches for directly produced stop pairs each decaying into one jet originating from
a $b$ and one jet from a light quark with the data collected in 2012 at 
$\sqrt{s}=8$ \TeV~and in 2015 at $\sqrt{s}=13$ \TeV~lead to 
exclusion mass limits in the range 100-380 GeV by the CMS~\cite{Khachatryan:2014lpa} 
and ATLAS~\cite{Aad:2016kww,ATLAS-CONF-2016-022} collaborations. 

Both ATLAS~\cite{ATLAS-CONF-2016-037,Aad:2012ypy,ATLAS-CONF-2016-075,ATLAS-CONF-2016-057,ATLAS-CONF-2016-094} 
and CMS~\cite{Chatrchyan:2013xsw, Chatrchyan:2013fea, CMS:yut, Chatrchyan:2013gia, CMS:2013qda, Khachatryan:2016iqn,CMS-PAS-SUS-16-013,CMS-PAS-SUS-14-020} have also looked for
signatures of RPV scenarios through either gluino decays assuming 
baryon number violating couplings, or squark decays 
assuming lepton number violating bilinear and trilinear couplings. The ensuing mass limits for the gluino and first and second generation squarks range from 800 GeV up to 1.9 \TeV~depending on the model assumptions. 

It is important to keep in mind that the limits quoted above assume the RPV decays to
proceed through the shortest decay chains. In particular the ones on direct production of stops decaying through baryon number violating couplings, 
are derived under the assumption of 100\% decay into a bottom and a light quark.
These limits carry thus some model-dependence
irrespective of whether lepton number violating decays are ignored or not.
As observed in Ref.~\cite{Evans:2014gfa}, if the stop is not the LSP in parts of the 
parameter space motivated by natural SUSY, then its decays 
may become dominated by channels with higher $b$-quark multiplicities.  
In this case, a different experimental strategy is called for when looking for a signal or setting limits, thus putting into more perspective the meaning 
and reach of the present experimental limits on light stops.  
However, it is to be stressed 
that even in Ref.~\cite{Evans:2014gfa} a
100\% decay in the final states under consideration is assumed, this time not for the
decaying mother stop itself but for the subsequent decay of the intermediate on-shell 
chargino present 
in the decay chain. 
As noted in the introduction, such an assumption makes the processes and the experimental
limits insensitive to the magnitude of the relevant RPV couplings. Not only is it desirable to be able to set limits
on these couplings as well, but in fact, in the configurations where {\sl the LSP 
is neither a squark nor a slepton}, the branching fractions
of the various RPV decays of the latter depend necessarily on the
magnitudes of the RPV couplings. That this is to be expected on general grounds can be seen from the simple fact that in the limit of vanishing RPV couplings
the RPC theory should be recovered smoothly.
 Indeed, in this limit, of all the RPV signal processes only the ones that tend to the 
 RPC signals, i.e. containing an on-shell long-lived LSP in the decay chain, will survive. This implies that
when decreasing the RPV couplings a crossover in favor of the decays containing the LSP must occur at some point.
Moreover, in the regions where they become sizable, the latter channels tend to be less sensitive to the RPV couplings
since the LSP decays only through RPV channels, thus with branching ratio $1$ to the relevant final states.
The only limitation is that the RPV couplings should remain sufficiently large for the LSP to decay within the detector,
otherwise the RPC search limits become effective.

Put differently, assuming a branching ratio of $1$ for a given
decay channel implicitly entails a given range of the RPV couplings, that would further depend
on the mass spectrum and RPC couplings of the particles involved in the decay.
This observation has two consequences:
\begin{itemize}
\item[--]while all the quoted present experimental limits on RPV scenarios have obviously some 
model-dependence, the sensitivity to the RPV couplings exacerbates this model-dependence;
\item[--]higher jet and/or lepton multiplicity decays probe smaller (even tiny) RPV couplings
benefiting in the same time from a reduced SM background.
\end{itemize}

The aim of the subsequent sections is to demonstrate the above general features
quantitatively in the case of baryon  number violating RPV couplings $\lambda_{331}''$ or $\lambda_{332}''$ 
that trigger the decay of stops leading to $b$-quarks, light quarks and possibly leptons in the final states.

\subsection{mass spectrum \label{subsec:spectrum}}
In this section we describe the  
simplified working assumptions made in the paper:

\begin{itemize}
\item[\sl (i)] $\lambda_{33i}''$, with $i=1$ or $2$, is the only non-vanishing RPV coupling,
\item[\sl (ii)] the light part of the SUSY spectrum is composed of one stop, one chargino, two neutralinos and the 
lightest CP-even Higgs (referred to respectively as $\tilde{t}, \chi^+, \chi^0/\chi^0_2$
the lighter/heavier neutralino and $h^0$ the SM-like Higgs throughout the paper).
All other SUSY and Higgs particles, except possibly for the gluino, are assumed to be too heavy to be produced at the LHC, 
\item[\sl (iii)] the RPV-MSSM-LSP is the lightest neutralino $\chi^0$.
\end{itemize}

A few comments are in order here. Assumptions {\sl (i), (ii), (iii)} are not mandatory for the validity of the general 
message we convey in this paper regarding the final-state-dependent sensitivity to the RPV couplings. They serve as a 
concrete illustration in one possible physically interesting configuration.  
Assumption {\sl (i)} can be seen as an idealization of some generic 
assumptions such as MFV
where baryon number violating RPV couplings containing 1st and 2nd generation indices are suppressed 
with respect to $\lambda_{332}''$ (or $\lambda_{331}''$) \cite{Nikolidakis:2007fc,Csaki:2011ge}. Alternatively, it could result from a
dynamical collective effect due to the running of several RPV couplings from a common value at some very high scale down to 
the electroweak scale where $\lambda_{332}''$ becomes much larger than the other couplings 
\cite{Mambrini:2001sj,Mambrini:2001ic}. In fact, our analysis does not depend crucially on the single
RPV coupling dominance assumption: indeed, combined with assumption {\sl (ii)}, assumption {\sl (i)} is not particularly 
restrictive given the hadronic final states and parameter space under consideration.  For one thing, $\lambda_{33i}''$ can 
be viewed as accounting for the combination $\sqrt{(\lambda_{332}'')^2 +(\lambda_{331}'')^2}$ since at present hadron colliders light $d$- and $s$-quark productions 
are indistinguishable.\footnote{This correspondence is valid up to indirect effects originating from RPV 
induced loop corrections to the $\tilde{t}$ mass \cite{Chamoun:2014eda}. These effects remain, however, negligibly small in the $\lambda_{33i}''$  range we consider.}
For another, most of the lepton number violating couplings in Eq.~(\ref{eq:WLNV}) 
do not contribute to the final states under 
consideration, or else are irrelevant due to the assumed heaviness of the squarks and sleptons. The only possible exception
is the set of $\lambda_{i j 3}'$ couplings that induce $\tilde{t}$ decays into bottom quark and a lepton. 
This channel would however be suppressed for a small left-handed component of the lightest stop, and in any
case can be vetoed as it leads to final states with leptons and no light quarks, different from the ones we study. 
Finally the baryon number violating couplings $\lambda_{132}'', \lambda_{232}''$
can in principle contribute to final states containing $b$- and light quarks through the flavor mixing of the 3rd generation
with the 1st and 2nd generation squarks (current states). However this mixing is very small for the SUSY spectrum we consider 
which suppresses the sensitivity to these couplings altogether. Thus most of the RPV couplings could still be non-vanishing
without affecting our analysis. Assumption {\sl (ii)} can be motivated on one hand by simplicity, with only  a small part of the MSSM spectrum to deal with phenomenologically, and on the 
other by the need to account for the light CP-even Higgs mass while keeping at 
a relatively moderate level the fine-tuning required to get the electroweak scale from the radiative electroweak symmetry breaking, see e.g. \cite{Papucci:2011wy}. It should be stressed however that the latter naturalness criterion being more a practical guide than a physics principle, the actual realisation of the low lying states of supersymmetry could well be through quite different configurations than the ones motivated by naturalness.

As concerns assumption {\sl (iii)}, obviously not motivated by dark matter issues
since the RPV-MSSM-LSP is unstable and assumed to decay promptly, its aim is to remain 
as close as possible to the conventional spectrum configurationns for which most of the present experimental bounds for RPC scenarios have
been established. In particular this allows  to relate in a well defined way to the latter bounds whenever 
$\lambda_{33i}''$ becomes too small for the $\chi^0$ to decay within the detector. Still it is important from a more 
general perspective to assess the dark matter candidates in the RPV context. We only note here that among the possible 
scenarios a light gravitino, being for that matter the true LSP (leaving the $\chi^0$ as the RPV-MSSM-LSP), can indeed 
provide a good metastable candidate even for moderately large RPV couplings of order $10^{-2}$ or larger, for sufficiently
heavy sfermions \cite{Lola:2007rw, Takayama:2000uz,Moreau:2001sr}. In fact, with assumption {\sl (i)}, a gravitino lighter than twice the $b$-quark mass
would be even totally stable. 

Besides assumptions
{\sl (i), (ii),(iii)}, we focus mainly, though not exclusively, 
on the MSSM parameter regions that are consistent with the following mass configuration:

\begin{eqnarray}
&&m_{\tilde{t}} \gtrsim m_{\chi^0_2} \gtrsim m_{\chi^+} \gtrsim m_{\chi^0} > m_t \ ,
\label{eq:spectconfig1}\\
&&m_{\tilde{t}} - m_{\chi^0}   < m_t \label{eq:spectconfig2} \ ,\\
&&m_{\tilde{t}} - m_{\chi^+}    > m_b   \label{eq:spectconfig3} \ .
\end{eqnarray}
Such a configuration has been already considered in Ref.~\cite{Evans:2014gfa} to illustrate the relevance of multi $b$-quark final states when an on-shell chargino is
present in the stop decay chain. In the present work we stress 
the relevance of the longer decay chain not considered previously, containing on-shell chargino and neutralino, and in particular the importance of the magnitude of $\lambda_{33i}''$ in selecting the stop decay channels that actually dominate. 
Note also the presence of two neutralinos in the low energy spectrum. This is
unavoidable when the chargino/neutralino light sector is assumed to be Higgsino-like
as we do: in the limit $M_1 \simeq M_2 \gg \mu \gg m_W$ and $\tan \beta \gg 1$ one finds $m_{\chi^0_2} - m_{\chi^+} \sim m_{\chi^+}- m_{\chi^0} \simeq \frac{5}{8} \frac{m_W^2}{M_1}$
up to loop corrections, 
which corresponds to a compressed spectrum satisfying the mass hierarchy in the chargino/neutralino sector as given in Eq.~(\ref{eq:spectconfig1}). 
However, as long as the configuration in Eq.~(\ref{eq:spectconfig2}) is satisfied the second neutralino, $\chi^0_2$, 
does not contribute significantly to the stop decay since it enters the decay chain 
only off shell, and is neglected throughout the study.

\section{Stop production and decays\label{sec:st_prod_dec}} 

\subsection{pair production \label{subsec:pair_production}}

The stop pair production at the LHC, $p p \to \tilde{t} \bar{\tilde{t}} + X$, proceeds mainly through gluon-gluon fusion QCD processes, see
\cite{Beenakker:1997ut, Beenakker:2010nq, Beenakker:2011fu} and references therein. While quark-anti-quark partonic 
contributions are subdominant at LHC energies, there could also be interesting
single, or same-sign pair, stop (associated) productions respectively through 
RPV quark-quark 
processes or QCD gluon-gluon processes \cite{Durieux:2013uqa}, \cite{Monteux:2016gag}. Some of these channels are 
suppressed in our case, either because $\lambda_{3 k i}''$ with $k \neq 3$ are 
assumed to be vanishing or due to the assumed heaviness of the gluino and first and second squark 
generations. The single stop production and decays can already constrain parts of the parameter space for a light LSP as shown in 
\cite{Monteux:2016gag}.
Note however that the corresponding production cross-section becomes subdominent as compared to the pair production when 
$\lambda_{3 3 i}''$ is
  taken $\lesssim$ $O(10^{-2})$ and 
  $m_{\tilde{t}} \gtrsim 500~\GeV$, and even totally suppressed for the much smaller values of $\lambda_{3 3 i}''$ 
  that we consider in this paper. 
  
 \subsection{RPV final states}
\label{sec:multi-jet_final_states}

Given the mass configurations described in Eqs.~(\ref{eq:spectconfig1}--\ref{eq:spectconfig3}), the leading RPV and RPC $\tilde{t}$ decays are respectively 
 $\tilde{t} \to \bar{b} \bar{d_i}$ and
$\tilde{t} \to \chi^+ b$, where $d_i$ with $i=1,2$, denotes respectively the $d$- and $s$-quark.  
Other decay channels such as 
$\tilde{t} \to \chi^0 t^* \to \chi^0 b f_1 \bar{f_1'}$ or
$\tilde{t} \to h^0 \tilde{t}^* \to h^0 \bar{b} \bar{d_i}$ 
(where $f_1$ and $\bar{f_1'}$ indicate SM fermions and the star off-shell states), 
are suppressed
by the off-shellness of the (s)top quark. Note also that a potential enhancement
of the Higgs channel by large soft-susy breaking trilinear coupling is suppressed
when the $\tilde{t}$ is essentially right-handed.
The subsequent leading RPV induced $\chi^+$ decays are 
$\chi^+ \to \tilde{t}^* \bar{b} \to \bar{b} \bar{b} \bar{d_i}$ and 
the much longer chain
 $\chi^+ \to \chi^0 W^{+ *} \to \tilde{t}^* \bar{t} (\tilde{\bar{t}}^* t)   W^{+ *} \to 
 \bar{b} \bar{d_i} \bar{t} (b d_i t)  f \bar{f}' $ with the top decaying ultimately
 to $b f_1 \bar{f}_1'$ and where we assumed $\chi^0$ decays through the shortest 
possible chain.
The latter decay, $\chi^0 \to \tilde{t}^* \bar{t}^{}(\tilde{\bar{t}}^* t)  \to \bar{b} \bar{d_i} \bar{b} (b d_i b)  f_1 \bar{f}_1'$,
is indeed dominant as a consequence of assumption {\sl (ii)} of Section \ref{subsec:spectrum}.
The other equally short chain
$\chi^0 \to \tilde{b}^* b \big(\tilde{\bar{b}}^* \bar{b}\big)\to  \bar{t}^{(*)} \bar{d_i} b \big(t^{(*)} d_i \bar{b}\big)$ is suppressed for sufficiently  
heavy  $\tilde{b}$. The longer chains  
$\chi^0 \to {\chi^+}^* W^{-(*)} \big({\chi^-}^* W^{+(*)}\big)\to \tilde{t}^* \bar{b}  
W^{-(*)} \big(\tilde{\bar{t}}^* b W^{+(*)}\big) \to  \bar{b} \bar{d_i} \bar{b} (b d_i b) f \bar{f}'$ or  $\chi^0 \to {\chi^+}^* W^{-(*)}
\big({\chi^-}^* W^{+(*)}\big) \to \tilde{\bar{b}}^* t  
W^{-(*)} \big(\tilde{b}^* \bar{t}  
W^{+(*)}\big)\to  t d_i t (\bar{t}  \bar{d_i} \bar{t}) f \bar{f}'$ are obviously even further suppressed.
 
 We have thus at hand the three different
 decay channels depicted in Figs.~\ref{fig:feyndiagrams} (a), (b) and (c). 
We refer to these respectively as \bftRP, \bfcRP and \bfRPl, 
 to stress the fact that channel (a) is the direct RPV stop decay,
channel (b) the shortest RPV cascade containing an (on-shell) 
chargino, and channel (c), defined as having an {\sl on-shell} $\chi^0$ intermediate state,
corresponds to the only surviving channel in the RPC limit
$\lambda_{33i}'' \to 0$. 
It is to be noted that the latter channel has not been 
 considered in \cite{Evans:2014gfa}.
 
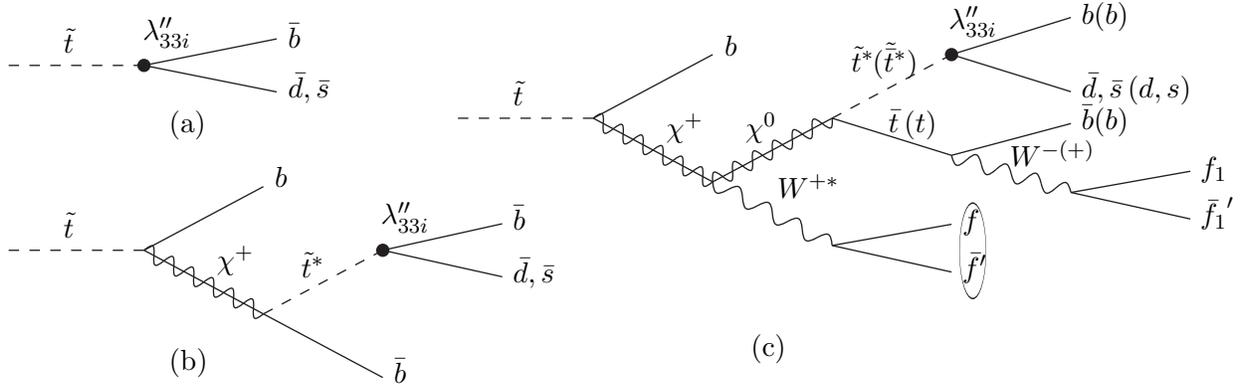
\begin{figure}[h]
\begin{picture}(300,150)(0,0)
\put(-150,100){\begin{picture}(300,100)(0,0)
\put(90, 35){$\tilde{t}$}
\DashLine(69, 30)(120, 30){4}
\put(175, 37){$\bar{b}$}
\Line(120, 30)(170, 40)
\put(175, 18){$\bar{d},\bar{s}$}
\Line(120, 30)(170, 20)
\put(120,40){$\lambda''_{33i}$}
\Vertex(120,30){2.5}
\put(130, 5){(a)}
\end{picture}}
%
\put(-150,30){\begin{picture}(300,100)(0,0)
\put(90, 35){$\tilde{t}$}
\DashLine(69, 30)(120, 30){4}
\put(170, 54){$b$}
\Line(120, 30)(165, 54)
\put(148, 23){$\chi^+$}
\Line(120, 30)(165, 6)
\Photon(120,30)(165, 6){3}{6}
\put(180, 20){$\tilde{t}^*$}
\DashLine(165, 6)(210, 30){4}
\put(260, 38){$\bar{b}$}
\Line(210, 30)(255, 40)
\put(260,  18){$\bar{d},\bar{s}$}
\Line(210, 30)(255, 20)
\put(210,40){$\lambda''_{33i}$}
\Vertex(210,30){2.5}
\put(215,  -20){$\bar{b}$}
\Line(165, 6)(210, -18)
\put(130, -15){(b)}
\end{picture}}

%
\put(20,80){\begin{picture}(300,100)(0,0)
\put(90, 35){$\tilde{t}$}
\DashLine(69, 30)(120, 30){4}
\put(170, 54){$b$}
\Line(120, 30)(165, 54)
\put(148, 23){$\chi^+$}
\Line(120, 30)(165, 6)
\Photon(120,30)(165, 6){3}{6}
\put(178, 23){$\chi^0$}
\Photon(165, 6)(210, 30){3}{6}
\Line(165, 6)(210, 30)
\put(218, 48){$\tilde{t}^*\!\!\!$}
\put(226, 48){$({\tilde{\bar t}^*})$}
\DashLine(210, 30)(255, 54){4}
\put(305, 66){$\bar{b} (b)$}
\Line(255, 54)(300, 68)
\put(305, 38){$\bar{d}, \bar{s} \, (d,s)$}
\Line(255, 54)(300, 40)
\put(232,  25){$\bar t$}
\put(238,  25){$(t)$}
\Line(210, 30)(255, 16)
\put(305, 26){$\bar{b} (b)$}
\Line(255,16)(300, 28)
\put(278, 12){$W^{-(+)}$}
\Photon(255, 16)(300, 2){-3}{4}
\Line(300, 2)(345, 10)
\put(350,  8){$f_1$}
\Line(300, 2)(345, -8)
\put(350,  -10){$\bar{f_1}'$}
\put(255,64){$\lambda''_{33i}$}
\Vertex(255,54){2.5}
\put(190, 0){$W^{+ *}$}
\Photon(165, 6)(210, -18){-3}{4}
\Line(210, -18)(255, -10)
\put(260,  -12){$f$}
\Line(210, -18)(255, -28)
\put(260,  -30){$\bar{f}'$}
\Oval(263,-20)(18,5)(0)
\put(180, -60){(c)}
\end{picture}}
\end{picture}
\caption{\small{Leading RPV stop decays assuming 
Eqs.~(\ref{eq:spectconfig1}-\ref{eq:spectconfig3}); (a): direct RPV stop decay (\bftRP),
(b): shortest RPV cascade containing an (on-shell) chargino (\bfcRP), (c): shortest 
RPV cascade containing an (on-shell) neutralino (\bfRPl); $f, f', f_1, f_1'$ denote SM fermions and the oval encircles fermions 
too soft to be detected.} 
\label{fig:feyndiagrams}}
\end{figure}

Note that because of the  nearly mass degenerate chargino and neutralino in our scenario, 
off-shell $W$ bosons from the \bfRPl stop decay chain are produced with a too small 
transverse momentum for their decay products to be reconstructed in Hight Energy Physics detectors. 
These are thus ignored in the following.\\

Since jets electric charges cannot be discriminated experimentally, 
we tag the various final states by their flavor content as follows:
\begin{itemize}
\item[$\bullet$] \bftRP $\equiv 1b1j$,
\item[$\bullet$] \bfcRP $\equiv 3b1j$,
\item[$\bullet$] \bfRPl $\equiv 1t2b1j$,
\end{itemize}
where $b$ ($t$) stands for the presence of a bottom-quark jet (top-quark) and $j$ indicates the presence of a light-quark jet. 
Since the \bfRPl channels are characterized by the presence
of a top quark in the decay chain followed by SM top decays, we have indicated only the presence of the top quark.
We are thus left effectively with six different categories
of final states corresponding to the decays of the produced stop and anti-stop as summarized in 
Table~\ref{tab:finalstates}. Final states with the same particle content (but opposite charges) 
are not duplicated in the table. We however continue to indicate explicitly the charges for definiteness 
when discussing the analytical structure of the cross-sections in Section \ref{subsec:NWA}. 
 
\begin{table}[!h]
\begin{center}
\renewcommand{\arraystretch}{1.5}
\begin{tabular}{|c||c|c|c|} \hline\hline
\backslashbox[0mm]{~~~~~$\bar{\tilde{t}}$ \\~~}{\\~~ \\$\tilde{t}$~~~~~} &  \bftRP & \bfcRP & \bfRPl\\ \hline \hline
 \bftRP   &  $2b2j$ &  $4b2j$ & $1t3b2j$\\  \hline
 \bfcRP  &  & $6b2j$  & $1t5b2j$  \\  
 \cline{1-1} \cline{3-4}
 \bfRPl   &  \multicolumn{1}{c}{~}  & & $2t4b2j$  \\ 
 \hline\hline
\end{tabular}
\end{center}
\caption{ \label{tab:finalstates} \small{The various final states corresponding to 
different contents of heavy ($b,t$) quarks, and light ($d$,$s$) quarks denoted 
generically by $j$, 
originating from the stop--anti-stop RPV decays; similar final states 
corresponding to interchanging the stop and anti-stop decays leading to the same 
particle content (irrespective of the electric charges) are listed only once.}}
\end{table}

\subsection{The $\lambda_{33i}''$ range}
\label{sec:RPVrange}
There exists a large set of upper bounds on the RPV couplings 
(see \cite{Barbier:2004ez} for a detailed discussion), some of which involve  $\lambda_{33i}''$.
Together with assumption {\sl (i)} of Sec.~\ref{subsec:spectrum}, we allow in the sequel
$\lambda_{33i}''$ to vary in the range
 
\begin{equation}
 10^{-7} \lesssim |\lambda_{33i}''| \lesssim 10^{-1} \ . 
\label{eq:RPVrange}
\end{equation}
Experimental upper bounds on $\lambda_{331}''$ and $\lambda_{332}''$ are typically weaker than the ones involving
only first and second generation, let alone the bounds on the lepton number violating couplings. Moreover, most of these bounds 
are on products of $\lambda_{33i}''$ with other RPV couplings. Such bounds can thus be easily satisfied through 
assumption {\sl (i)} of Sec.~\ref{subsec:spectrum}. 
There are also upper bounds set directly on $\lambda_{332}''$ and/or $\lambda_{331}''$, coming from constraints on 
the $Z$-boson hadronic width, neutron--anti-neutron oscillations and single nucleon decays: the first is ${\cal O}(1)$, the 
second and the third are model-dependent and are made easily compatible with the upper bound in Eq.~(\ref{eq:RPVrange}) for squark masses 
$\gtrsim 500$~GeV (even more so for single nucleon decays assuming a gravitino mass 
$\gg1$~eV or an axion scale 
$\gtrsim 10^{10}$~GeV). Likewise, the upper bound in Eq.~(\ref{eq:RPVrange}) can be easily made compatible 
with bounds on the product $|\lambda_{331}'' (\lambda_{332}{''})^*|$ obtained from $K^0 - \bar{K^0}$ mixing for squark masses
in the few hundred GeV range. All in all, the upper bound of Eq.~(\ref{eq:RPVrange})
is only taken as a working assumption and could in principle be somewhat larger.  
Note however that values of $\lambda_{33i}''$ much larger than $10^{-1}$ would lead to too large and 
negative loop corrections to the squared stop mass \cite{Chamoun:2014eda}.   

The lower bound in Eq.~(\ref{eq:RPVrange}) is an estimate of the magnitude of 
$\lambda_{33i}''$ that guarantees decays within the detector. Since in the 
configuration under study the lightest stop is not the 
lightest MSSM particle, one should consider not only the lifetime due to direct 
RPV two-body decay of the stop, Fig.~\ref{fig:feyndiagrams}(a), but also that of the daughter chargino due to its decay as given in Fig.~\ref{fig:feyndiagrams}(b),
or the neutralino due to its decay as given in Fig.~\ref{fig:feyndiagrams}(c).  
In the absence of any prior about which channel among the \bftRP, \bfcRP or \bfRPl
is dominant one should consider the most conservative bound, i.e. the longest decay
length. The various $c\tau$'s are (approximately) given by
\begin{equation}
c\tau_{\tilde{t} \to b d_i}{\rm [meter]} \simeq \frac{8.3\times 10^{-18}}{|\lambda_{33i}''|^2} \Big(\frac{600 \GeV}{m_{\tilde{t}}}\Big),
\label{eq:ctau-stop}
\end{equation}
for the direct RPV stop decay, where $d_{1,2}$ denote the first and second generation down quarks, and

\begin{eqnarray}
c\tau_{\chi^0 \to t b d_i}{\rm [meter]} &\simeq& \frac{
2.6 \times 10^{-16}}{\alpha_{\chi^0} 
|\lambda_{33i}''|^2}  \Big(\frac{m_{\tilde{t}}}{600 \GeV} \Big)^4 \Big(\frac{500 \GeV}{m_{\chi^0}}\Big)^5 
\left((1 - r^4 ) \, 
        (1 - 8 r^2 + r^4 ) - 
       24 \, r^4 \log r
\right)^{-1}, \nonumber \\ 
      \label{eq:ctau-neutralino} 
\end{eqnarray}
for the Higgsino component of the neutralino RPV decay  where we defined  
$\displaystyle \alpha_{\chi^0} \equiv \frac{g_{\chi^0}^2}{4 \pi}$, $g_{\chi^0}$ denoting 
the $\chi^0-\tilde{t}-t$ coupling, and  $\displaystyle r \equiv \frac{m_t}{m_{\chi^0}}$ where $m_t$ is the top mass, 
and neglected  $b$- and light quark masses.\footnote{In deriving these expressions 
we included consistently the color factors, 
averaged over the spin of the decaying particle and assumed the lightest 
stop to be essentially right-handed. (Note that some simple formulae for the neutralino decay length 
in the literature, e.g. Eq.(7.6) of Ref.~\cite{Barbier:2004ez}, assume a pure photino content and do not apply in our case.)
 We also rely on the simplifying assumption of  
 instantaneous decay at the mean lifetime, and 
travel of the decaying particle close to ($70$\% of) the speed of light in the laboratory frame. 
A more accurate evaluation of the decay lengths should take into account boost factors from the actual 
mass and energy distributions of the decaying particles produced at various energies at the partonic level, as well
as their lifetimes distribution.}
In Eq.~(\ref{eq:ctau-neutralino}) we approximate the stop propagator by a point interaction
which leads to an overestimate of the decay length and thus to a safe 
conservative bound, but we provide the exact integral over the three-body phase space
taking into account the matrix element spinorial structure of the final state.
The $c\tau$ corresponding to the chargino decay $\chi^+ \to b b d_i$ is given by
$2 \times c\tau_{\chi^0 \to t b d_i}$ in the limit $m_t \to 0$ and with the proper
substitution of chargino mass and coupling, where the global factor two difference
between the two $c\tau$'s is due to the majorana nature of $\chi^0$.
From Eqs.~(\ref{eq:ctau-stop}, \ref{eq:ctau-neutralino}) one has generically the hierarchy
\begin{equation}
c\tau_{\tilde{t} \to b d_i} \ll c\tau_{\chi^+ \to b b d_i} \lesssim
c\tau_{\chi^0 \to t b d_i} \label{eq:ctau-hierarchy} \ ,
\end{equation}
if $m_{\tilde{t}} > m_{\chi^0} \simeq m_{\chi^+} \lesssim 550 $~\GeV~and $\alpha_{\chi^0}, \alpha_{\chi^+} < 1$. The lower bound for $|\lambda_{33i}''|$ is 
thus determined by the decay length of the neutralino provided that
it corresponds to values of $|\lambda_{33i}''|$ for which the stop decays containing
a neutralino indeed dominate.

With a fiducial region of $c\tau \lesssim 3$~meters and taking $m_{\tilde{t}}= 600 $~\GeV, one has from Eq.~(\ref{eq:ctau-stop}) 
the lower bound $|\lambda_{33i}''| \gtrsim 1.6 \times 10^{-9}$, while varying $m_{\chi^0} \simeq m_{\chi^+}$ in the range $(600 ~\text{GeV} -m_t)$  to $600 $~\GeV, one obtains from Eq.~(\ref{eq:ctau-neutralino}) 
with a typical $\alpha_{\chi^0} \simeq 10^{-2}$ the stronger bound $|\lambda_{33i}''| \gtrsim (0.8$ -- $2.4) \times 10^{-7}$. Of course, lighter stop and neutralino lead to more
stringent lower bounds, e.g. $m_{\tilde{t}} = 400 $~\GeV~and $m_{\chi^0} = m_{\tilde{t}} - m_t$ would require
$|\lambda_{33i}''| \gtrsim 3.4 \times 10^{-6}$. 
However, a stop that light becomes barely compatible with our assumption that it is heavier than a chargino, since such a 
low mass configuration
would start conflicting with limits on rare $B$-decays (see also the discussion in Section \ref{subsec:LE_constraints}).

When $m_{\chi^0} \simeq m_{\chi^+} \gtrsim 560 $~\GeV~but still smaller than the 
stop mass, the 3-body phase space reduction 
in the $\chi^0$ decay width as compared to that in the $\chi^+$ decay width, 
does not compensate anymore for the factor two difference between the two
widths. As a result, 
the hierarchy of the chargino and neutralino $c\tau$'s is reversed with
respect to Eq.~(\ref{eq:ctau-hierarchy}).   
However, the relevant lower bound
for $|\lambda_{33i}''|$ is still determined by the decay length of the 
neutralino. Indeed the chargino 
becomes detector-stable typically also for $|\lambda_{33i}''| = {\cal O}(10^{-7})$, where, as shown in the following Sections, the stop decay channels not containing 
a neutralino become highly suppressed. 

Finally, note that we neglect altogether
the gravitationally induced direct stop decay into a top-quark and a gravitino. This
channel could lead to large missing energy in the final state. However, it is
Planck scale suppressed unless the gravitino mass is in the deep
sub-eV range \cite{Aad:2015zva}. As noted previously in this section, a gravitino much lighter than $1$\eV~is disfavored
by proton decay bounds, otherwise $\lambda''_{331 }$ and $\lambda''_{332}$ would have to be typically much smaller than ${\cal O}(10^{-7})$ 
where the LHC exclusion limits on RPC  signatures apply. This suggests a rather heavy gravitino, for which stop decays with 
missing energy are not significant, and which is moreover welcome in scenarios of gravitino 
dark matter. One should however keep in mind that such stringent individual upper bounds  on 
$\lambda''_{331 }$ and $\lambda''_{332}$ from
proton decay \cite{Choi:1998ak}, can be relaxed through possible destructive interference if the two RPV couplings
are allowed to be simultaneously non-vanishing, thus bringing them again within the lower part of the 
range given in  Eq.~(\ref{eq:RPVrange}).\footnote{In such configurations where the decay into gravitinos can be comparable
to the RPV decays, one could make use of the very different scaling in $m_{\tilde{t}}$ in the $c\tau$'s, namely $m_{\tilde{t}}^4$
for the \bfRPl~decay, as compared to $m_{\tilde{t}}^{-5}$ or $m_{\chi^0}^{-5}$ for the stop or the neutralino decaying into gravitinos, 
to extract information from limits on both prompt decays and displaced vertices, see e.g. \cite{Liu:2015bma}.}

More generally, recasting experimental LHC limits on long-lived particle searches \cite{Aad:2015rba,CMS:2014wda} as done in 
\cite{Csaki:2015uza,Monteux:2016gag},  
constrains the various $c\tau$'s to be in the millimeter range. Although the latter studies do not compare directly to ours, as they scan
different mass spectra configurations, a $c\tau \simeq 3$mm for a decaying chargino LSP of $600$\GeV~\cite{Csaki:2015uza} would 
increase the lower bound in Eq.~(\ref{eq:RPVrange}) to $\simeq  2.5 \times 10^{-6}$.

\section{Narrow Width Approximation \label{subsec:NWA}}

A key point is the relative magnitudes of the various cross-sections and 
their sensitivities to 
$\lambda''_{33i}$. By looking at Fig.~\ref{fig:feyndiagrams}, one could naively expect the six channels listed in  
Table~\ref{tab:finalstates} to all scale similarly with $(\lambda''_{33i})^4$. If this were the case, then the relative magnitudes of the corresponding cross-sections would 
not to be affected by $\lambda''_{33i}$, and  
the longer chains would yield smaller cross-sections due to phase space effects 
as well as to matrix elements suppression by other couplings and intermediate propagators. 
There is in fact much more to it if one takes into account total widths and branching ratios of 
the unstable intermediate particles. This section is devoted to an 
analytical study of these features.
To help understand the sensitivity to the RPV coupling we derive 
the expressions for the cross-sections of the various stop decay channels relying on 
the narrow width approximation (NWA), see e.g. \cite{DeWit:1986it}. It is well-known that 
the NWA is not
always quantitatively reliable. In particular it can fail
not only when couplings are large leading to
large widths, but also
for mass configurations similar to the ones we are considering in this paper, even for small
couplings, that is when daughter and parent particles are very close in mass and the
effective center of mass energy at the partonic level is of the same order as 
(twice) the parent 
particle mass \cite{Berdine:2007uv,Kauer:2007zc, Kauer:2007nt,Uhlemann:2008pm}. 
The quantitative analysis in the subsequent Sections will thus not rely on this 
approximation. Nonetheless, the NWA renders reasonably well the qualitative behavior, 
providing a physical understanding of the effects. Moreover in the configurations where the 
NWA is expected to be valid, a very good quantitative agreement with the 
numerical simulation based on exact matrix element calculation gives a significant
cross-check of the results.

Following the discussion in Section~\ref{sec:multi-jet_final_states}, the predominant 
decay chain for the RPV-MSSM-LSP is $\chi^0 \to \tilde{t}^* \bar{t}^{}(	\tilde{\bar{t}}^* t)  \to \bar{b} \bar{d_i} \bar{b} (b d_i b)  f_1 \bar{f}_1'$. 
We can thus take, irrespective of the mass hierarchy involving $\tilde{t}$ and $\chi^+$:
\begin{equation}
BR\big(\chi^0 \to 	\tilde{t}^* \bar{t}^{}(	\tilde{\bar{t}}^* t)  \to \bar{b} \bar{d_i} \bar{b} (b d_i b)  f_1 \bar{f}_1'\big) \approx 1 \ .
\label{eq:Br_neutralino}
\end{equation}
To be specific we first derive the various expressions under the assumptions
$\lambda_{332}''\neq 0, \lambda_{331}''=0$, and $d_i=s$ (i.e. $i=2$).  
Defining 
\begin{eqnarray}
\Gamma_{\tilde{t}\text{-RPV}} &\equiv& \Gamma (\tilde{t} \to \bar{b} \bar{s}) 
\label{eq:tRP_def} \\
\Gamma_{\chi\!\text{-RPV}} &\equiv& \Gamma (\tilde{t} \to \bar{b} \bar{s} \bar{b} b)  \\
\Gamma_{\text{RPC-like}}  &\equiv& \Gamma (\tilde{t} \to \bar{b} \bar{s} \bar{b} (b s b) f_1 
\bar{f_1'} b f \bar{f'}),
\end{eqnarray}
the NWA allows to write,  
\begin{eqnarray}
\Gamma_{\chi\!\text{-RPV}} & \simeq & \Gamma (\tilde{t} \to \chi^+ b) \times BR(\chi^+ \to \bar{b} \bar{s} 
\bar{b})  
\label{eq:cRP} \\
\Gamma_{\text{RPC-like}}  & \simeq & \Gamma (\tilde{t} \to \chi^+ b) \times BR(\chi^+ \to \bar{b} \bar{s} 
\bar{b} (b s b) f_1 \bar{f_1'}  f \bar{f'}) \label{eq:RPC0} \\
& \simeq &   \Gamma (\tilde{t} \to \chi^0 f \bar{f'} b) \times BR(\chi^0 \to \bar{b} 
\bar{s} \bar{b} (b s b)f_1 \bar{f_1'}) \nonumber \\
& \simeq &   \Gamma (\tilde{t} \to \chi^0 f \bar{f'} b)
\label{eq:RPC} 
\end{eqnarray}
where we made use of Eq.~(\ref{eq:Br_neutralino}) when writing Eq.~(\ref{eq:RPC}). 
Moreover, the fact that $\chi^+$ decays with branching ratio $\simeq 1$   
into $\bar{b} \bar{s} \bar{b}$ and $\bar{b} \bar{s} \bar{b} (b s b) f_1 \bar{f_1'}  f \bar{f'}$ leads through 
Eqs.~(\ref{eq:cRP}, \ref{eq:RPC0}) to
\begin{equation}
 \Gamma_{\chi\!\text{-RPV}} + \Gamma_{\text{RPC-like}}  \simeq \Gamma (\tilde{t} \to \chi^+ b) \simeq  ``\lambda_{332}''{\rm -independent}" \ . \label{eq:cRP+RPC}
\end{equation}
A residual sensitivity to $\lambda_{332}''$ in $\Gamma_{\chi\!\text{-RPV}} + \Gamma_{\text{RPC-like}}$ would still come from loop contributions to the
stop mass itself that enters $\Gamma (\tilde{t} \to \chi^+ b)$.
However this higher order effect is essentially screened for the range
$\lambda_{332}'' \lesssim 0.1$ under consideration.
Therefore, the only significant dependence on the RPV coupling in the stop total
width\footnote{ 
neglecting flavor violating transitions such as $\tilde{t} \to \chi^+ s$
and the decay channels 
$\tilde{t} \to \chi^0 t^* \to \chi^0 b f_1 \bar{f_1'}$ or
$\tilde{t} \to h^0 \tilde{t}^* \to h^0 \bar{b} \bar{s}$ as noted in
Section~\ref{sec:multi-jet_final_states}.},  
$\Gamma_{\tilde{t}\text{-RPV}} + \Gamma_{\chi\!\text{-RPV}} + \Gamma_{\text{RPC-like}}$, 
originates from the two body stop decay which can be parametrized as follows, 
\begin{equation}
\Gamma_{\tilde{t}\text{-RPV}}= (\lambda_{332}'')^2 \times \Gamma_1(\tilde{t} \to \bar{b} \bar{s}), \label{eq:stop2body}
\end{equation}
with the notation
\begin{equation}
\Gamma_1 \equiv \Gamma_{|\lambda_{332}''=1}\ .
\end{equation}
We now show that the longest decay chain width $\Gamma_{\text{RPC-like}}$ is not always negligible with respect to $\Gamma_{\chi\!\text{-RPV}}$ or $\Gamma_{\tilde{t}\text{-RPV}}$ and can even overpower these.
The relative magnitude of $\Gamma_{\text{RPC-like}}$ and  $\Gamma_{\chi\!\text{-RPV}}$ is controlled by that of $BR(\chi^+ \to \bar{b} \bar{s} \bar{b})$ and 
$BR(\chi^+ \to \bar{b} \bar{s} \bar{b} (b s b) f_1 \bar{f_1'}  f \bar{f'})$ through 
Eqs.~(\ref{eq:cRP}, \ref{eq:RPC0}), where the relative magnitude of the latter branching
ratios depends on the value of $\lambda_{332}''$. 
Indeed,
on the one hand the NWA and Eq.~(\ref{eq:Br_neutralino}) imply that 
\begin{equation}
\Gamma (\chi^+ \to \bar{b} \bar{s} \bar{b} (b s b) f_1 \bar{f_1'} f \bar{f'}) =
\Gamma (\chi^+ \to \chi^0 f \bar{f'}) \times BR(\chi^0 \to \bar{b} \bar{s} \bar{b} (b s b) f_1 \bar{f_1'}) \simeq \Gamma (\chi^+ \to \chi^0 f \bar{f'}), \label{eq:cGammalong}
\end{equation}  
showing that $\Gamma (\chi^+ \to \bar{b} \bar{s} \bar{b} (b s b) f_1 \bar{f_1'} f \bar{f'})$ is essentially $\lambda_{332}''$ 
independent and is {\sl identical} to the $\chi^+$ width of the RPC case $\Gamma (\chi^+ \to \chi^0 f \bar{f'})$. On the other hand,
since the stop is {\sl off-shell} in the decay $\chi^+ \to \bar{b} \bar{s} \bar{b}$, obviously the corresponding width
scales with $(\lambda_{332}'')^2$,  
\begin{equation}
\Gamma (\chi^+ \to \bar{b} \bar{s} \bar{b}) = (\lambda_{332}'')^2 \times \Gamma_1 (\chi^+ \to \bar{b} \bar{s} \bar{b}) \label{eq:cGammashort} \ .
\end{equation}
Let us now define the following two ratios, 
\begin{eqnarray}
r_1  & \equiv & \frac{\Gamma_1 (\tilde{t} \to \bar{b} \bar{s})}{\Gamma (\tilde{t} \to \chi^+ b)} \ ,\label{eq:r1} \\
r_2 & \equiv & \frac{\Gamma_1 (\chi^+ \to \bar{b} \bar{s} \bar{b})}{\Gamma (\chi^+ \to \bar{b} \bar{s} \bar{b} (b s b) f_1 \bar{f_1'} f \bar{f'})}=\frac{\Gamma_1(\chi^+ \to \bar{b} \bar{s} \bar{b})}{\Gamma 
(\chi^+ \to \chi^0 f \bar{f'})} \ , \label{eq:r2}
 \end{eqnarray}
 that are essentially 
 $\lambda_{33i}''$ independent (apart from a very small sensitivity in the loop 
correction to the stop mass, as noted previously), and determined mainly by the RPC parameters of the MSSM.
The dependence of the chargino decay branching ratios on $\lambda_{332}''$ follows then 
 easily from Eqs.~(\ref{eq:cGammalong}, \ref{eq:cGammashort}, \ref{eq:r2}), 
 \begin{eqnarray}
 BR(\chi^+ \to \bar{b} \bar{s} \bar{b}) &=& \frac{ r_2 \times (\lambda_{332}'')^2}{ 1 + r_2 \times (\lambda_{332}'')^2} \label{eq:Brchar1} 
 \ , \\
BR(\chi^+ \to \bar{b} \bar{s} \bar{b} (b s b) f_1 \bar{f_1'} f \bar{f'}) &=& \frac{1}{ 1 + r_2 \times (\lambda_{332}'')^2} \ . \label{eq:Brchar2}
\end{eqnarray}
 It is clear from these expressions that for sufficiently 
 small $\lambda_{332}''$ the \bfRPl decay $\chi^+ \to \bar{b} \bar{s} \bar{b} (b s b) f_1 \bar{f_1'} f \bar{f'}$ becomes comparable
 or even dominates the R{PV} decay
 $\chi^+ \to \bar{b} \bar{s} \bar{b}$. Upon use
 of Eqs.~(\ref{eq:cRP}, \ref{eq:RPC0}, \ref{eq:stop2body}) the same conclusion holds 
 for the stop widths: the size of $\lambda_{332}''$ controls the relative magnitudes of $\Gamma_{\tilde{t}\text{-RPV}}$, $\Gamma_{\chi\!\text{-RPV}}$ and $\Gamma_{\text{RPC-like}}$,  the latter becoming largely dominant for a very small
 RPV coupling!

We note in passing that the form of Eq.~(\ref{eq:RPC}) might wrongly suggest
that $\Gamma_{\text{RPC-like}}$ is $\lambda_{332}''$
independent. In fact the $\lambda_{332}''$ dependence in $\Gamma (\tilde{t} \to \chi^0 f \bar{f'} b)$ is encoded in the
total width of $\chi^+$, or equivalently in $BR(\chi^+ \to \bar{b} \bar{s} 
\bar{b} (b s b)f_1 \bar{f_1'}  f \bar{f'})$. This should be
contrasted with $\Gamma(\chi^+ \to \bar{b} \bar{s} 
\bar{b} (b s b) f_1 \bar{f_1'}  f \bar{f'})$ which is independent of 
$\lambda_{332}''$.  

 Using the above results, it is now straightforward to express the stop decay branching ratios, and the stop pair production and decay cross-sections, in terms of $\lambda_{332}''$, $r_1$ 
 and $r_2$. Before doing so, we note first that
 all the above steps remain valid if  $\lambda_{332}''$ is replaced
 by $\lambda_{331}''$ and  the $s$- replaced by the $d$-quark, but also when 
 both couplings
 $\lambda_{331}''$ and $\lambda_{332}''$ are simultaneously non-vanishing.  
 Since the difference between the $d$- and $s$-quark masses is irrelevant, the 
 ratios $r_1$ and $r_2$ are essentially unchanged
 when replacing the $s$- by a $d$-quark. 
 The general case, summing up the $s$ and $d$ contributions, is thus obtained by simply replacing $\lambda_{332}''$ by $\lambda_
 {33i}''$ 
 with
 \begin{equation}
 \lambda_{33i}'' \equiv \sqrt{(\lambda_{332}'')^2 +(\lambda_{331}'')^2} \label{eq:lambdaeff} \ ,
 \end{equation}
 in the above formulae.
  Putting everything together one finds the following general
 form for the stop pair production and decay cross-sections:
\begin{itemize}
\item \bftRP--\bftRP $\equiv 2b2j$,
\begin{eqnarray}
\sigma (2b2j) &  \simeq &  \sigma(p p \to \tilde{t} \bar{\tilde{t}}) \times 
BR( \tilde{t} \to \bar{b} \bar{d_i} ) \times BR(\bar{\tilde{t}} \to b d_i) \nonumber \\
& \simeq & \sigma(p p \to \tilde{t} \bar{\tilde{t}}) \times 
\frac{r_1^2 \times (\lambda_{33i}'')^4}{\big(1 + r_1 \times (\lambda_{33i}'')^2\big)^2} \ , \label{eq:four_jets}
\label{eq:2b2j}
\end{eqnarray}
with
\begin{equation}
BR( \tilde{t} \to \bar{b} \bar{d_i}) = \frac{\Gamma_{\tilde{t}\text{-RPV}}}{\Gamma_{\tilde{t}\text{-RPV}}
+ \Gamma_{\chi\!\text{-RPV}} + \Gamma_{\text{RPC-like}}} \ . \nonumber 
\end{equation}
\item \bftRP--\bfcRP $\equiv 4b2j$,
\begin{eqnarray}
\sigma (4b2j)
&\simeq& \sigma(p p \to \tilde{t} \bar{\tilde{t}}) \times 
\big( BR( \tilde{t} \to \bar{b} \bar{d_i} \bar{b} b) 
\times BR(\bar{\tilde{t}} \to b d_i)  \nonumber \\
&& \;\;\;\;\;\;\;\;\;\;\;\;\;\;\;\;\;\;\;\; + \,BR(\bar{\tilde{t}} \to b d_i b \bar{b}) 
\times BR(\tilde{t} \to \bar{b} \bar{d_i}) \big) \nonumber \\
&\simeq& 2 \times \sigma(p p \to \tilde{t} \bar{\tilde{t}}) \times 
BR( \tilde{t} \to \bar{b} \bar{d_i} \bar{b} b) 
\times BR(\bar{\tilde{t}} \to b d_i) \nonumber \\
& \simeq & \sigma(p p \to \tilde{t} \bar{\tilde{t}}) \times
\frac{2 r_1 r_2  \times (\lambda_{33i}'')^4}{
 \big(1 + r_1 \times (\lambda_{33i}'')^2\big)^2  \big(1 + r_2 \times (\lambda_{33i}'')^2\big)} \ , \label{eq:six_jets}
 \end{eqnarray}
with
\begin{equation}
BR( \tilde{t} \to \bar{b} \bar{s} \bar{b} b) = \frac{\Gamma_{\chi\!\text{-RPV}}}{\Gamma_{\tilde{t}\text{-RPV}}
+ \Gamma_{\chi\!\text{-RPV}} + \Gamma_{\text{RPC-like}} } \ . \nonumber
\end{equation}
\item \bfcRP--\bfcRP $\equiv 6b2j$,
\begin{eqnarray}
\sigma (6b2j)
& \simeq &  \sigma(p p \to \tilde{t} \bar{\tilde{t}}) \times 
BR( \tilde{t} \to \bar{b} \bar{d_i} \bar{b} b) \times BR(\bar{\tilde{t}} \to b d_i b \bar{b}) \nonumber \\
& \simeq & \sigma(p p \to \tilde{t} \bar{\tilde{t}}) \times
 \frac{r_2^2  \times (\lambda_{33i}'')^4}{
 \big(1 + r_1 \times (\lambda_{33i}'')^2\big)^2  \big(1 + r_2 \times (\lambda_{33i}'')^2\big)^2} \ . \label{eq:eight_jets}
\end{eqnarray}
\item \bfRPl--\bftRP  $\equiv 1t3b2j$,
\begin{eqnarray}
\sigma (1t3b2j)
&\simeq& \sigma(p p \to \tilde{t} \bar{\tilde{t}}) \times 
\big( BR(\tilde{t} \to \bar{b} \bar{d_i} \bar{b} (b d_i b) f_1 
\bar{f_1'} b f \bar{f'}) 
\times BR(\bar{\tilde{t}} \to b d_i)  \nonumber \\
&& \;\;\;\;\;\;\;\;\;\;\;\;\;\;\;\;\;\;\;\; + \,
BR({\tilde{t}} \to j \bar{b}) 
\times BR(\bar{\tilde{t}} \to b d_i b (\bar{b} \bar{d_i} \bar{b}) \bar{f}_1 
f_1' \bar{b} \bar{f}  f') \big) \nonumber \\
&\simeq& 2 \times \sigma(p p \to \tilde{t} \bar{\tilde{t}}) \times 
BR(\tilde{t} \to \bar{b} \bar{d_i} \bar{b} (b d_i b) f_1 
\bar{f_1'} b f \bar{f'}) 
 \times BR(\bar{\tilde{t}} \to b d_i) \nonumber \\ 
 & \simeq &  \sigma(p p \to \tilde{t} \bar{\tilde{t}}) \times
\frac{2 r_1  \times (\lambda_{33i}'')^2}{
 \big(1 + r_1 \times (\lambda_{33i}'')^2\big)^2  \big(1 + r_2 \times (\lambda_{33i}'')^2\big)} \ , \label{eq:eightprime_jets}
 \end{eqnarray}
with 
 \begin{equation}
BR( \tilde{t} \to \bar{b} \bar{d_i} \bar{b} (b d_i b) f_1 
\bar{f_1'} b f \bar{f'}) = \frac{\Gamma_{\text{RPC-like}}}{\Gamma_{\tilde{t}\text{-RPV}}
+ \Gamma_{\chi\!\text{-RPV}} + \Gamma_{\text{RPC-like}} } \ . \nonumber
\end{equation}

\item \bfRPl--\bfcRP $\equiv 1t5b2j$,
\begin{eqnarray}
\sigma (1t5b2j)
&\simeq& \sigma(p p \to \tilde{t} \bar{\tilde{t}}) \times 
\big( BR(\tilde{t} \to \bar{b} \bar{d_i} \bar{b} (b d_i b) f_1 
\bar{f_1'} b f \bar{f'}) 
\times BR(\bar{\tilde{t}} \to b d_i b \bar{b})  \nonumber \\
&& \;\;\;\;\;\;\;\;\;\;\;\;\;\;\;\;\;\;\;\; + \,
BR(\tilde{t} \to \bar{b} \bar{d_i} \bar{b} b) 
\times BR(\bar{\tilde{t}} \to b d_i b (\bar{b} \bar{d_i} \bar{b}) \bar{f}_1 
f_1' \bar{b} \bar{f}  f') \big) \nonumber \\
&\simeq& 2 \times \sigma(p p \to \tilde{t} \bar{\tilde{t}}) \times 
BR(\tilde{t} \to \bar{b} \bar{d_i} \bar{b} (b d_i b) f_1 
\bar{f_1'} b f \bar{f'}) 
 \times BR(\bar{\tilde{t}} \to b d_i b \bar{b}) \nonumber \\
& \simeq &  \sigma(p p \to \tilde{t} \bar{\tilde{t}}) \times
\frac{2 r_2  \times (\lambda_{33i}'')^2}{
 \big(1 + r_1 \times (\lambda_{33i}'')^2\big)^2  \big(1 + r_2 \times (\lambda_{33i}'')^2\big)^2}  \ . \label{eq:ten_jets} 
 \end{eqnarray}
\item \bfRPl--\bfRPl  $\equiv 2t4b2j$,
\begin{eqnarray}
\sigma (2t4b2j)
&\simeq& \sigma(p p \to \tilde{t} \bar{\tilde{t}}) \times 
BR(\tilde{t} \to \bar{b} \bar{d_i} \bar{b} (b d_i b) f_1 
\bar{f_1'} b f \bar{f'}) 
\times BR(\bar{\tilde{t}} \to b d_i b (\bar{b} \bar{d_i} \bar{b}) \bar{f}_1 
f_1' \bar{b} \bar{f}  f') \nonumber \\  \label{eq:2t4b2j}
& \simeq  & \sigma(p p \to \tilde{t} \bar{\tilde{t}}) \times
 \frac{1}{
 \big(1 + r_1 \times (\lambda_{33i}'')^2\big)^2  \big(1 + r_2 \times (\lambda_{33i}'')^2\big)^2} \label{eq:twelve_jets} \ .
\end{eqnarray}
\end{itemize}
We have replaced $s$ by $d_i$ in the above expressions to stress the fact that these are valid 
either for the case of $s$ alone, or for the case of $d$ alone, or else for the sum of the two, depending on the values
of $\lambda_{331}'', \lambda_{332}''$ in Eq.~(\ref{eq:lambdaeff}). 

The analytical form of Eqs.~(\ref{eq:four_jets} -- \ref{eq:twelve_jets}) 
illustrate 
clearly the deviation from the naive expectation that all cross-sections would scale with 
$(\lambda_{33i}'')^4$.  One sees that such scaling is generically modified by the 
\bfRPl~component. Moreover, even for the \bftRP~and \bfcRP contributions 
different final state cross-sections can  have various sensitivities to 
$\lambda_{332}''$ depending on the following possible regimes:

\begin{equation}
r_a \ll (\lambda_{33i}'')^{-2}, \; r_a \sim (\lambda_{33i}'')^{-2}, \;
r_a \gg (\lambda_{33i}'')^{-2}, \; \; \; (a=1,2) \ . \label{eq:regimes}
\end{equation}
These regimes are triggered by the interplay between the RPV and RPC sectors. For instance
the magnitude of $r_1$ is controlled by the degree of mass degeneracy between
the stop and the chargino. Similarly, the degeneracy between the chargino and neutralino
masses implies typically a large $r_2$. Perhaps the most striking feature that comes out of
the NWA expressions is that the
variation of $\lambda_{33i}''$ over several orders of magnitude, 
within the range given in Eq.~(\ref{eq:RPVrange}),  
triggers the dominance of very different final states without reducing the total cross-sections.
In particular, while the \bftRP--\bftRP clearly dominates for relatively large values
of $\lambda_{33i}''$, the \bfRPl--\bfRPl becomes dominant for very small values of this 
coupling. Furthermore, one can easily determine from Eqs.~(\ref{eq:four_jets}, \ref{eq:six_jets},\ref{eq:eight_jets}, \ref{eq:ten_jets},
\ref{eq:twelve_jets}) the scaling relations

\begin{eqnarray}
\frac{\sigma(2b2j) \cdot \sigma(6b2j)}{[\sigma(4b2j)]^2}=\frac{1}{4}
\label{eq:b-SR}, \\
\frac{\sigma(6b2j) \cdot \sigma(2t4b2j)}{[\sigma(1t5b2j)]^2}=
 \frac{1}{4} \ .
\label{eq:t-SR}
\end{eqnarray}
We refer to these two scaling relations respectively as b-SR and t-SR,
where the first one involves shorter decay chains with no top-quark final
states and the second longer decay chains with top-quark final states. 
These scaling relations lead also to
\begin{equation}
\frac{\sigma(2t4b2j)}{\sigma(2b2j)} = \Big(\frac{\sigma(1t5b2j)}{\sigma(4b2j)} \Big)^2 \ .
\end{equation}

To summarize, we derived in this Section analytical expressions for the cross-sections 
with all possible stop decay final states,  
in a form that untangles the dependence on 
the RPV $\lambda_{33i}''$ coupling from that on the MSSM mass spectrum and RPC 
couplings encoded in the $r_a$ ratios Eqs.~(\ref{eq:r1}, \ref{eq:r2}). Moreover these 
expressions imply scaling relations among the cross-sections independently of the 
couplings and masses. Given the complexity of the long chain decays, these analytical results will prove very 
useful, even though established within the approximation of narrow width, when interpreting 
the results  and assessing the validity of the exact matrix element numerical computation
in Section~\ref{sec:results}.

\section{Benchmark points and constraints}

In order to estimate the cross-sections for the processes of interest, we interfaced several software packages as discussed in the following.
Firstly we used the {\sc Sarah}~\cite{Staub:2011dp} {\sc Mathematica}~\cite{ram2010} package to generate model files in 
UFO format compatible with the {\sc MadGraph5$\_$aMC@NLO} \cite{Alwall:2014hca} Monte Carlo generator. 
Then we used {\sc Sarah} to implement 
the MSSM trilinear RPV model in {\sc SPheno} \cite{Porod:2011nf} so as to calculate the entire 
SUSY mass spectrum and couplings.
We adopted a bottom-bottom approach, where the values of
the supersymmetric and soft SUSY breaking parameters are provided directly at the electroweak scale.
This approach has the benefit of being simple without sacrificing the typical supersymmetric 
correlations among various low energy states masses and couplings, and of being
model-independent in view of our present ignorance of how supersymmetry is realized
at high scales.


Using the low scale MSSM option of the {\sc SPheno} code,  we performed a scan over the SUSY input parameters 
to determine benchmark points that are consistent with our spectrum assumptions discussed in Sections 
\ref{subsec:spectrum} and \ref{sec:st_prod_dec}, as well as with constraints from the available physical observables.
We generated several mass spectra in different regions of the relevant MSSM 
parameter space, fixing the EWSB scale to  $Q_{EWSB} = 1$~\TeV~and including 1-loop 
corrections to all SUSY particle masses and 2-loop corrections to the lightest CP-even Higgs 
mass. For each given parameter point we used {\sc HiggsBounds} \cite{Bechtle:2008jh, Bechtle:2013gu} and {\sc HiggsSignals} \cite{Bechtle:2013xfa} to 
confront the Higgs sector computed by {\sc SPheno} with existing measurements and exclusion limits. 
Moreover we accounted for the low energy flavor constraints coming from the recent measurements of $B^0$ decaying into a pair of muons \cite{Aaij:2012nna,Aaij:2013aka,Chatrchyan:2013bka}. For given values of the soft SUSY 
breaking parameters in the stop and gaugino sectors satisfying these constraints, a further scan over the $\mu$ parameter was
performed such that the lighter chargino and neutralinos remain Higgsino-like and the resulting masses reproduce the hierarchy given by Eqs.~(\ref{eq:spectconfig1} -- \ref{eq:spectconfig3}). 
 For the remainder of this paper we choose two benchmark
sets of input parameters as given in Table~\ref{tab:input_parameters}, corresponding to 
two stop mass values $m_{\tilde{t}}= 600$~\GeV~and $1$~\TeV. 
The values we take in Table~\ref{tab:input_parameters} 
should be understood as given at the EWSB scale. Note that we have put to zero several
of these parameters (see last line of Table~\ref{tab:input_parameters}), in particular
the off-diagonal components in flavor space of soft masses keeping up with our MFV assumption,
and the soft SUSY breaking trilinear couplings $T_{33i}''$ associated with $\lambda_{33i}''$
as they involve only scalar states and thus would not contribute to our study at leading order.

The large mass splitting between the two stop states in accordance with 
assumption {\sl (ii)} of Sec.~\ref{subsec:spectrum} is achieved through the large
numerical difference between $(m_{\tilde{Q}})_{33}$ and $(m_{\tilde{U}})_{33}$
rather than through a large off-diagonal component of the mass matrix. 
The mixing between the light and heavy stops is thus very small, therefore the
lighter stop, essentially right-handed, has its baryon number violating RPV decay 
controlled mainly by the magnitude of $\lambda_{33i}''$. Note also that the values of $m_{\tilde{t}}$,
respectively 600~\GeV ~and 1~\TeV ~in the two benchmark scenarios still vary slightly 
by about
-0.5\% +1.5\% around the central value due on one hand to the sensitivity to $\mu$ 
through the mixing in the stop sector, though suppressed by the moderately large value of $\tan \beta$,
and on the other hand to the sensitivity to $\lambda_{33i}''$ through loop corrections
\cite{Chamoun:2014eda}. 

Including $1$- and $2$-loop corrections from the RPC sector, the lighter CP-even Higgs mass remains essentially at 125 \GeV, as extra $1$-loop corrections
from the RPV sector \cite{Dreiner:2014lqa} which have also been included, are 
negligible in
the scanned $\lambda_{33i}''$ range given in Eq.~(\ref{eq:RPVrange}). The small variation in the
mass splitting among the light chargino and neutralinos is a residual effect of the small
$\mu/M_1$ and $\mu/M_2$ ratios as already noted in Sec.~\ref{sec:st_prod_dec}.
All other states are very heavy (between 1.5 and 3 \TeV) and do not affect our study.
Since we rely on the low scale MSSM option the renormalization group running of couplings and masses involves only the range between $m_Z$ and the EWSB scale. This allows to treat consistently the gauge and 
Yukawa couplings extracted at the $m_Z$ scale and the input SUSY parameters $\mu, \tan\beta$ and the (tree-level) CP-odd
neutral Higgs mass $m_A$ defined at the EWSB scale. In particular we make no theoretical assumptions relating the RPV-MSSM parameters at very high scales that would have induced correlations at low scales through the renormalization group evolution.
In this context assumption {\sl (i)} of Sec.~\ref{subsec:spectrum} with values
in the range defined in Eq.~(\ref{eq:RPVrange}) should be viewed as defined at the 
EWSB scale.
The running of $\lambda_{33i}''$ from the EWSB to the $m_{\tilde{t}}$ or $m_{\chi^+}$ scales 
where the various stop decay channels are evaluated, remains very small and it is neglected in our 
study.
Note however, that $\lambda_{332}''$ affects the running of the top-quark 
Yukawa coupling between the EWSB scale and $m_Z$.  
Similarly, there are no high scale assumptions about the soft SUSY
breaking masses and trilinear couplings.

\begin{table}[!h]
\begin{center}
\renewcommand{\arraystretch}{1.5}
\begin{tabular}{|c|c|c|} \hline\hline
Benchmark points &  1 & 2 \\ \hline
 $\tan \beta$   &  \multicolumn{2}{c|}{10} \\  \hline
 $M_1$  &  \multicolumn{2}{c|}{2.5 \TeV}    \\  \hline
 $M_2$  &  \multicolumn{2}{c|}{1.5 \TeV}    \\  \hline
 $M_3$  &  \multicolumn{2}{c|}{1.7 \TeV}    \\   \hline
 $(m_{\tilde{Q}})_{33}$  & \multicolumn{2}{c|}{2 \TeV~}  \\ \hline
 $(m_{\tilde{U}})_{33}$  & 570 \GeV &  964 \GeV \\ \hline
 $(m_{\tilde{D}})_{33}= (m_{\tilde{U}})_{ii}=(m_{\tilde{D}})_{ii}=(m_{\tilde{E}})_{ii}=(m_{\tilde{Q}})_{ii}=
 (m_{\tilde{L}})_{ii}$, $i=1,2$   & \multicolumn{2}{c|}{3 \TeV}\\ \hline
 $(T^u)_{33}$  & -2100 \GeV &  -2150 \GeV \\ \hline
 $m_A$  &    \multicolumn{2}{c|}{2.5 \TeV}\\ \hline
 $\mu$  &  400-650 \GeV  &  750-1000 \GeV \\ \hline
 $\lambda_{33i}''\equiv \sqrt{(\lambda_{332}'')^2 +(\lambda_{331}'')^2}$ &  \multicolumn{2}{c|}{$10^{-7} - 10^{-1}$}   \\ \hline 
  $T^l, T^d, 
   (T^u)_{ij}, (m_{\tilde{Q},\tilde{U},\tilde{D},\tilde{L},\tilde{E}})_{ij}, T_{33i}''$
   , $i \neq j=1,2,3$, $(T^u)_{ii}, i=1,2$&    \multicolumn{2}{c|}{0}\\ \hline \hline
\end{tabular}
\end{center}
\caption{\small{Two lists of benchmark SUSY parameters defined at the low scale $Q^2_{EWSB}=1$ \TeV${}^2$ taken 
as input for {\sc SPheno}.
All other non-listed supersymmetric or soft SUSY breaking parameters are either computed
from the input, such as $m_{H_{1,2}}^2$, or irrelevant to the present study, such
as $\lambda_{ijk}, \lambda_{ijk}', \mu_i, T_{ijk}, T_{ijk}', B_i, \tilde{m}_{1 i}$ for all three generations, and $T_{ijk}''$ for $i,j=1,2$. We also take 
$m_b(m_b)_{\overline{\text{MS}}} = 4.18 \GeV$ and 
         $m_t(\text{pole}) = 173.5 \GeV$. See \cite{Porod:2011nf} for
         the values of the other SM input parameters.}}
\label{tab:input_parameters}
\end{table}

\begin{table}[!h]
\begin{center}
\renewcommand{\arraystretch}{1.5}
\begin{tabular}{|c|c|c|} \hline\hline
Benchmark points &  1 & 2 \\ \hline
 $m_{\tilde{t}}$ & {$\sim$} 600 \GeV  & {$\sim$} 1 \TeV \\ \hline
 $m_{\chi^+}$   & {$\sim$ 400-650 \GeV}   & {$\sim$ 750-1000 \GeV}\\ \hline
$m_{\chi^+} - m_{\chi^0}$  & \multicolumn{2}{c|}{{$\sim$ 1.5-2.5 \GeV}} \\ \hline
$m_{\tilde{t}} - m_{\chi^+}$ & {$\sim -45$ -- $200$ \GeV} & $\sim 1$ - $245$ \GeV\\ \hline
$m_{\chi^0_2} - m_{\chi^+}$   & \multicolumn{2}{c|}{{$\sim$ 4-5 \GeV}} \\ \hline
$m_{\chi^0_{3}} \sim m_{\chi^+_2}$, $m_{\chi^0_4}$ & \multicolumn{2}{c|}{{$\sim$ 1.5 \TeV, $\sim$ 2.5 \TeV}}\\ \hline
 $m_{h^0}$  &   \multicolumn{2}{c|}{$\sim$ 125 \GeV}   \\ \hline
 $m_A \approx m_{H^0} \approx m_{H^\pm}$  &  \multicolumn{2}{c|}{{$\sim$ 2.5 \TeV}} \\ \hline
 $M_{\tilde{g}}$  &  \multicolumn{2}{c|}{{$\sim$ 1.87 \TeV}}    \\  \hline
 $M_{\tilde{t} 2} \approx M_{\tilde{b} 1}$  & \multicolumn{2}{c|}{{$\sim$ 2 \TeV}}  \\  \hline
$M_{\tilde{b} 2} \approx M_{\tilde{u} 1,2} \approx M_{\tilde{d} 1,2}$   & \multicolumn{2}{c|}{{$\sim$ 3 \TeV}}  \\  \hline
 $M_{\tilde{l} 1,2}, M_{\tilde{\nu} 1,2}$  &  \multicolumn{2}{c|}{{$\sim$ 3 \TeV}}  \\  \hline\hline
$(g-2)_\mu^{\rm SUSY}$ & 3 $-$ 3.3 ~$\times 10^{-11}$ & 3.2 $-$ 3.3 ~$\times 10^{-11}$\\ \hline 
$\delta \rho^{\rm SUSY}$ & 5.7 $-$ 5.9 ~$\times 10^{-5}$ & $\sim$5.5 ~$\times 10^{-5}$\\ \hline
$BR(B \to X_s \gamma)/BR(B \to X_s \gamma)^{SM}$ & 0.89 $-$ 0.92 & 0.95 $-$ 0.96\\ \hline
$BR(B^0_s \to \mu \mu)$ & 3.36 $-$ 3.39 ~$\times10^{-9}$ & 3.38 $-$ 3.40 ~$\times10^{-9}$\\ \hline
$BR(B^0_d \to \mu \mu)$ & 1.08 $-$ 1.09 ~$\times 10^{-10}$ & $\sim$ 1.09 ~$\times 10^{-10}$ \\ \hline\hline 
\end{tabular}
\end{center}
\caption{\small{Two lists of benchmark observables generated with {\sc SPheno} corresponding to the input 
of Table~\ref{tab:input_parameters} and taken as input for {\sc MadGraph5$\_$aMC@NLO}. 
Pole masses are evaluated at one-loop order except for the lightest CP-even Higgs which includes the 
2-loop corrections. 
}} \label{tab:parameters}
\end{table}

\subsubsection{Low energy constraints \label{subsec:LE_constraints}}
A large number of low energy and precision observables can be very sensitive to 
BSM physics. Among these, the LEP/SLC electroweak precision observables, 
the leptons anomalous magnetic moments and electric dipole moments 
as well as low energy quark or lepton number violating processes. 
In Table~\ref{tab:parameters} we give the values in our two benchmark points of only a few of them.\footnote{For more details on the level of 
accuracy used see \cite{Porod:2011nf} and references therein.}

The anomalous magnetic moment of the muon is a very important test bed for virtual effects from BSM physics as it is one of
the most accurately measured quantities in particle physics; for a review see e.g. Ref.~\cite{Miller:2012opa}. At the one-loop level $(g-2)_\mu$ receives 
contributions from the purely SUSY neutralino/smuon and chargino/muonic-sneutrino RPC sectors.  In our benchmark points
the smuon sector is very heavy and the chargino/neutralino relatively heavy as well, leading to the small contribution reported
in Table~\ref{tab:parameters} given the chosen moderate value of $\tan \beta$. 
Other possible one-loop effects from nonzero $\lambda, \lambda'$ RPV couplings, or from CP-violating phases \cite{Ibrahim:1999hh}, are absent in our scenario.
Moreover,  two-loop RPC SUSY corrections \cite{Stockinger:2006zn}, 
are not expected to be significant in our case even for a relatively light stop, due to the moderate values of the $\mu$
parameter and $\tan \beta$.  The $3.6\sigma$ discrepancy  
$\Delta \left(\frac12 (g-2)_\mu\right) = 288(63)(49) \times 10^{-11}$, \cite{Agashe:2014kda}, 
between the experimental measurement and the 
theoretical SM predictions is thus too large to be accounted for by our benchmark points, leaving open the issue of the
uncertainties on the theoretical estimates of the SM hadronic contributions.

Virtual corrections to the $\rho$ parameter originate from the squark and slepton
left-handed states. They tend to be suppressed for heavy states 
as a result of decoupling but can be enhanced by mass splitting between up and down 
flavors as a result of custodial symmetry breaking \cite{Cho:1999km}. In our 
benchmark scenario where the lighter stop is mainly right-handed and all other squark
and slepton states heavy and almost degenerate, no sizable effects on 
$\delta \rho$ are expected from these sectors even for a relatively light $\tilde{t}$. The resulting range for $\rho \simeq 1 + \delta \rho^{\rm SUSY}$ obtained in our scan 
remains consistent within $2\sigma$ with the experimental value \cite{Agashe:2014kda}.

The $B$-meson  radiative inclusive decay $B \to X_s \gamma$ is sensitive to 
virtual effects from various sectors of the MSSM associating the charged Higgs
to the top quark, the up squarks to the charginos and the down squarks to the neutralinos
or to the gluino \cite{Bertolini:1990if}. Only the $\chi^+$--$\tilde{t}$ loops are
sizable 
in our case as the stop is much lighter than all other squarks and the gluino. Moreover it 
is mainly right-handed and the chargino higgsino-like, thus further leading to a ${\cal 
O}(m_t/M_W)$ enhancement in the amplitude.
The charged Higgs yields like-wise suppressed contributions due to its very heavy mass.
Taking into account the  recent update for the SM theoretical prediction 
$BR(B \to X_s \gamma) = (3.36 \pm 0.23) 
\times 10^{-4}$ \cite{Misiak:2015xwa}, our scan remains within $1.2\sigma$ from the 
combined experimental value $BR(B \to X_s \gamma) = (3.43 \pm 0.21 \pm 
0.07)\times10^{-4}$\cite{Amhis:2014hma}. It is however interesting to note that keeping
only the right-handed stop and Higgsino-like contributions, the mass ratio dependence 
in the loop functions favor, for lighter stops, heavier charginos  in order 
to cope with the  
$BR(B \to X_s \gamma)$ constraints.  As a consequence a sufficiently light stop would require  
a reduced mass splitting with respect to the lighter chargino, eventually even forbidding
the hierarchy given in Eq.~(\ref{eq:spectconfig1}) and favoring a stop MSSM-LSP. 
The latter would imply a stop decaying 100\% into $b$+jet final states, giving support to the model-independence of the  present exclusion limits based on this assumption, 
as long as the ensuing bounds remain low enough. 
For instance we find that a lower bound of $0.89$ on $BR(B \to X_s \gamma)/BR(B \to X_s \gamma)^{SM}$ as adopted e.g. in \cite{Dreiner:2013jta} would typically require $m_{\tilde{t}} \gtrsim 400$~\GeV.
Still, a more quantitative study is
needed as mass degeneracy between the stop and chargino could still be allowed 
favoring the third regime of Eq.~(\ref{eq:regimes}) and thus final states with
\bfcRP or \bfRPl components.
For instance relaxing the lower bound to $\sim 0.84$ would allow lighter
non-LSP stops, e.g. $m_{\tilde{t}} \lesssim 385$ \GeV, with $m_{\chi^+} \gtrsim 198$ \GeV.

Finally, regarding the $B^0$ decay into a pair of muons, LHCb~\cite{Aaij:2012nna,Aaij:2013aka} and CMS~\cite{Chatrchyan:2013bka} have
recently reported observation of such decays,  
with the combined fits leading to $BR(B^0_s \to \mu \mu) = (2.8 {}^{+0.7}_{-0.6}) \times 10^{-9}$ 
and $BR(B^0_d \to \mu \mu) = (3.9 {}^{+1.6}_{-1.4}) \times 10^{-10}$ that are compatible with
the SM at $2\sigma$-level  \cite{CMS:2014xfa}.
Our benchmark numbers are consistent with the updated SM theoretical predictions
$BR(B^0_s \to \mu \mu) = (3.65 \pm 0.23) \times 10^{-9}$ and 
$BR(B^0_d \to \mu \mu) = (1.06 \pm 0.09) \times 10^{-10}$ \cite{Bobeth:2013uxa}.

\vspace{.5cm}
\section{Cross-sections and uncertainties}
\label{sec:results}

Using the spectrum calculator and event generator tools as described in the previous Sections we have computed the 
total cross-section and decays of a pair of stops in $pp$ collisions at $\sqrt{s} = 14$ \TeV~ for the two
benchmark points given in Table~\ref{tab:input_parameters} and the various combinations of final states given in 
Table~\ref{tab:finalstates},  except for the $1t3b2j$ final state since it remains subdominant everywhere in the considered 
$\lambda_{33i}''$ range. In Figs.~\ref{fig:bench11}  and \ref{fig:bench31} we illustrate the sensitivity to the magnitude of 
$\lambda_{33i}''$, and in Figs.~\ref{fig:bench12} and~\ref{fig:bench32} the
sensitivity to the stop-chargino mass splitting for the two benchmark points. Before commenting these results, we
discuss first the various theoretical uncertainties.


\begin{figure*}[!h]
    \begin{center}
      \subfloat[]{\includegraphics[width=0.41\textwidth]{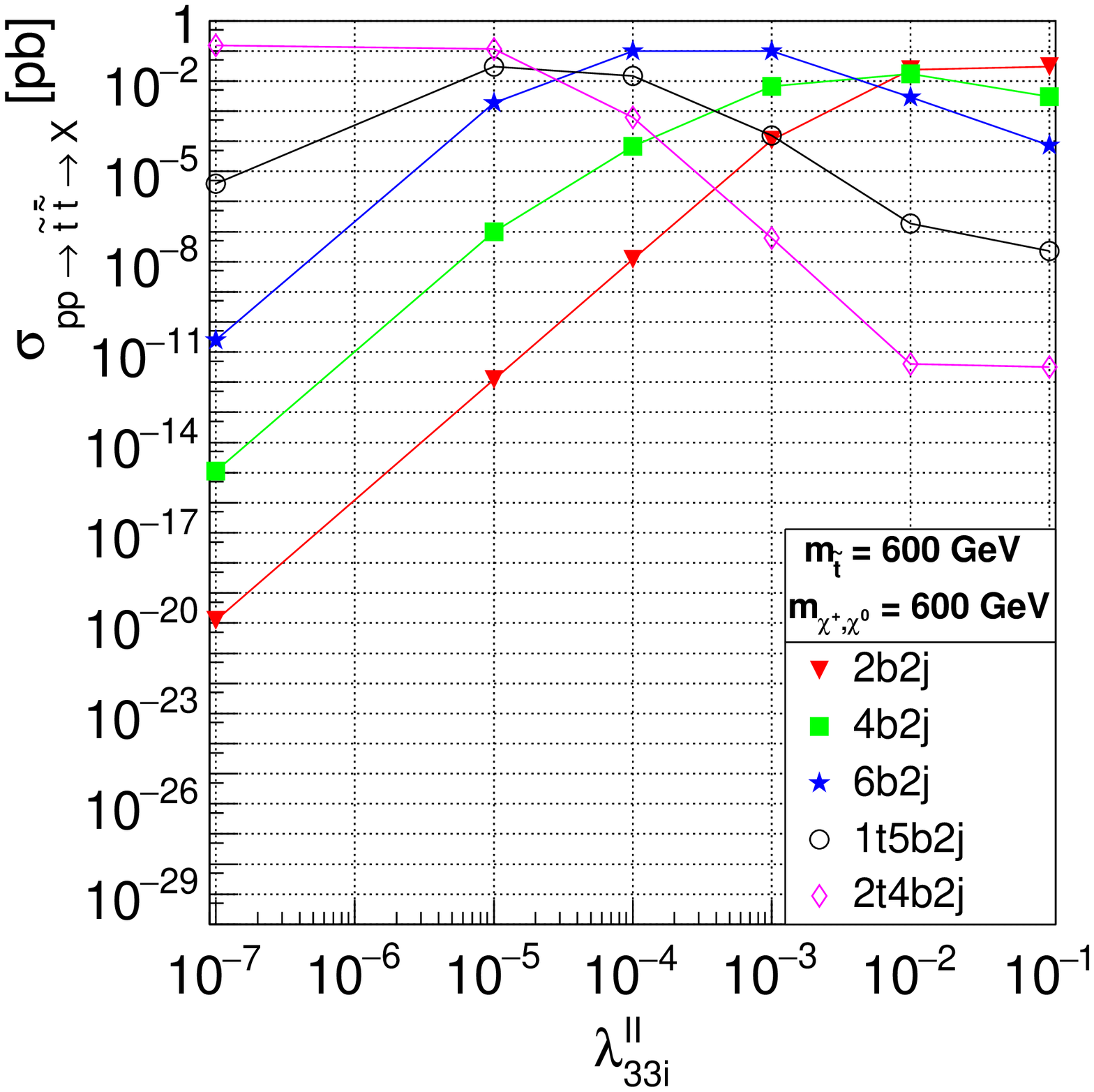}}
      \subfloat[]{\includegraphics[width=0.41\textwidth]{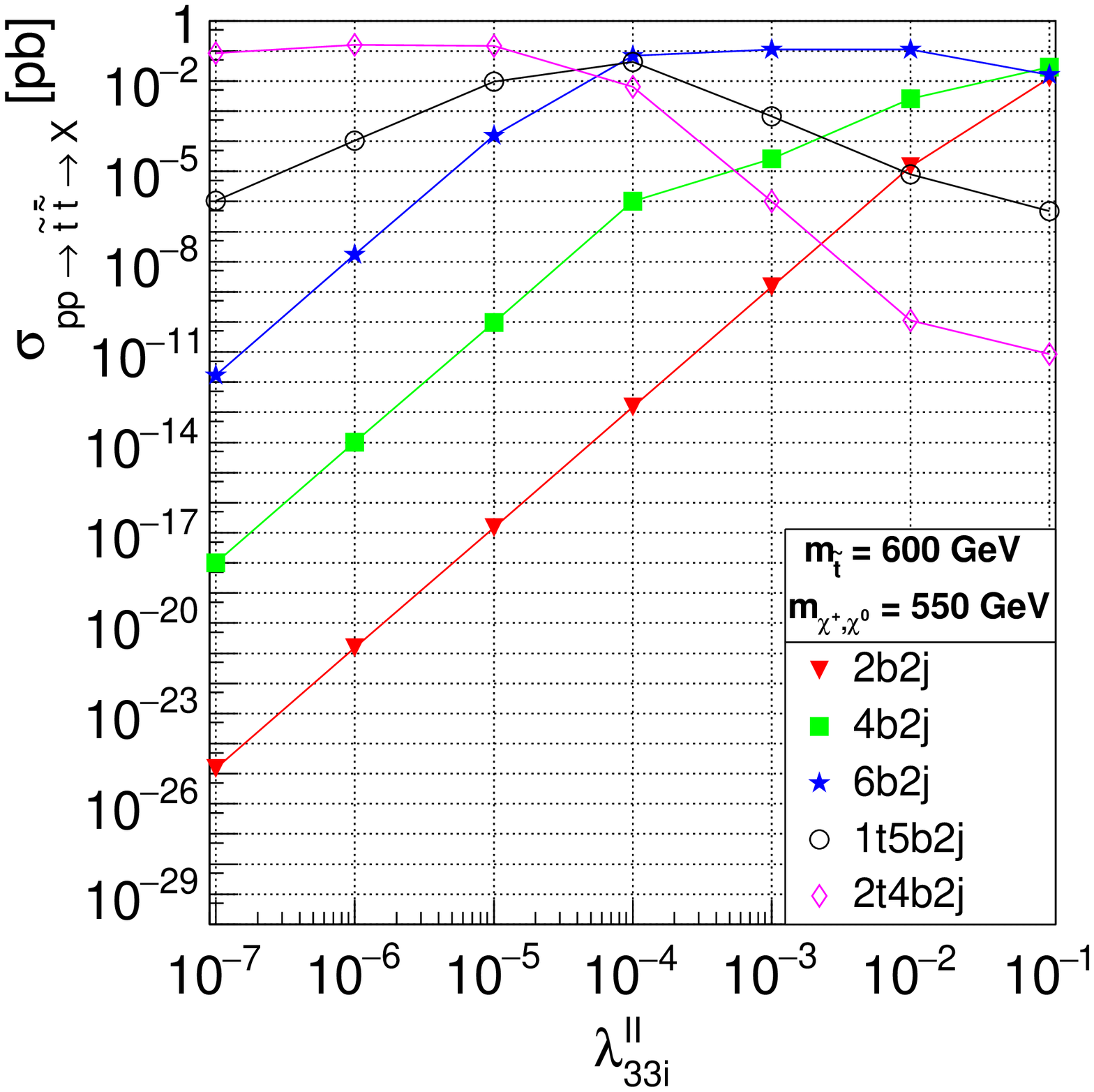}}\\
      \subfloat[]{\includegraphics[width=0.41\textwidth]{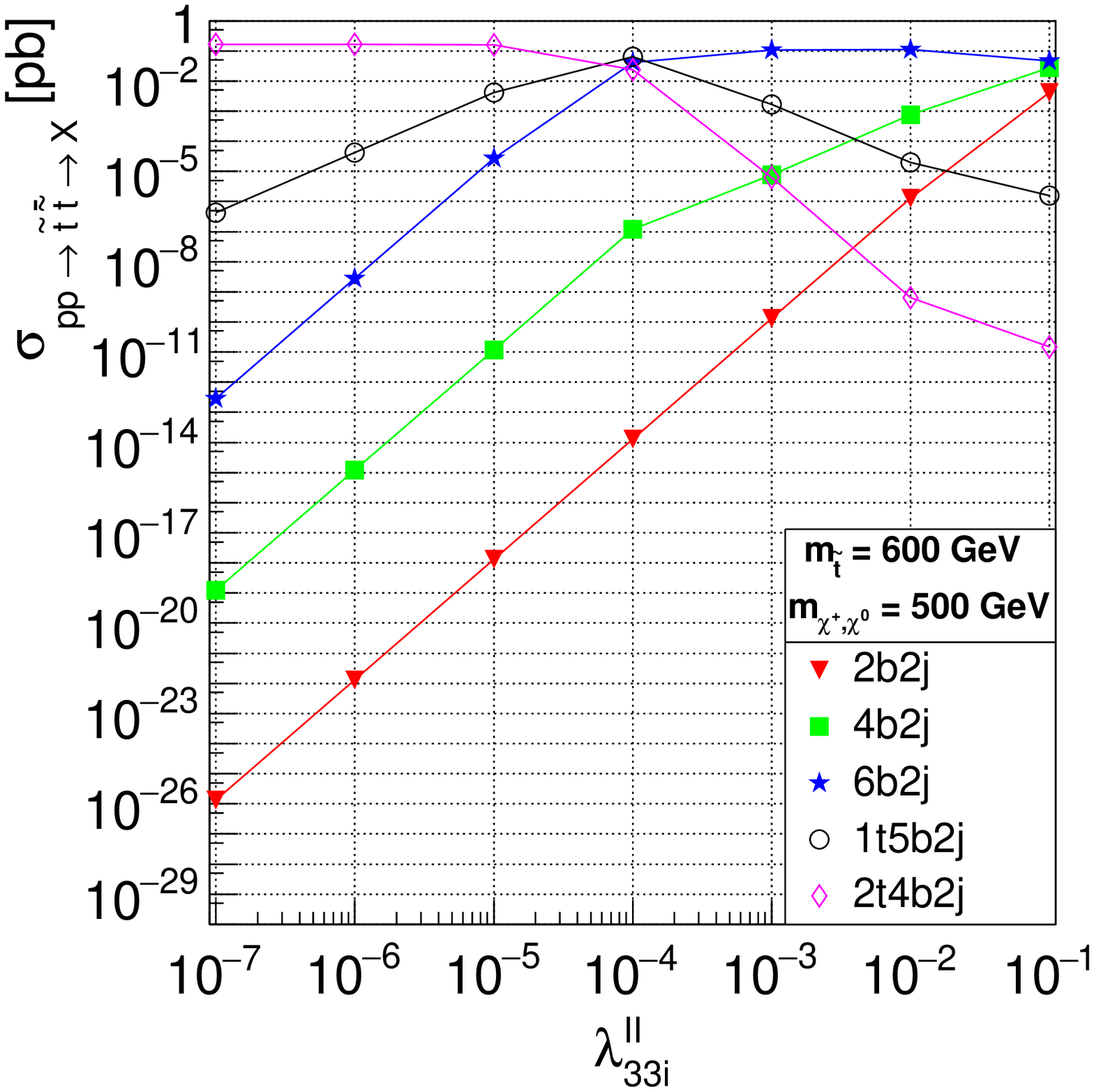}}
      \subfloat[]{\includegraphics[width=0.41\textwidth]{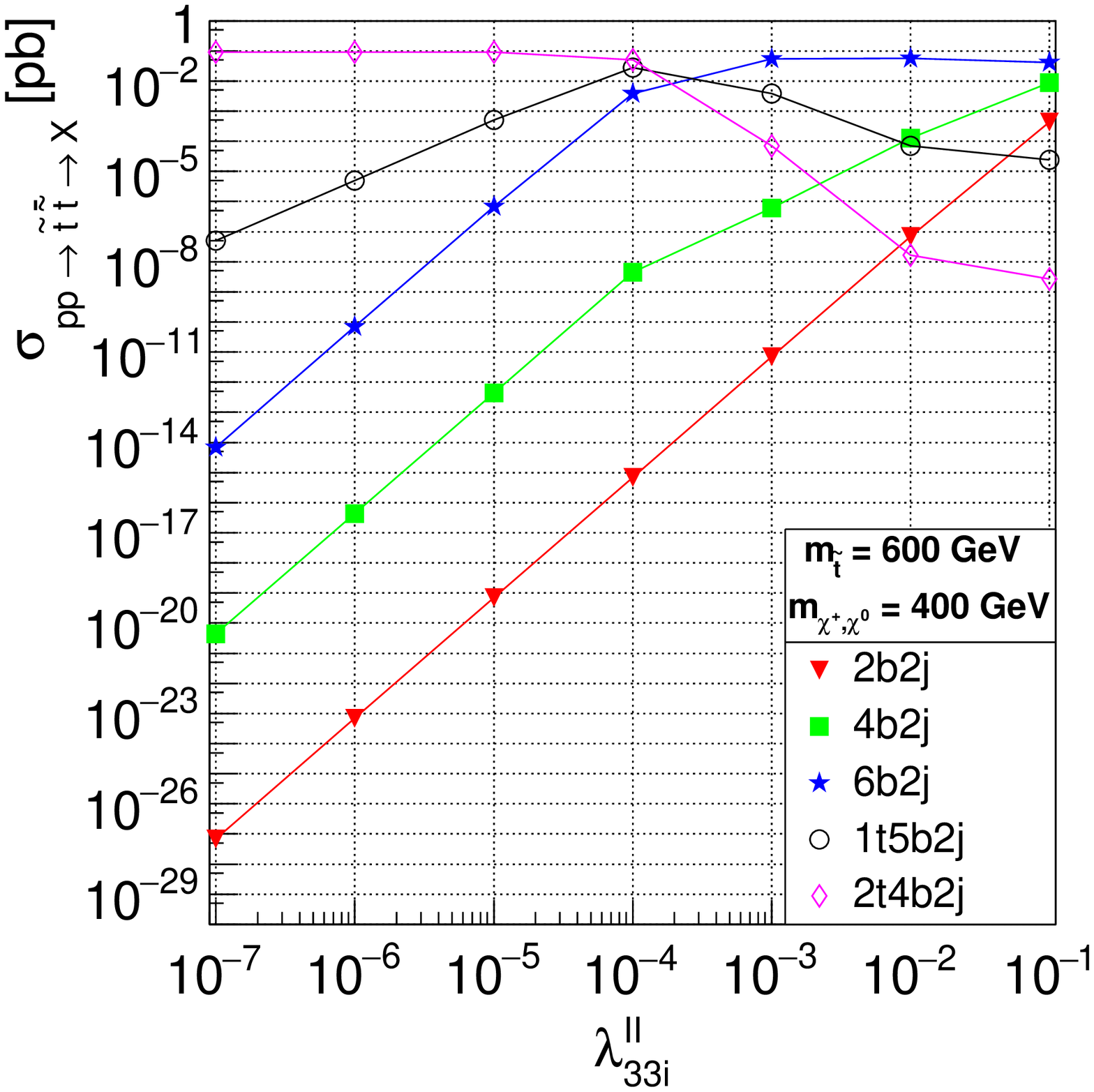}}\\
\end{center}
\caption{\label{fig:bench11} \small{Benchmark 1: production cross-section for $\sigma ( p p \rightarrow \tilde{t} \bar{\tilde{t}} 
\rightarrow X)$ at $\sqrt{s}=14$ \TeV, where $X= 2b2j$ (red triangles), $4b2j$ (green squares), $6b2j$ (blue stars), $1t5b2j$ (black empty circles) and $2t4b2j$ (pink diamonds), as a function of $\lambda_{33i}''$ and for $m_{\tilde{t}} - m_{\chi^+} \simeq 0$ \GeV (a), $50$ \GeV (b), $100$ \GeV (c) and $200$ \GeV (d). See Tabs.~\ref{tab:input_parameters} and \ref{tab:parameters} for the low-energy values of the MSSM parameters.}}
\end{figure*}


\begin{figure*}[!ht]
    \begin{center}
      \subfloat[]{\includegraphics[width=0.42\textwidth]{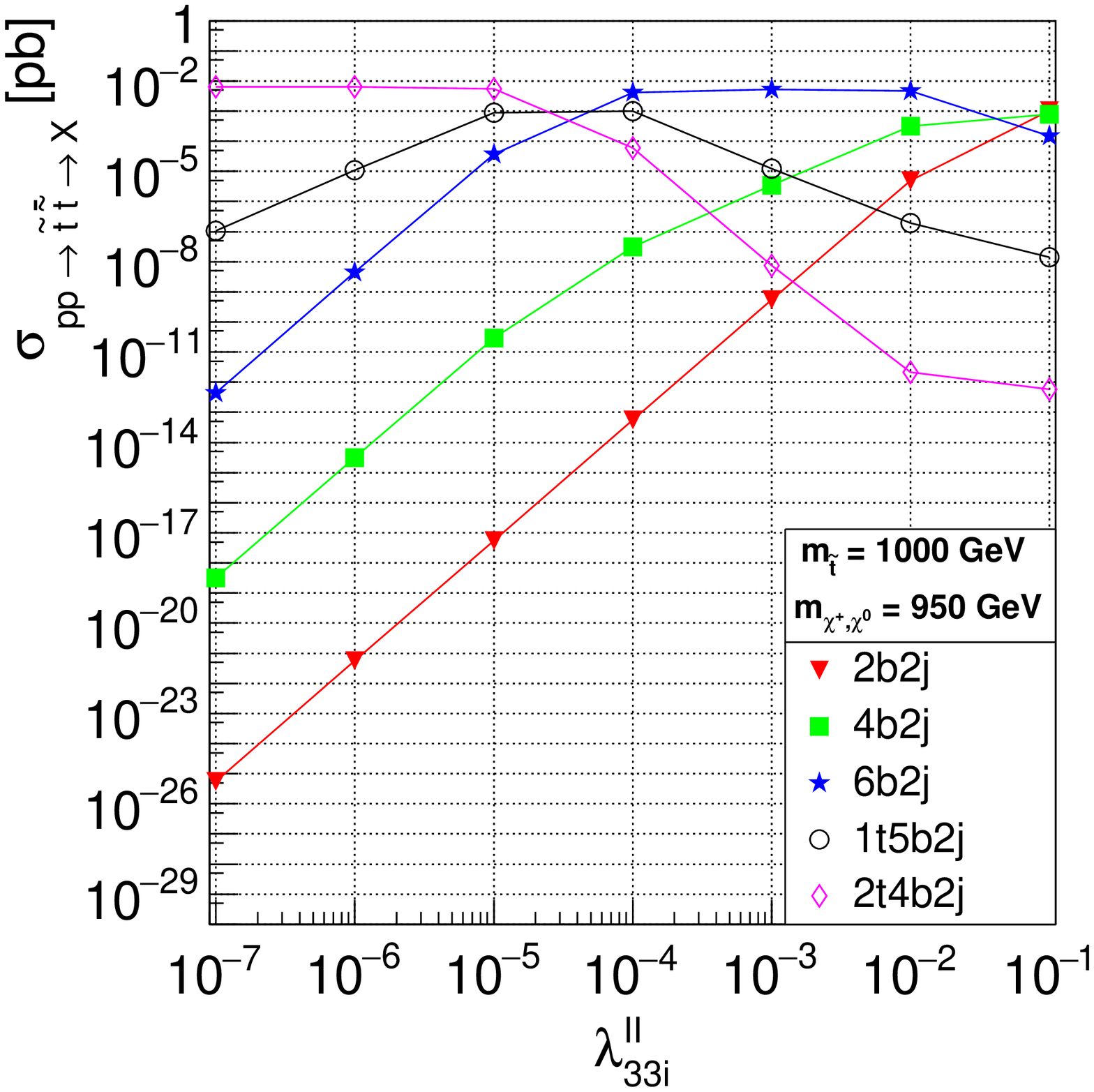}}
      \subfloat[]{\includegraphics[width=0.42\textwidth]{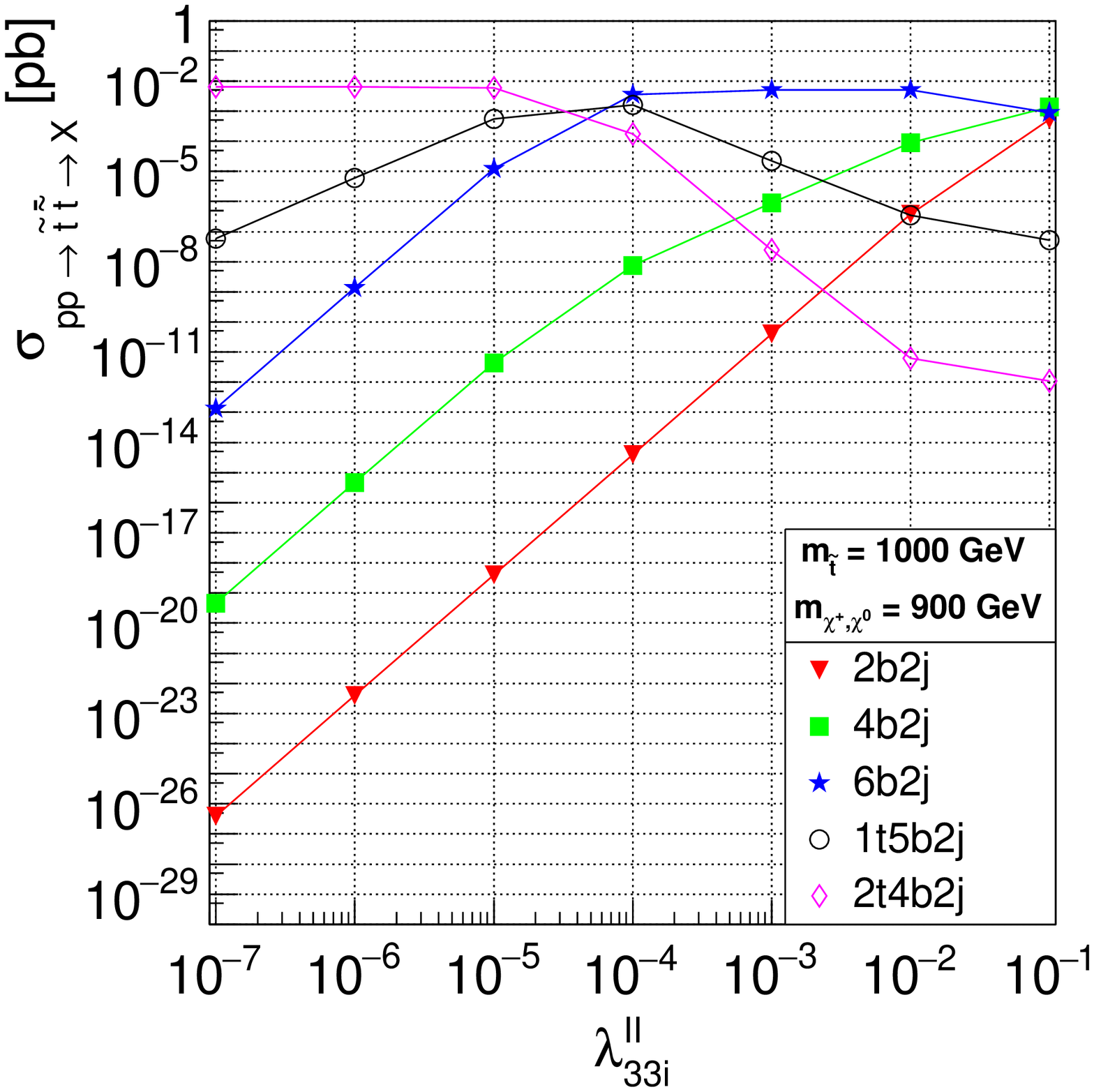}}\\
      \subfloat[]{\includegraphics[width=0.42\textwidth]{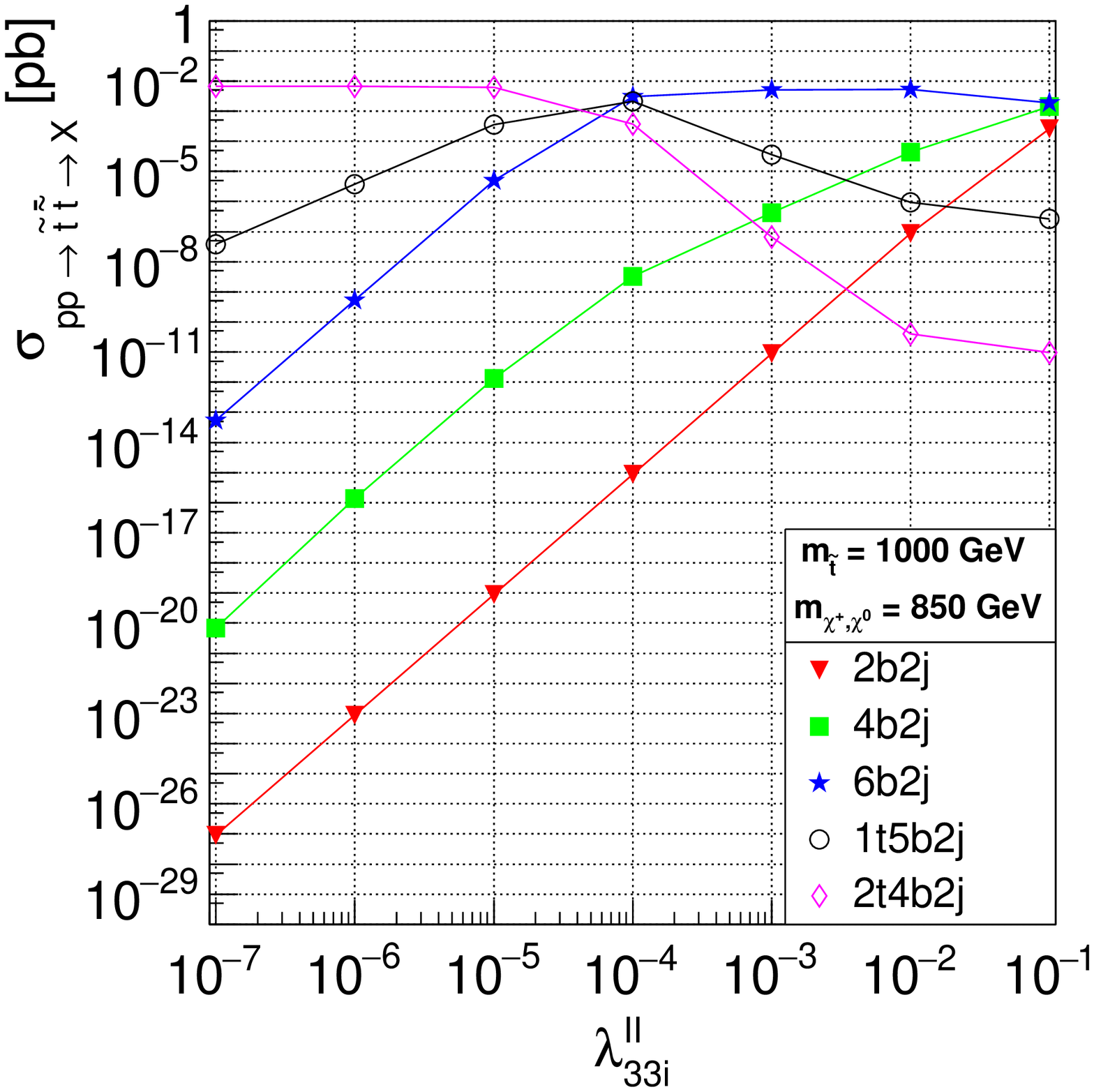}}
      \subfloat[]{\includegraphics[width=0.42\textwidth]{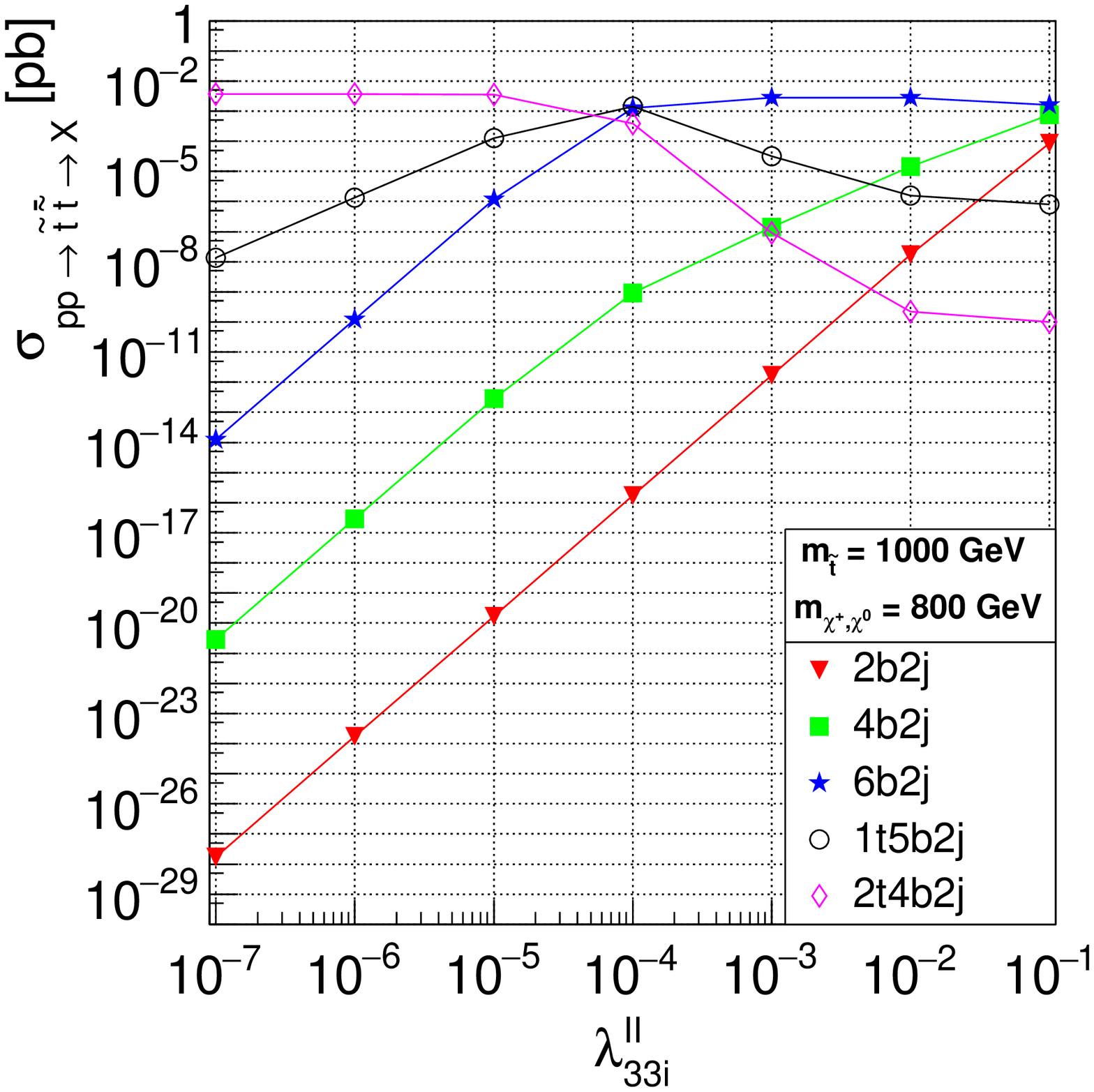}}\\
\end{center}
\caption{\label{fig:bench31} 
\small{Benchmark 2: production cross-section for $\sigma ( p p \rightarrow \tilde{t} \bar{\tilde{t}} 
\rightarrow X)$ at $\sqrt{s}=14$ \TeV, where $X= 2b2j$ (red triangles), $4b2j$ (green squares), $6b2j$ (blue stars), $1t5b2j$ (black empty circles) and $2t4b2j$ (pink diamonds), as a function of $\lambda_{33i}''$ and for 
$m_{\tilde{t}} - m_{\chi^+} \simeq 50$ \GeV (a), $100$ \GeV (b), $150$~\GeV (c) and $200$ \GeV (d). See Tabs.~\ref{tab:input_parameters} and \ref{tab:parameters} for the low-energy values of the MSSM 
parameters.}}
\end{figure*}


\begin{figure*}[!ht]
    \begin{center}
      \subfloat[]{\includegraphics[width=0.42\textwidth]{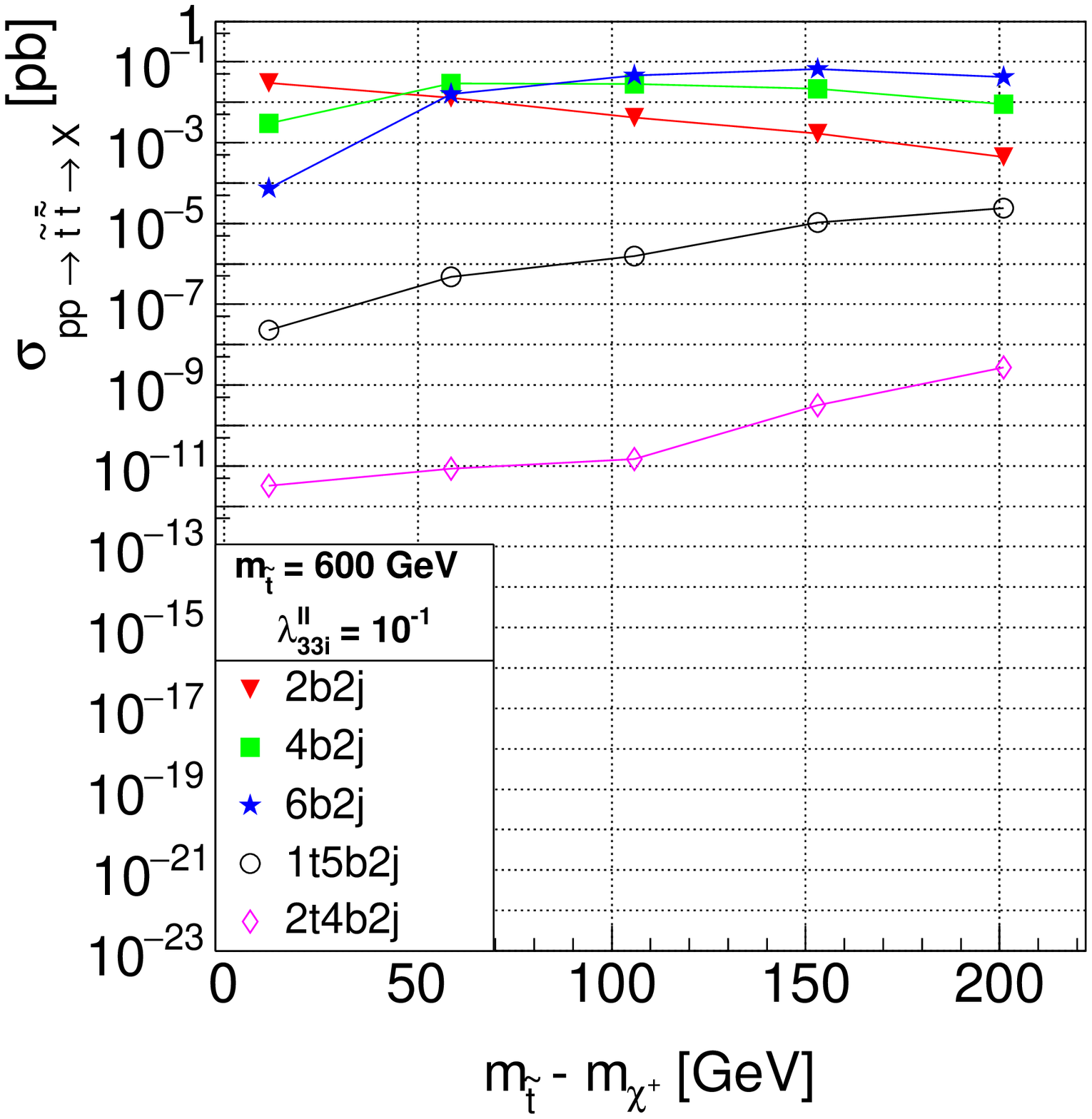}}
      \subfloat[]{\includegraphics[width=0.42\textwidth]{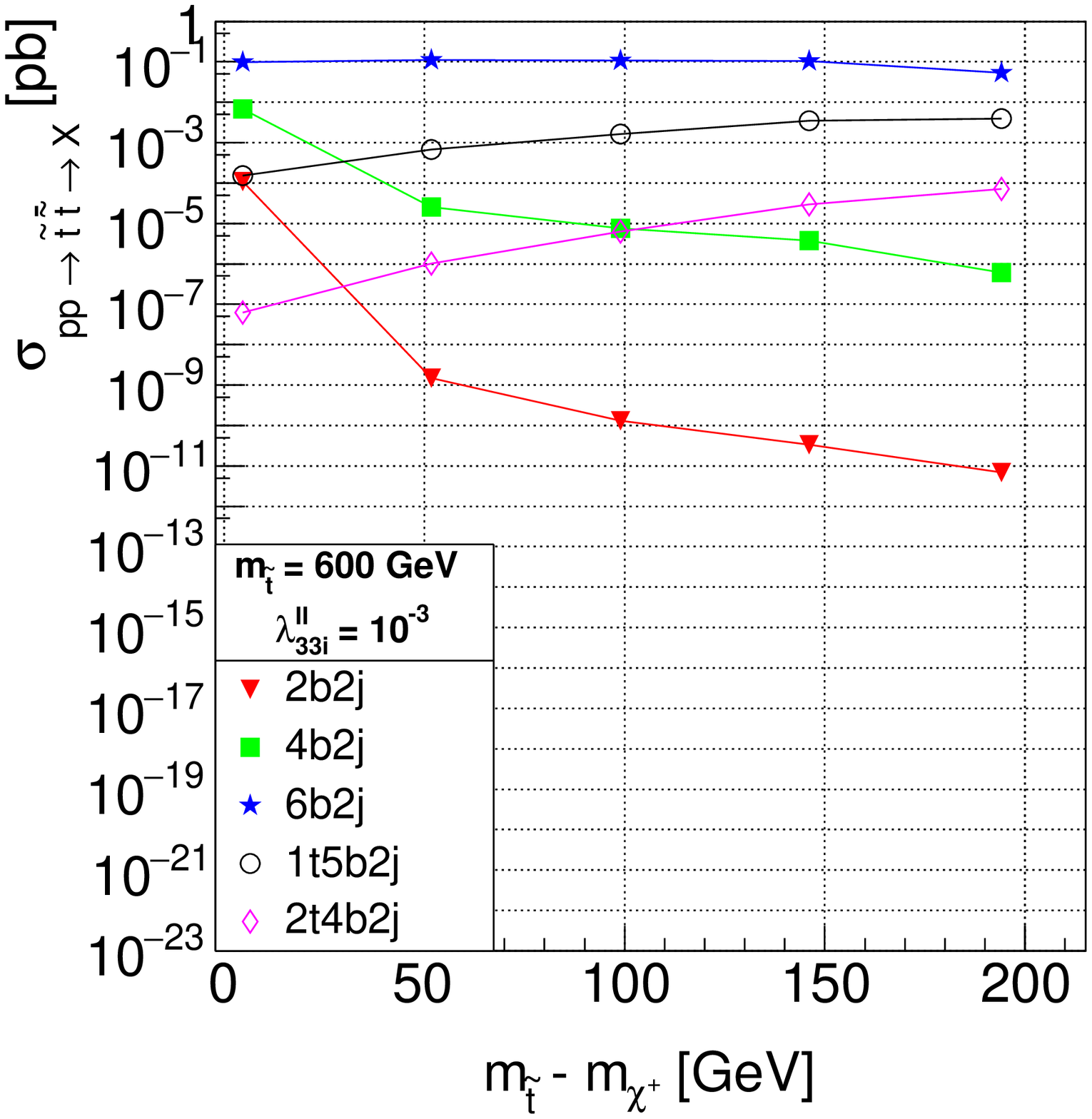}}\\
      \subfloat[]{\includegraphics[width=0.42\textwidth]{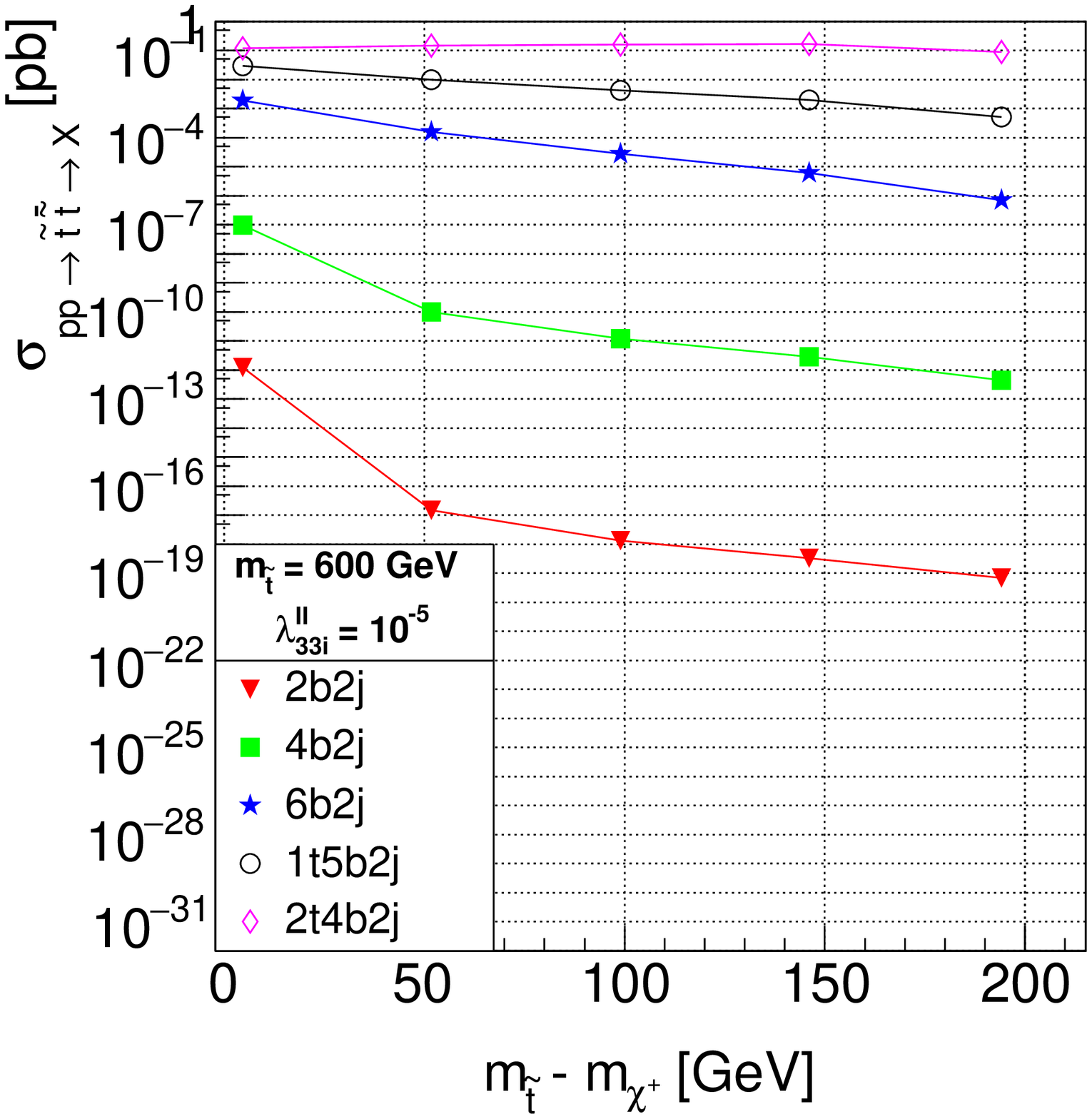}}
      \subfloat[]{\includegraphics[width=0.42\textwidth]{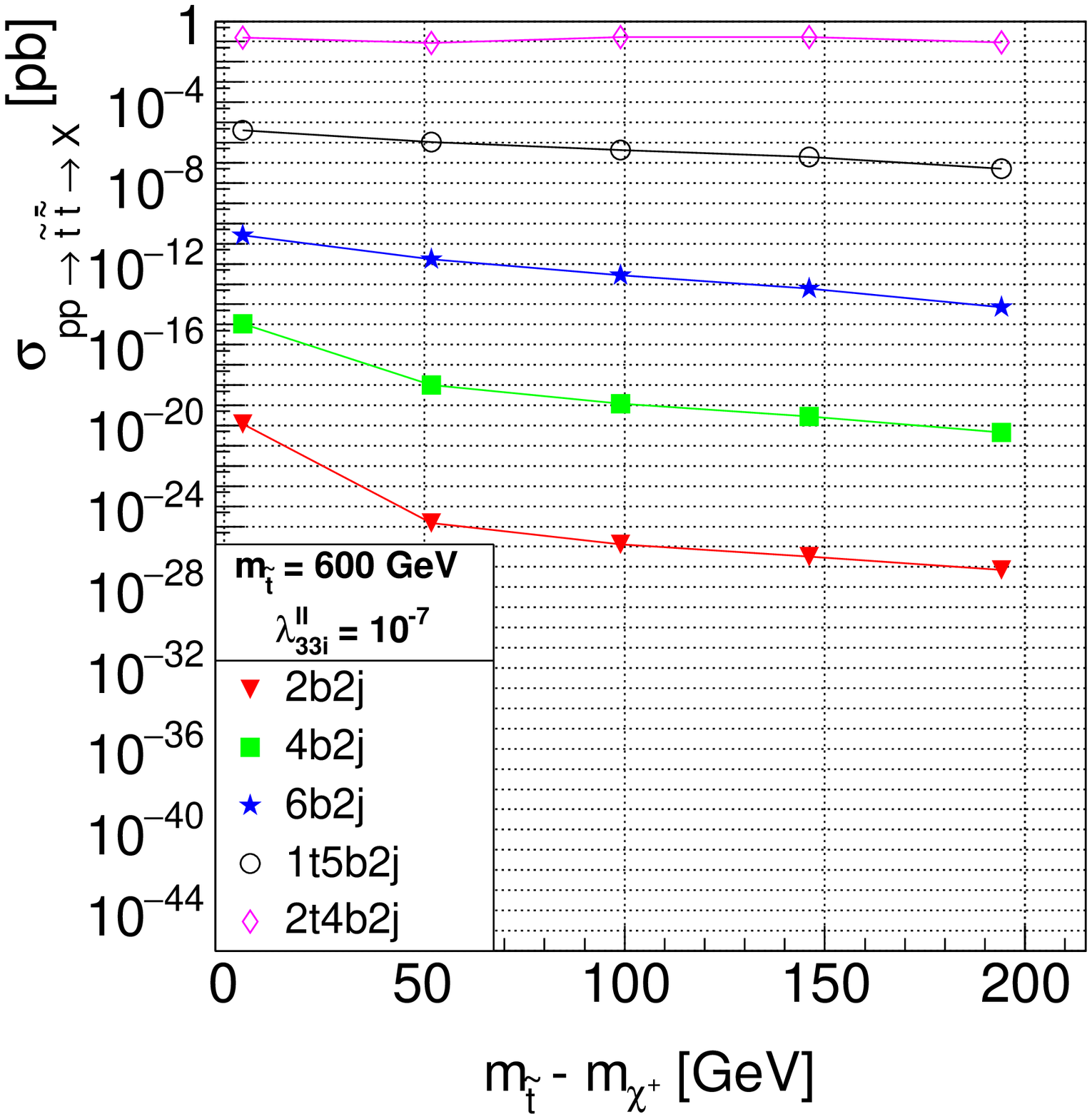}}\\
\end{center}
\caption{\label{fig:bench12} 
\small{Benchmark 1: production cross-section for $\sigma ( p p \rightarrow \tilde{t} \bar{\tilde{t}} 
\rightarrow X)$ at $\sqrt{s}=14$ \TeV, where $X= 2b2j$ (red triangles), $4b2j$ (green squares), $6b2j$ (blue stars), $1t5b2j$ (black empty circles) and $2t4b2j$ (pink diamonds), as a function of $m_{\tilde{t}} - m_{\chi^+}$ and for 
$\lambda_{33i}'' = 10^{-1}$ (a), $10^{-3}$ (b), $10^{-5}$ (c) and $10^{-7}$ (d). 
See Tabs.~\ref{tab:input_parameters} and \ref{tab:parameters} for the low-energy values of the MSSM parameters.}}
\end{figure*}

\begin{figure*}[!ht]
    \begin{center}
      \subfloat[]{\includegraphics[width=0.42\textwidth]{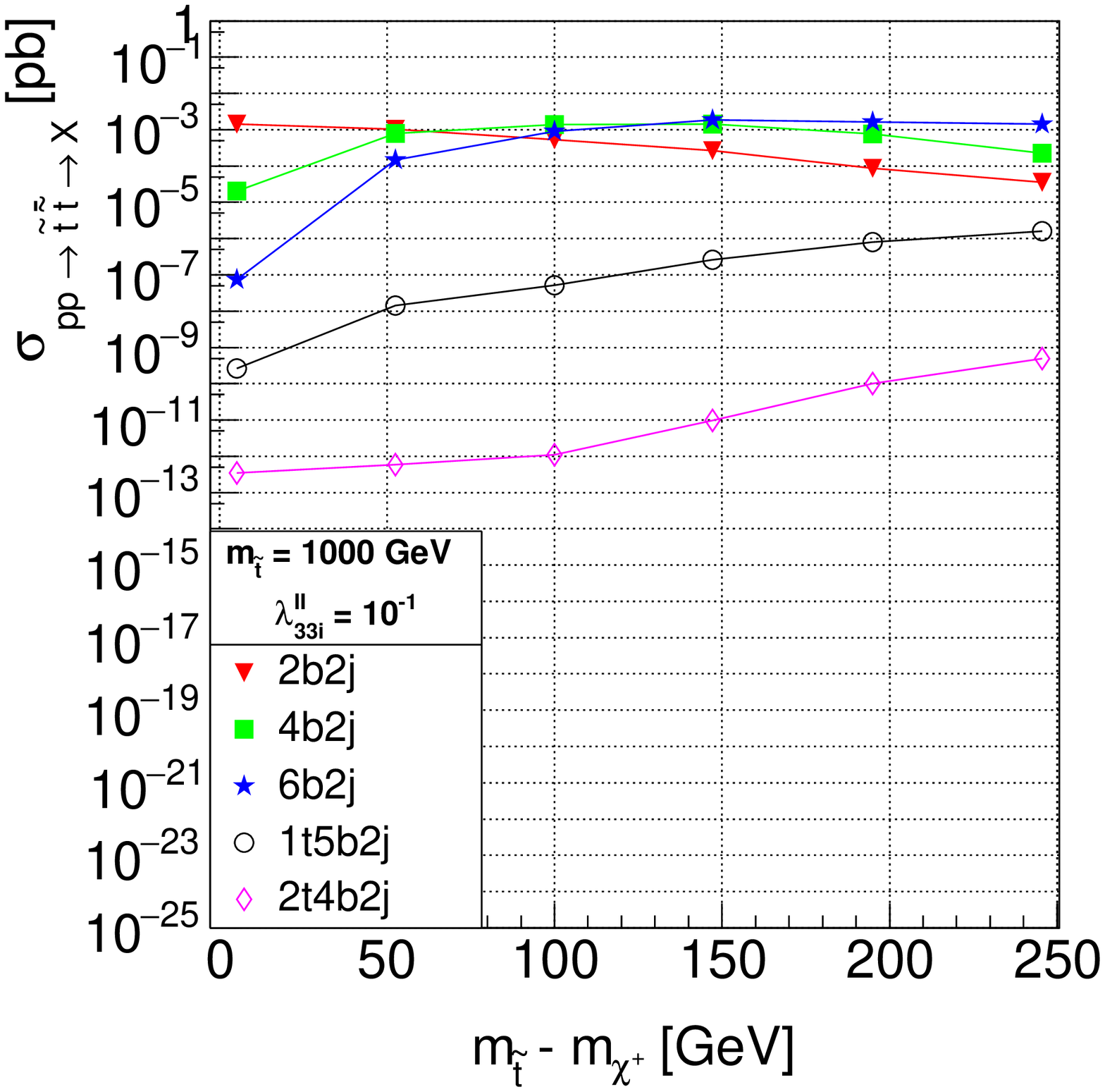}}
      \subfloat[]{\includegraphics[width=0.42\textwidth]{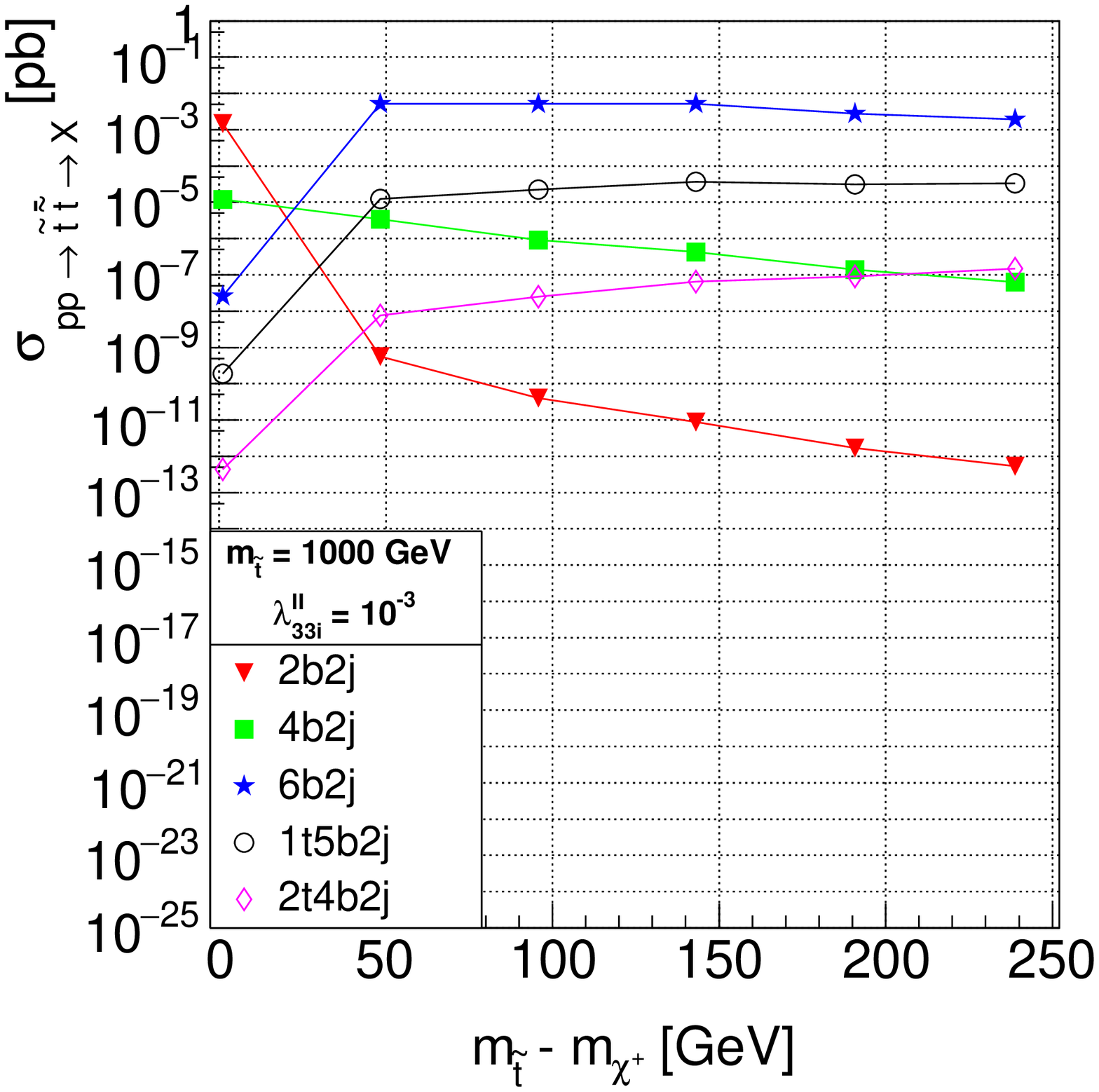}}\\
      \subfloat[]{\includegraphics[width=0.42\textwidth]{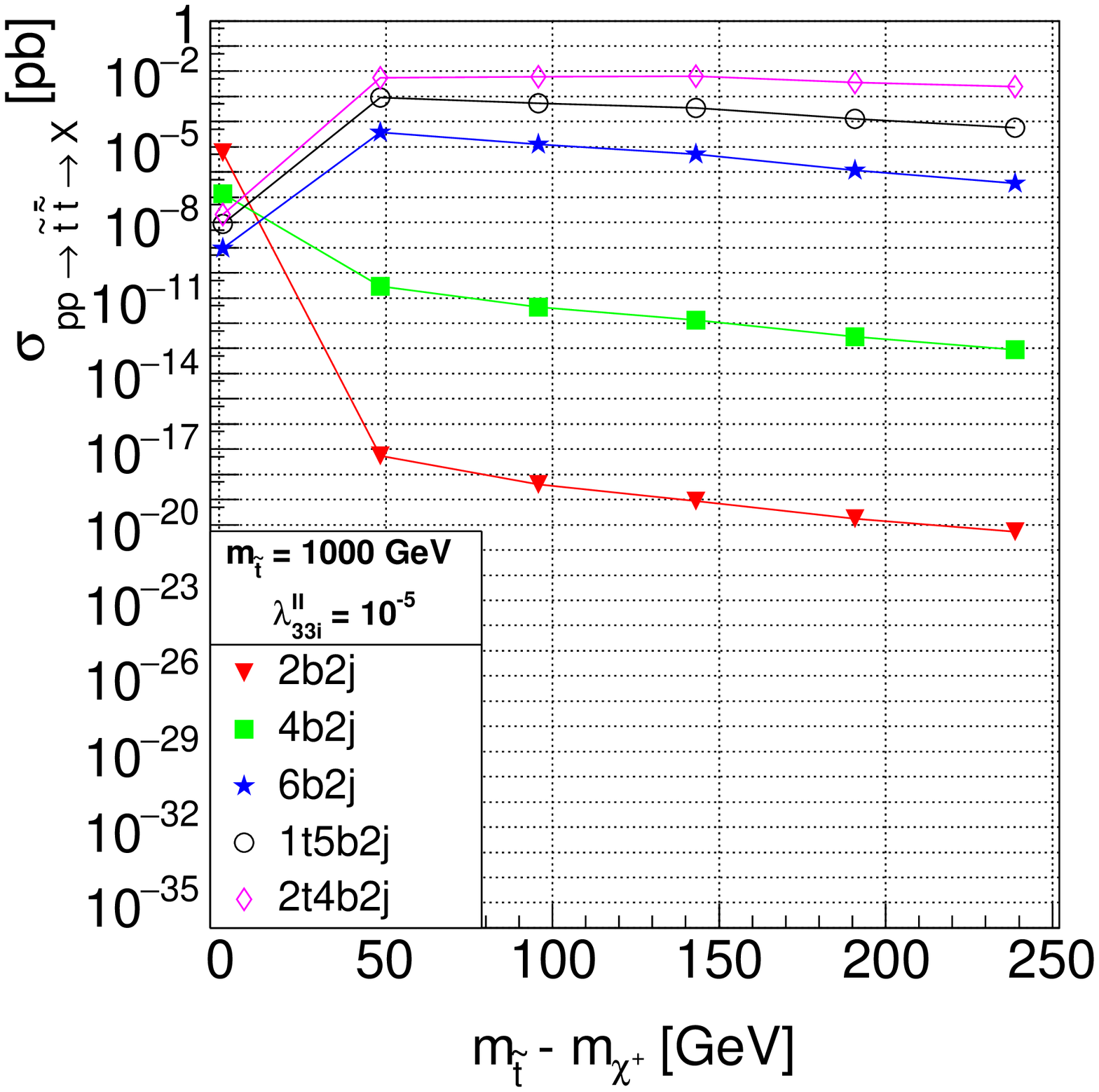}}
      \subfloat[]{\includegraphics[width=0.42\textwidth]{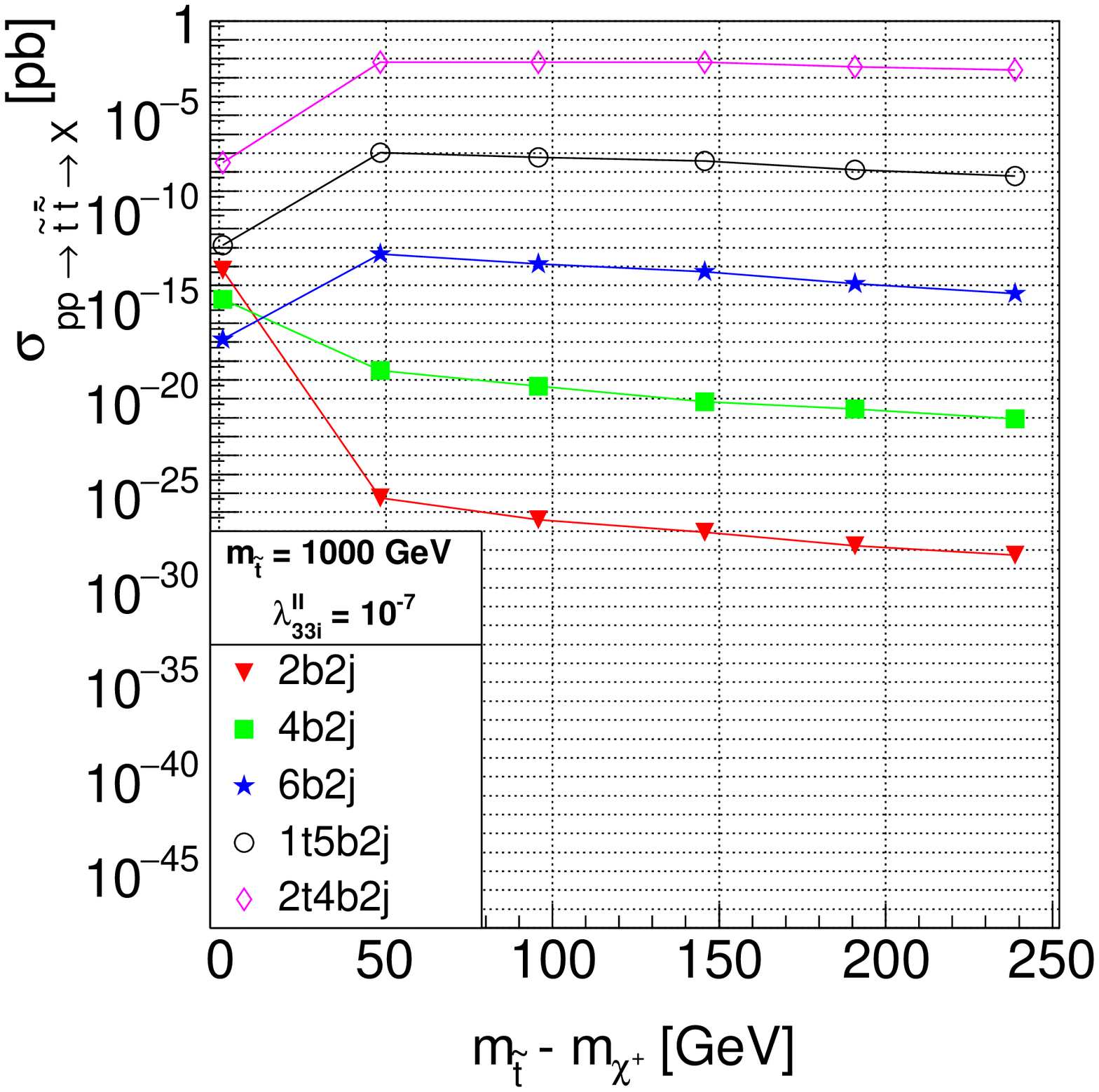}}\\
\end{center}
\caption{\label{fig:bench32} 
\small{Benchmark 2: production cross-section for $\sigma ( p p \rightarrow \tilde{t} \bar{\tilde{t}} 
\rightarrow X)$ at $\sqrt{s}=14$ \TeV, where $X= 2b2j$ (red triangles), $4b2j$ (green squares), $6b2j$ (blue stars), $1t5b2j$ (black empty circles) and $2t4b2j$ (pink diamonds), as a function of $m_{\tilde{t}} - m_{\chi^+}$ and for 
$\lambda_{33i}'' = 10^{-1}$ (a), $10^{-3}$ (b), $10^{-5}$ (c) and $10^{-7}$ (d). 
See Tabs.~\ref{tab:input_parameters} and \ref{tab:parameters} for the low-energy values of the MSSM parameters.
}}
\end{figure*}

\subsection{Theoretical uncertainties}
\label{subsec:theouncert}

Besides the BSM uncertainties which cannot be really quantified and are somewhat fixed through the choice of the MSSM parameters, 
there are other theoretical inputs, whose uncertainties must be taken into account
when quoting the expected cross sections for a given process. 
Since we 
are interested in the evaluation of the total cross-sections involving the SUSY-QCD process
of stop pair production followed by SUSY-EW decays through various short and long chains, 
we choose to generate the $p p \to \tilde{t} \bar{\tilde{t}}$ 
processes at the leading order (LO) accuracy level. 
SUSY-QCD calculations up to next-to-leading-order (NLO) as well as resummed 
soft gluons at next-to-leading-logarithmic (NLL) level for the partonic  stop pair 
production cross-section in proton-(anti)proton collisions are well-known, see \cite{Borschensky:2014cia} for a recent appraisal. These calculations contribute to reducing 
scale uncertainties and typically lead to an increase of the cross-section above LO results \cite{Kane:1982hw,Harrison:1982yi,Dawson:1983fw}, especially 
near the partonic stop pair production threshold.  On the other hand, Parton 
Distribution Functions (PDFs) have been recently supplemented by soft gluon threshold
resummation at the NLO accuracy \cite{Bonvini:2015ira}. Using these
PDFs consistently in conjunction with the resummed partonic matrix
element calculations, showed a partial cancellation of the above mentioned
threshold effects bringing them closer to the fixed order results. One thus expects the cross-section for heavy stop pair production to be 
well approximated by fixed order NLO results.
Moreover, the latter corrections are in turn expected to be moderate for our benchmark points with very heavy 
colored SUSY states. In fact comparing for instance the NLO-NLL results in the decoupled 
gluon/squarks limits given in \cite{Borschensky:2014cia} to the LO results we find an
increase of the former in excess of 30\% for $m_{\tilde{t}} = 600$ \GeV~at $\sqrt{s}=14$ \TeV.
However, due to the above mentioned partial cancellation the effect would be smaller
for production cross-sections dominated by stops almost at rest
when NLL contributions are consistently included also in the PDFs. 
The difference between NLO and LO
production cross-sections would thus be within the uncertainties related to 
scale variation or to the choice of PDF sets {(discussed below)}. Another reason to stick 
consistently to LO accuracy for the production cross-section in the present study, is that 
the dominant virtual QCD corrections to the stop decay chains are not readily 
available at the level of matrix element calculations for the considered channels. 
Moreover, even though some of these corrections could partly cancel in branching 
ratios, the latter entail the NWA which, as pointed out in Section~\ref{subsec:NWA} and discussed quantitatively in Section~\ref{sec:final_states_lambda}, is not always a good 
approximation to the full matrix element calculations. 

We now turn to the uncertainties from the 
PDFs and from the factorization and renormalization scales, evaluated for the $2b2j$ and $6b2j$ final state 
processes at the center of mass energy of 14~\TeV ~using {\sc MadGraph5$\_$aMC@NLO}. 

\subsubsection{Systematic uncertainty from scale variation}
In order to evaluate the scale uncertainty, we vary the renormalization and factorization scales independently with respect to the fixed scales central values $\mu_{R} = \mu_{F} = m_{\tilde{t}}$.  We 
choose values within the range $m_{\tilde{t}}/2 < \mu_{R}$, $\mu_{F}< 2m_{\tilde{t}}$.
The computation is performed using the NNPDF23LO1 set \cite{Rojo:2015nxa} 
for three different stop mass points corresponding to $m_{\tilde{t}}$ = 600, 800 and 1000 \GeV, three different stop-chargino mass splitting equal to 50, 100 and 150 \GeV~and three different values of the coupling $\lambda_{33i}''$ = 10$^{-1}$, 10$^{-3}$ and 10$^{-6}$.  
At a given $m_{\tilde{t}}$, we take the scale uncertainty to be the largest difference in cross section relative to the central value. We 
note that the fractional scale uncertainty for both $2b2j$ and $6b2j$ processes is approximately $^{-25\%}_{+40\%}$, independently from the stop mass, the stop-chargino mass splitting and the $\lambda_{33i}''$ value.

\subsubsection{Systematic uncertainty from PDF}
Systematic uncertainties due to PDFs are evaluated by computing the cross sections of the two processes $2b2j$ and $6b2j$  at the center of 
mass energy of 14 \TeV~with {\sc MadGraph5$\_$aMC@NLO} using two different PDF sets: NNPDF23LO1 \cite{Rojo:2015nxa}, CTEQ6L \cite{Pumplin:2002vw}. 
The estimation of these uncertainties is performed similarly to the evaluation of the scale uncertainty, 
for three values of the stop mass, $m_{\tilde{t}}$ = 600, 800 and 1000 \GeV, three stop-chargino mass splitting equal to 50, 100 and 150 
\GeV~and three RPV $\lambda_{33i}''$ couplings corresponding to 10$^{-1}$, 10$^{-3}$ and 10$^{-6}$. The result appears to be slightly dependent on $m_{\tilde{t}}$. The resulting relative variation in cross sections is found to be around 24\% for $m_{\tilde{t}}$ = 600 \GeV, 28\% for $m_{\tilde{t}}$ = 800 \GeV ~and 32\% for $m_{\tilde{t}}$ = 1 \TeV.\\

Finally, we note that both PDF and scale uncertainties associated with the $2b2j$ final 
state process are consistent within 2\% with the ones found for the $6b2j$ final state: this 
result allows us to assume the same order of magnitude for the uncertainty associated with the 
other RPV-processes listed in Sec.~\ref{sec:multi-jet_final_states}.

\subsection{\label{sec:final_states_lambda} Final states sensitivity to $\lambda_{33i}''$}
As can be seen from Figs.~\ref{fig:bench11} and \ref{fig:bench31},
the various cross-sections vary over several orders of magnitude
due to a very high sensitivity to $\lambda_{33i}''$. 
The extreme values of $\lambda_{33i}''$ feature
a reversed hierarchy of the contributions of the different
final states.
The most striking aspect is that the busiest $2t4b2j$ final state  
dominates for extremely small values ${\cal O}(10^{-7}$~--~$10^{-5})$ of 
$\lambda_{33i}''$ while the $2b2j$, $4b2j$ and $6b2j$ final states  
dominate for $\lambda_{33i}''$ of  ${\cal O}(10^{-3}$~--~$10^{-1})$, yet with 
comparable cross-sections  of order a few tens to a hundred femtobarns.
Moreover, as shown on Figs.~\ref{fig:bench12}(a) and~\ref{fig:bench32}(a), the relative contributions of the dominant 
$2b2j$, $4b2j$ and $6b2j$ final states for large $\lambda_{33i}''$ depend also on the stop
chargino mass splitting, typically with a (reversed) hierarchy given by the $b$-quark 
multiplicity. The $4b2j$ channel can be comparable to the two other channels but is rarely 
dominant. The
$2b2j$ will always eventually dominate for sufficiently large 
$\lambda_{33i}''\gtrsim 10^{-2}$ (e.g.
for $\lambda_{33i}'' \gtrsim 10^{-1}$ not shown on the figures, its dominance
prevails for small to moderate ranges of mass splitting). In contrast, the 
$6b2j$ channel dominates in a range of intermediate values of $\lambda_{33i}''\gtrsim 10^{-3}$
when the mass splitting is moderate to large. 

These features illustrate clearly the
complementarity of the different final states in view of extracting information in the RPV-coupling/mass-splitting parameter space. 
The general trend of the sensitivity to $\lambda_{33i}''$ can be understood  
qualitatively from the NWA expressions, Eqs.~(\ref{eq:four_jets}) through 
(\ref{eq:twelve_jets}).
From the asymptotic behavior of these NWA expressions at small $\lambda_{33i}''$, in the regime
$r_1 \times (\lambda_{33i}'')^2 \ll 1$ and 
$r_2 \times (\lambda_{33i}'')^2 \ll 1$, one sees that all the topless final state
channels scale with $(\lambda_{33i}'')^4$, the 
channels with one top scale with $(\lambda_{33i}'')^2$ and the channel with two top-quarks  
tends to be constant in $\lambda_{33i}''$, which explains the tremendous orders
of magnitude difference in the cross-sections and the dominance of the \bfRPl
channel. If the other extreme of asymptotically large $\lambda_{33i}''$ were allowed,
i.e. $\lambda_{33i}'' \gtrsim 1$, $r_1 \times (\lambda_{33i}'')^2 \gg 1$ and 
$r_2 \times (\lambda_{33i}'')^2 \gg 1$, then only the $2b2j$ would survive, becoming
almost $\lambda_{33i}''$-independent, the other channels scaling with 
increasing inverse powers of $\lambda_{33i}''$ for increasing multiplicity of $b$- 
and $t$-quarks in the final state. In fact, if one remains in the domain of 
moderate values of $\lambda_{33i}''$ the behavior becomes more 
sensitive to $r_1, r_2$. 
In our scenario $r_2$ is always very large, typically several 
orders of magnitude  larger than $r_1$,  due to the smallness of the decay width 
of $\Gamma (\chi^+ \to \chi^0 f_2' \bar{f_2})$ and to the fact that $\chi^+$ and $\chi^0$ are almost degenerate. For the considered range of $\lambda_{33i}''$ we are always
in the regime $r_2 \times (\lambda_{33i}'')^2 \gg 1$. 
Similarly, $r_1$ can become equally large but only in corners of the parameter space
where the stop is almost degenerate with the chargino as seen from
Eq.~(\ref{eq:r1}).
This allows to understand the relative magnitudes of the various cross-sections
shown on the figures. For instance the ratio 
$\sigma(6b2j)/\sigma(2b2j)$ scales with $r_1^{-2}  (\lambda_{33i}'')^{-4}$ and is
indeed (much) larger than $1$ even at the upper edge of the domain of Eq.~(\ref{eq:RPVrange}), 
except when $r_1$ becomes large due to small stop-chargino mass splitting, eventually
reversing the hierarchy between the two cross-sections consistently with the numerical
behavior shown on Figs.~\ref{fig:bench12}(a) and~\ref{fig:bench32}(a). One can understand
similarly the behavior of $\sigma(4b2j)$ that is bounded essentially between the 
$2b2j$ and $6b2j$ cross-sections irrespective of the mass splitting. Note however
that $\sigma(4b2j)/\sigma(6b2j)$ scales with $2 r_1 (\lambda_{33i}'')^2$, so that
the $4b2j$ channel can come to dominate over all the other channels for moderate mass
splitting and a $\lambda_{33i}''$ somewhat larger than the range we consider for
the analysis. Turning to the final states containing one or two top-quarks, their tiny contribution in the upper part of the $\lambda_{33i}''$ range, cf. Figs.~\ref{fig:bench11} and  \ref{fig:bench31}, is due to the size of $r_2$. For instance
$\sigma(2b2j)/\sigma(2t4b2j)$ scales with $r_1^2 r_2^2 (\lambda_{33i}'')^8$, but the
large suppression for $\lambda_{33i}''\lesssim 0.1$ is compensated for by a very
large value of $r_2 \approx {\cal O}(10^7)$ as a consequence of the compressed light 
chargino/neutralino sector.

We turn now to a quantitative discussion of the comparison between the full
matrix element calculation and the NWA. Given the huge difference in the scaling of the various cross-sections and the variations over several 
orders of magnitudes, this comparison is an important cross-check of the results.
We indeed find that the NWA works reasonably well in configurations where
it is expected to do so~\cite{Berdine:2007uv,Kauer:2007zc,
Kauer:2007nt,Uhlemann:2008pm}. We check first the scaling relations
t-SR and b-SR given in Eqs.~(\ref{eq:b-SR},~\ref{eq:t-SR}), as these provide global 
tests that do not require the knowledge of the stop production cross-section
nor the $r_1,r_2$ ratios. A systematic test of t-SR and b-SR using all the 
cross-sections in Table~\ref{tab:bench1-7} and in Table~\ref{tab:bench2-7} gave a 
relative deviation of $10\%$ or more from these scaling relations only in 
$\lesssim 9\%$ of the cases, while a deviation of $\lesssim 5 \%$ obtains in 
$\sim 80 \%$ of the cases and a deviation of $\lesssim 1 \%$ in $\sim 66 \%$ of 
the cases. 
It is also instructive to identify the configurations where the NWA fails badly.
We find that deviations of more than $30\%$, reaching up to $135\%$, occur in 
less than $5\%$ of the cases and only for t-SR that involves long chain decays. 
These correspond to points of benchmark 1 having large values of $\lambda_{33i}''$ and very small stop-chargino mass splitting, such as for $\lambda_{33i}''=10^{-1}$ 
and $m_{\tilde{t}} - m_{\chi^+} = 59$ and 11~\GeV~and for 
$\lambda_{33i}''=10^{-2}$ 
and $m_{\tilde{t}} - m_{\chi^+} = 5$~\GeV, shown in Table~\ref{tab:bench1-7}. 
Such large deviations are in accord with the general expectations \cite{Berdine:2007uv}.
We have also checked the NWA for individual cross-sections. This allowed
to disentangle the reasons for the differences from the results of the full matrix element calculations. Very good quantitative agreement is observed for the 
shortest decay chains, and for small values of $\lambda_{33i}''$ and/or large mass splitting for longer decay chains. The cross-sections given by 
Eqs.~(\ref{eq:four_jets} --\ref{eq:twelve_jets})
reproduce globally the behavior shown in Figs.~\ref{fig:bench11} and \ref{fig:bench31}.  

We discuss now three spectrum configurations that are outside one or the other of the
assumptions given in Eqs.~(\ref{eq:spectconfig1} -- \ref{eq:spectconfig3}). 
 The values of  $m_{\tilde{t}} - m_{\chi^+} = -36$ or $-43$~\GeV, shown in Table~\ref{tab:bench1-7}, correspond to points violating Eq.~(\ref{eq:spectconfig1}) with an MSSM-LSP stop. 
As expected, in this case
the $2b2j$ channel largely dominates independently of the magnitude of $\lambda_{33i}''$, the next-to-leading channel, $4b2j$, being two to three orders 
of magnitude smaller.
It is however noteworthy that the $2b2j$ channel can be dominant even when 
Eq.~(\ref{eq:spectconfig1})
is satisfied, provided that 
the positive mass splitting $m_{\tilde{t}} - m_{\chi^+}$ remains sufficiently small
and $\lambda_{33i}''$ sufficiently large. One sees this tendency from the
$m_{\tilde{t}} - m_{\chi^+} =5$ and $11$~\GeV~points
in Table~\ref{tab:bench1-7} for benchmark 1. 
For instance,  the $2b2j$ channel 
can still be an order of magnitude greater than the total of the 
remaining channels for a stop/chargino mass splitting in excess of $10 \GeV$,
as illustrated for $m_{\tilde{t}} - m_{\chi^+} = 11$~\GeV~and $\lambda_{33i}''=10^{-1}$. A smaller mass splitting, at the edge of 
 the validity of Eq.~(\ref{eq:spectconfig3}), leads to even larger effects, as one can see in Table~\ref{tab:bench2-7} by looking at 
 the point $m_{\tilde{t}} - m_{\chi^+} = 5$~\GeV~and $\lambda_{33i}''=10^{-1}$.
In this case the $2b2j$ channel dominates the other channels by almost two orders of magnitude.
A larger mass splitting 
would require larger values of $\lambda_{33i}''$ to ensure the dominance of the 
$2b2j$ channel.
In fact there is a correlation between the mass splitting and the size of the RPV coupling that
can be understood in terms of the NWA cross-section of  Eq.~(\ref{eq:four_jets}): the 
$2b2j$ channel
becomes dominant, with a branching ratio close to one, when
$r_1 \times (\lambda_{33i}'')^2 \gg 1$, say ${\cal O}(10)$ or larger. Indeed, the ratio $r_1$ becomes large for small stop/chargino mass splitting due to phase-space
suppression  of the width $\Gamma (\tilde{t} \to \chi^+ b)$, see Eq.~(\ref{eq:r1}), 
implying that $2b2j$ can dominate for moderately small $\lambda_{33i}''$.
More generally the regime where $2b2j$ dominates is characterized roughly by
$|\lambda_{33i}''| \gtrsim 3 \times r_1^{-1/2}$. The present LHC limits 
 \cite{Aad:2016kww,ATLAS-CONF-2016-022,Khachatryan:2014lpa} where the $2b2j$ dominance
is assumed, can thus be interpreted as excluding either scenarios where the stop is 
the MSSM-LSP, or the domain delineated by the above relation in scenarios where a chargino and 
a neutralino are lighter than the stop.

If Eq.~(\ref{eq:spectconfig3}) is not satisfied but the mass splitting still larger than the $s$-quark or $d$-quark 
masses then the \bfcRP and \bfRPl decays occur dominantly through the LFV 
channel $\tilde{t} \to s(d) \chi^+$ (recall that we assume MFV). 
The effect is thus noticeable for the small 
values of $\lambda_{33i}''$ where the \bfcRP or \bfRPl decays are expected to dominate. 
This is illustrated for all values of  $\lambda_{33i}''$ with mass splitting of $1 \GeV$ in Table~\ref{tab:bench2-7}.
There are two effects: for $\lambda_{33i}''$
in the intermediate range $10^{-4}$--$10^{-2}$, the $2b2j$ channel becomes largely dominant
over the $6b2j$ and $1t5b2j$ channels contrary to the typical cases with larger 
stop/chargino mass splitting.
In this intermediate $\lambda_{33i}''$ range
the LFV channels with smaller $b$-quark multiplicity and larger light jet multiplicity 
such as $4b4j$ and $1t3b4j$ final states have cross-sections comparable to that of the 
$2b2j$ channel given the size of the corresponding CKM mixing angles. In contrast, in the range $ 10^{-7} \lesssim \lambda_{33i}''\lesssim 10^{-5}$
the cross-sections for all the final states listed in Table~\ref{tab:finalstates} become
suppressed as can be seen in the corresponding blocks of  Table~\ref{tab:bench2-7} 
and mass splitting of $1 \GeV$ indicating that the dominant channel corresponds now to the LFV \bfRPl--\bfRPl final state
$2t2b4j$. The study of final states with more light quarks and less 
$b$-quark multiplicity
can thus be motivated in the context of an inclusive search comprising the very narrow part of the parameter space
having an extremely compressed $\tilde{t}/\chi^+$ spectrum.

Last but not least, we consider the case where Eq.~(\ref{eq:spectconfig2}) is 
not satisfied. The decay channel $\tilde{t} \to t \chi^0 (\chi^0_2)$ is now open leading
to $4t2b2j$ final states. A detailed study of this channel is outside the scope of the 
present paper and we do not give here the corresponding cross-section. Its is however
interesting to note the indirect effect of this channel on the cross-sections
given in Tables~\ref{tab:bench1-7} and \ref{tab:bench2-7}. Indeed, the expected drop
of the latter when the top-neutralino channel sets in is found to remain relatively
moderate. For instance, comparing the points $m_{\tilde{t}} - m_{\chi^+} = 146$~\GeV~
and $194 \GeV$ of Table~\ref{tab:bench1-7} one sees that the drop in the leading
cross-sections $6b2j, 1t5b2j$ and $2t4b2j$ is by a factor of order $2$--$2.5$
or less, 
depending on the magnitude of $\lambda_{33i}''$.
Similar effects are found for benchmark 2, as seen from a comparison of the points 
$m_{\tilde{t}} - m_{\chi^+} = 143$~\GeV~and $239 \GeV$ of Table~\ref{tab:bench2-7}.
This suggests that the final states considered in the present study can still contribute to signatures outside the specific mass configurations that we relied on.

To conclude this section, we stress the main point of the analysis: if part of the
chargino/neutralino sector is lighter than the lightest stop, 
channels with different jet multiplicities probe dominantly different ranges of the 
RPV coupling. 
This is due to a distinct dependence on 
$\lambda_{33i}''$ of the various decay widths and branching ratios, thus triggering the dominance of different channels for different values of this coupling. 
We depict this general feature 
schematically in Fig.~\ref{fig:lambda_sensitivity} for a typical configuration, keeping in mind 
that the actual dominance ranges can change depending on the masses and RPC couplings. 
The analytical expressions for the cross-sections in terms of the RPV coupling and decay widths in the RPC sector given in Sec.~\ref{subsec:NWA}
allow a clear qualitative understanding of these features.
\begin{figure}[h]
\begin{picture}(60,60) 
\thicklines \put(-180,20){\vector(1,0){380}}
\put(-200,10){$\lambda_{33i}''$}
\put(-170,10){$\lesssim 10^{-5}$}
\put(-140,20){\line(0,1){10}}
\put(-170,35){\RPlRPl}
\put(-80,10){$\sim 10^{-4}$}
\put(-65,20){\line(0,1){10}}
\put(-75,35){\RPlcRP}
\put(-5,10){$\sim 10^{-3}$}
\put(10,20){\line(0,1){10}}
\put(0,35){\cRPcRP}
\put(70,10){$\sim 10^{-2}$}
\put(85,20){\line(0,1){10}}
\put(75,35){\cRPtRP}
\put(160,10){$\gtrsim 10^{-1}$}
\put(155,20){\line(0,1){10}}
\put(155,35){\tRPtRP}
 \end{picture}
 \caption{\small{Schematic illustration of $\lambda_{33i}''$ as a  
 {\sl signature generator}; different magnitudes of this coupling favor different
 final states.}}
\label{fig:lambda_sensitivity} 
\end{figure}
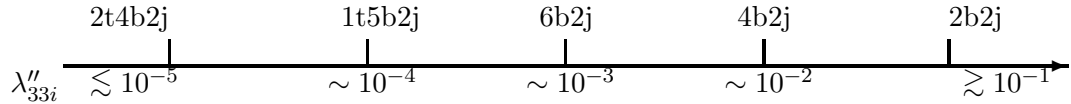
The RPV coupling thus plays the role of a {\sl signature generator}.

\section{RPV Final States and SM background: a discussion \label{sec:discussion}}

The LHC is currently in its Run 2 data taking period, which started in 2015 and it is now providing proton-proton collisions at $\sqrt{s}=13$~\TeV\ to both ATLAS and CMS, continuously improving in the delivered peak luminosity. With the current schedule, 100 fb$^{-1}$ of data and a possible further push in the center-of-mass energy to $\sqrt{s}=14$~\TeV, both ATLAS and CMS will be able to carry the BSM searches which are the core of the LHC Run 2 physics program. In what follows we give a brief overview of how the different RPV signatures, which are the focus of this paper, are, or can be looked for at the LHC. For the different final states treated, $2b2j$, $4b2j$, $6b2j$, $1t5b2j$ and $2t4b2j$, we either review the current experimental analyses, or, for those channels where no experimental analyses have been performed yet, we propose, based on similar existing analyses, a search strategy with a list of SM backgrounds which could impact their sensitivities.

\subsection{$2b2j$}

Direct production of stop quarks with a subsequent RPV decay into two jets has been searched for at LEP and Tevatron, where a 95\% upper limit on the mass of such particles was set to respectively 82.5 \GeV~\cite{Heister:2002jc} and 100~\GeV~\cite{Aaltonen:2013hya}. As pointed out by Ref.~\cite{Bai:2013xla} first searches for stop production at the LHC did not succeed on being sensitive to any stop mass until the trigger strategy changed from using high transverse momentum ($p_T$) multi-jet triggers, which had the effect of shaping the background towards high masses, to triggering on the totality of the hadronic energy deposited in the calorimeter (ATLAS~\cite{ATLASstopRpV}), a variable less correlated to the two masses of the di-jet resonances, or lower $p_T$ jets (CMS~\cite{Khachatryan:2014lpa}).
Both ATLAS and CMS have looked for stop production pairs final state where the two stops decay into $\bar{b}\bar{s}+bs$. The two stop-quark resonances are identified as wide hadronic calorimeter ``fat" jets with a cone size, $R$, of the order of $1,1.5$~\cite{Sapeta:2015gee}. Given that the characteristic distance between the two particles stemmed from a resonance, $\Delta R$, is of the order $2 m/p_T$, where $m$ is the stop-quark mass and $p_T$ its transverse momentum, this kind of signature allows to access a relatively low mass spectra, where most of the center of mass energy goes to the boost of the produced resonance pair.  

The main challenge for hadronic jet based searches is to understand the normalizations and shapes of the multijet background, this has been shown to be possible using data-driven techniques~\cite{Aad:2016zqi}. The discrimination between signal and background is done by exploiting kinematic quantities such as the value of the reconstructed fat jet masses, which is the same for the two fat jets from stop pair production, and other jet substructure properties such as the difference in $p_T$ between the two subjets identified by un-doing the last step of the fat jet clustering, more pronounced in multijet events (see Ref.~\cite{ATLAS-CONF-2016-022} and reference therein). 
Also fundamental to reduce multi-jet background, $b$-tagging algorithms are used to identify the presence of jets issued from the hadronization of $b$-quarks~\cite{ATL-PHYS-PUB-2015-022,Chatrchyan:1494669}.
After bump hunting, the two LHC experiments could exclude at 95\% Confidence Level stop quark production for masses up to 
345~\GeV (ATLAS)~\cite{ATLAS-CONF-2016-022} and 385~\GeV (CMS)~\cite{Khachatryan:2014lpa}. 
This final state is sensitive not only to the value of the stop mass in the case the stop is the LSP, but if the lightest neutralino and the lightest chargino are lighter than the stop quark, then this class of analyses could be sensitive to the hardest part of the  $\lambda_{332}''$ spectrum considered, as shown in Fig.~\ref{fig:bench11}, see also Section~\ref{sec:final_states_lambda}.

\subsection{$4b2j$ and $6b2j$}

As discussed in section~\ref{sec:final_states_lambda} the topless multi-$b$-jet signatures saturate the stop quark branching ratio for intermediate values of $\lambda_{33i}''\gtrsim 10^{-3}$. The  $4b2j$ and $6b2j$ signatures have the highest cross sections at large $\lambda_{33i}''$ value when the differences between the stop and the chargino/neutralino masses are maximal, see Figs~\ref{fig:bench11} and \ref{fig:bench12}. In this scenario, for low stop masses, such that $2 m/p_T$ is $O(1)$, the same strategy as the searches in the $2b2j$ final state can be used, where two structured large sized hadronic jets of particles are produced back to back. This facilitates the task of eliminating the combinatorial background that arises when the presence of multiple reconstructed objects in the final state does not allow to assign them to one of the particle originating the decay. This results in a poor reconstruction of the resonant peaks. 
Moreover the presence of resonances within the fat jets helps discriminating against the background when using the value of the reconstructed invariant mass of the stop and chargino candidates and of more specific jet substructure related quantities such as the $k_{t}$ splitting scale and $n$-subjettiness (see Ref.~\cite{ATLASboostop} and references therein). It has also been recently suggested~\cite{Hamaguchi:2015uqa} that jet reconstruction techniques based on a mass-jump clustering algorithm with variable size can be used to reconstruct multi-jet resonance in very busy environment as the one produced by boosted stop squarks decay into $4b2j$ or $6b2j$ final states. 

In the case of resolved regime, where most jets from the stop and chargino decay are reconstructed, the signal is characterized by events with high jet and $b$-jet multiplicity. If no $b$-tagging is required at the analysis level, the signal, even for stop masses of about one \TeV, although it would present very high jet multiplicity, would be still swamped by the presence of a large multi-jet background~\cite{Aad:2015mzg}. 

When the $b$-jet identification is used, the physics processes that could mimic RPV stop signal, include any resonant multi-$b$-jet production such as $t\bar{t}$ + $X$, abundantly produced in the $\sqrt{s}=14$~\TeV~proton-proton collisions at the LHC. Background processes to this final state include $t\bar{t}$ plus light and heavy flavored jets, $t\bar{t}$ plus vector boson, and $t\bar{t}H(\rightarrow b\bar{b})$ production, where both top quarks decay fully hadronically. The inclusive $t\bar{t}$ cross section is known at NNLO in QCD including resummation of soft gluon terms at next-to-leading-logarithmic (NNLL)~\cite{ref:xs5}; at $\sqrt{s}=14$~\TeV, $\sigma_{t\bar{t}}=954^{+23}_{-34}$(scale)$^{+16}_{-18}$(pdf) pb, 45.7\% of which decays fully hadronically~\cite{Agashe:2014kda}. At the analysis level when asking for more than 2 $b$-tagged jets it is more likely to select events from processes where extra heavy flavors are produced. For $\sqrt{s}=14$~\TeV,  $\sigma_{t\bar{t} + b\bar{b}}$ is known at NLO to be 2.63$^{+86}_{-70}$(scale) pb~\cite{Bevilacqua:2009zn}, $\sigma_{t\bar{t} \rightarrow Z}$ is also known at NLO with a value of 1057$^{+110}_{-104}$(scale)$^{+20}_{-25}$(pdf) fb~\cite{Maltoni:2015ena} while $\sigma_{t\bar{t} \rightarrow W}=769^{+228}_{-170}$(scale)$^{+54}_{-61}$(pdf) fb~\cite{Campbell:2012dh}. As for the associated top and Higgs production, $t\bar{t}H$ cross section is known at NNLO in QCD and EW plus resummation of soft gluon at NNLL; at $\sqrt{s}=14$~\TeV, $\sigma_{t\bar{t}H(H \rightarrow b\bar{b})} = 625^{+29}_{-42}$(scale)$^{+14}_{-14}$(pdf) fb~\cite{Kulesza:2015vda}. 
Recent LHC analyses at $\sqrt{s}=8$~\TeV~\cite{Aad:2016zqi} show how by just selecting a high multi-jet ($\approx$8) and $b$-jet multiplicity ($\approx$4) the background composition is made at about 80\% of multijets. This background has little resemblance with multi-resonant production and can be discriminated using multi-variate analysis which exploits different energy regime, event shape, using quantities such as centrality, aplanarity and the mass of the reconstructed top quark candidates. Such kind of analysis needs to control the uncertainties on the main top-like background and at the same time removing as much as possible multi-jet background.

Despite the large background from multi-jet events, given the large energy deposit in the hadronic calorimeter from the decay products of the pair of massive resonances, and the large presence of jets coming from $b$-quark, as also stated in Ref.~\cite{Evans:2014gfa}, this channel is very promising. Searches at hadron colliders for gluino pair production and subsequent RPV decay into $tbs$, where similar final states are investigated, could already show sensitivity to this channel~\cite{Khachatryan:2016iqn}.

\subsection{$1t5b2j$ and $2t4b2j$}

Signatures with decays into top quarks saturate the branching ratio for  $\lambda_{33i}''<10^{-5}$, as already 
discussed in Section~\ref{sec:final_states_lambda}. 
These signatures are interesting since the presence of a lepton from the top quark decay can be easily identified at trigger level and used to eliminate the otherwise overwhelming multi-jet background, such as in the fully hadronic signatures. After selecting at least six $b$-tagged jets in addition to at least two light-jets and one lepton, for the case of $1t5b2j$, or two leptons, for the case of $2t4b2j$, the main irreducible background arises from $t\bar{t}+jets$ and $t\bar{t}H(\rightarrow b \bar{b})$ + jets. For simplicity in this discussion we limit ourself to analyzing the dileptonic top quark decay for the $2t4b2j$ final state to allow discussing the backgrounds composition to both final states. In this case, the main background for both final states comes from $t\bar{t}+b\bar{b}b\bar{b}$ + jets. The LO cross section for $t\bar{t}+b\bar{b}b\bar{b}$ at $\sqrt{s}=14$~\TeV, estimated with 
{\sc MadGraph5$\_$aMC@NLO} using LHAPDF 6.1.6~\cite{Buckley:2014ana}, is 290$^{+400}_{-160}$(scale)$^{+90}_{-50}$(pdf) fb, comparable with the signal cross section in the low $\lambda_{33i}''$ regime, see for example $\lambda_{33i}''<10^{-5}$ in Table~\ref{tab:bench1-7}. 
The presence of neutrinos and the large jet multiplicity present in this final state makes it difficult to reconstruct completely the final state, i.e. to assign unambiguously reconstructed leptons and jets to the stop and anti-stop decays. This effect weakens the power of distributions such as the invariant mass of the reconstructed stop and chargino candidates, to discriminate signal and background events. 
The large energy deposited in the detectors, equal approximately to twice the stop mass, could have the role of the missing transverse envergy for RPC searches, to discriminate signal against the softer top quark pair production, using for example the transverse energy of the event ($H_{T}$). On the other hand the softer part of the $H_{T}$ distribution can be useful to  control the effect of major systematic uncertainties, especially the large theoretical uncertainties on $t\bar{t}+b\bar{b}b\bar{b}$ cross section, on the LHC sensitivity for this channel. This class of final states can use the analysis techniques developed for $ttH(H\rightarrow b \bar{b})$ searches and that need to be extended to higher jet multiplicity. 

As for the case of $4b2j$ and $6b2j$, both ATLAS and CMS searches for gluino pair production and subsequent RPV decay to a pair of top quarks and jets through the R-parity violating decay of either the neutralino into three quarks or the top squark into SM quarks~\cite{Khachatryan:2016iqn,ATLAS-CONF-2016-094}, could already be reinterpreted as limits on $\lambda_{33i}''$ using this channel.

\section{Conclusion \& outlook}
\label{sec:conclusion}

The ever stronger exclusion limits on SUSY particle masses from negative searches at the
LHC seem to disfavor, if not to rule out, low energy supersymmetry as the correct theory
beyond the SM. However, one should not lose sight of the distinction between SUSY as a 
general framework and its various possible model realizations. Only a class of the 
latter, leading to RPC signatures with striking missing energy, is being heavily 
excluded by the LHC. If RPV baryon number violating couplings are allowed, a class of 
signatures, generally with high jet and lepton multiplicity and no missing energy is 
expected at the LHC. Searches for such RPV signals usually assume that the shortest 
chain particle decays involving RPV vertices have 100\% $BR$. 
Previous works have already 
pointed out, for a particular case involving decaying stops with  
$\lambda_{331}'', \lambda_{332}'' \neq 0$, the existence of a region in the mass 
parameter space for which longer decay chains and 
richer final states than the plain $\tilde{t} \to b s, b d$ can originate.

In this paper we described how different stop-pair final states arise when different values of the 
RPV coupling and different supersymmetric particle mass splittings are considered. This
is exhaustively investigated for the case of  
proton--proton collisions at center-of-mass energy of $\sqrt{s}=14$~\TeV. After having 
defined a set of working assumptions concerning the mass hierarchy 
and the allowed range of $\lambda_{33i}''$, we examined the sensitivity of the stop 
decay branching ratios to $\lambda_{33i}''$, first analytically by means of the NWA 
approximation showing that the variation of $\lambda_{33i}''$ over several orders of 
magnitude triggers the dominance of very different final states, then numerically 
relying on automated matrix element calculations.

Using for the latter a bottom-bottom approach in the phenomenological MSSM, we generated
the full mass spectrum and couplings and identified two benchmark points taking into 
account all possible constraints ranging from the measured Higgs mass to 
the experimental low energy constraints. For these two benchmark points we estimated the
cross sections for the relevant final states 
differing by the number of heavy and light flavored quarks ($2b2j$, $4b2j$, $6b2j$, 
$1t5b2j$ and $2t4b2j$), as a function of 
$\lambda_{33i}''$ and the stop/chargino mass splitting, confirming numerically what is 
seen analytically with the NWA approximation. 
Finally we discussed the phenomenology of the RPV stop production and decays and its 
rich experimental signatures, stressing that {\sl the smaller the 
values of $\lambda_{33i}''$ the larger the quark mutliplicity of the dominant final 
states.} Some of these final states having so far not been extensively looked at experimentally, we 
briefly discussed how they can be searched for at the LHC. 

While other studies, including single stop resonant
and associate productions as well as the increaslingly strict limits on displaced vertices and long-lived particles,
contribute to narrowing down the viable RPV scenarios, significant parts of the parameter space
remain to be explored.
As such, an exciting possibility still lies ahead, that a  
light part of the MSSM spectrum, a key issue for the naturalness of SUSY, may be 
stashed in the present and future LHC data. \\

\begin{acknowledgements}  We thank Florian Staub for very useful discussions and 
substantial help in the
use of {\sc Sarah} and implementation of the {\sc SPheno} package for the RPV-MSSM. 
We also would like to thank Stephane Lavignac for helpful discussions,  
as well as Jared Evans and 
  Angelo Monteux for the various comments that helped improve the present paper.

This work has been carried out thanks to the support of the OCEVU Labex (ANR-11-LABX-0060) and the 
A$\star$MIDEX project (ANR-11-IDEX-0001-02) funded by the ``Investissements d'Avenir" French government program managed by the ANR.
\end{acknowledgements}


\bibliographystyle{utphys} 
\bibliography{references-v4}

\newpage
\begin{table}[h]
\begin{center}
\tiny{ 
\begin{tabular}{|c|c|c|c|c|c|c|c|}
\hline \hline
 $\lambda_{33i}''$ & $\mu$ [GeV] & $ m_{\tilde{t}} - m_{\chi^{+}} [GeV] $ & $ \sigma(2b2j) $ [pb] &  $ \sigma(4b2j) $ [pb] & $\sigma(6b2j)$ [pb] & $\sigma(1t5b2j)$ [pb] & $\sigma(2t4b2j)$ [pb]\\ 
\hline \hline 
\multirow{6}{*}{$10^{-1}$} & 
400 &  201 &  4.38 $ \cdot 10^{-4}$  & 8.80 $ \cdot 10^{-3}$  & 4.19 $ \cdot 10^{-2}$  & 2.42 $ \cdot 10^{-5}$  & 2.70 $ \cdot 10^{-9}$  \\ 
& 450 &  153 &  1.67 $ \cdot 10^{-3}$  & 2.16 $ \cdot 10^{-2}$  & 6.66 $ \cdot 10^{-2}$  & 1.05 $ \cdot 10^{-5}$  & 3.21 $ \cdot 10^{-10}$  \\ 
& 500 &  106 &  4.17 $ \cdot 10^{-3}$  & 2.83 $ \cdot 10^{-2}$  & 4.63 $ \cdot 10^{-2}$  & 1.56 $ \cdot 10^{-6}$  & 1.47 $ \cdot 10^{-11}$  \\ 
& 550 &  59 &  1.26 $ \cdot 10^{-2}$  & 2.87 $ \cdot 10^{-2}$  & 1.60 $ \cdot 10^{-2}$  & 4.79 $ \cdot 10^{-7}$  & 8.45 $ \cdot 10^{-12}$  \\ 
& 600 &  11 &  3.01 $ \cdot 10^{-2}$  & 2.98 $ \cdot 10^{-3}$  & 7.29 $ \cdot 10^{-5}$  & 2.30 $ \cdot 10^{-8}$  & 3.23 $ \cdot 
10^{-12}$  \\ 
& 650 &  -36 &  3.08 $ \cdot 10^{-2}$  & 7.02 $ \cdot 10^{-5}$  & 5.19 $ \cdot 10^{-8}$  & 6.36 $ \cdot 10^{-10}$  & 2.65 $ \cdot 10^{-12}$  \\ 

\hline \hline 
\multirow{6}{*}{$10^{-2}$} & 
400 &  194 &  6.92 $ \cdot 10^{-8}$  & 1.28 $ \cdot 10^{-4}$  & 5.72 $ \cdot 10^{-2}$  & 6.99 $ \cdot 10^{-5}$  & 1.67 $ \cdot 10^{-8}$  \\ 
& 450 &  146 &  3.30 $ \cdot 10^{-7}$  & 3.88 $ \cdot 10^{-4}$  & 1.09 $ \cdot 10^{-1}$  & 4.93 $ \cdot 10^{-5}$  & 4.35 $ \cdot 10^{-9}$  \\ 
& 500 &  100 &  1.30 $ \cdot 10^{-6}$  & 7.68 $ \cdot 10^{-4}$  & 1.09 $ \cdot 10^{-1}$  & 1.97 $ \cdot 10^{-5}$  & 6.37 $ \cdot 10^{-10}$  \\ 
& 550 &  52 &  1.44 $ \cdot 10^{-5}$  & 2.57 $ \cdot 10^{-3}$  & 1.12 $ \cdot 10^{-1}$  & 7.86 $ \cdot 10^{-6}$  & 1.14 $ \cdot 10^{-10}$  \\ 
& 600 &  5 &  2.40 $ \cdot 10^{-2}$  & 1.71 $ \cdot 10^{-2}$  & 2.96 $ \cdot 10^{-3}$  & 1.82 $ \cdot 10^{-7}$  & 4.10 $ \cdot 10^{-12}$  \\ 
& 650 &  -43 &  3.29 $ \cdot 10^{-2}$  & 6.68 $ \cdot 10^{-5}$  & 4.51 $ \cdot 10^{-8}$  & 5.88 $ \cdot 10^{-10}$  & 2.62 $ \cdot 10^{-12}$  \\ 

\hline \hline 
\multirow{6}{*}{$10^{-3}$} & 
400 &  194 &  6.96 $ \cdot 10^{-12}$  & 1.23 $ \cdot 10^{-6}$  & 5.40 $ \cdot 10^{-2}$  & 3.90 $ \cdot 10^{-3}$  & 7.06 $ \cdot 10^{-5}$  \\ 
 & 450 &  146 &  3.33 $ \cdot 10^{-11}$  & 3.77 $ \cdot 10^{-6}$  & 1.06 $ \cdot 10^{-1}$  & 3.53 $ \cdot 10^{-3}$  & 2.93 $ \cdot 10^{-5}$  \\ 
 & 500 &  99 &  1.32 $ \cdot 10^{-10}$  & 7.56 $ \cdot 10^{-6}$  & 1.09 $ \cdot 10^{-1}$  & 1.64 $ \cdot 10^{-3}$  & 6.28 $ \cdot 10^{-6}$  \\ 
 & 550 &  52 &  1.50 $ \cdot 10^{-9}$  & 2.58 $ \cdot 10^{-5}$  & 1.09 $ \cdot 10^{-1}$  & 6.71 $ \cdot 10^{-4}$  & 1.03 $ \cdot 10^{-6}$  \\ 
 & 600 &  5 &  1.08 $ \cdot 10^{-4}$  & 6.81 $ \cdot 10^{-3}$  & 9.82 $ \cdot 10^{-2}$  & 1.55 $ \cdot 10^{-4}$  & 6.16 $ \cdot 10^{-8}$  \\ 
 & 650 &  -43 &  3.29 $ \cdot 10^{-2}$  & 6.51 $ \cdot 10^{-5}$  & 4.35 $ \cdot 10^{-8}$  & 5.85 $ \cdot 10^{-10}$  & 2.62 $ \cdot 10^{-12}$  \\ 

\hline \hline 
\multirow{6}{*}{$10^{-4}$} & 
400 &  194 &  6.96 $ \cdot 10^{-16}$  & 3.27 $ \cdot 10^{-9}$  & 3.85 $ \cdot 10^{-3}$  & 2.76 $ \cdot 10^{-2}$  & 4.96 $ \cdot 10^{-2}$  \\ 
& 450 &  146 &  3.33 $ \cdot 10^{-15}$  & 1.62 $ \cdot 10^{-8}$  & 1.97 $ \cdot 10^{-2}$  & 6.57 $ \cdot 10^{-2}$  & 5.46 $ \cdot 10^{-2}$  \\ 
& 500 &  99 &  1.32 $ \cdot 10^{-14}$  & 4.68 $ \cdot 10^{-8}$  & 4.13 $ \cdot 10^{-2}$  & 6.30 $ \cdot 10^{-2}$  & 2.40 $ \cdot 10^{-2}$  \\ 
& 550 &  52 &  1.51 $ \cdot 10^{-13}$  & 2.05 $ \cdot 10^{-7}$  & 6.98 $ \cdot 10^{-2}$  & 4.28 $ \cdot 10^{-2}$  & 6.55 $ \cdot 10^{-3}$  \\ 
& 600 &  5 &  1.22 $ \cdot 10^{-8}$  & 6.87 $ \cdot 10^{-5}$  & 9.68 $ \cdot 10^{-2}$  & 1.50 $ \cdot 10^{-2}$  & 6.22 $ \cdot 10^{-4}$  \\ 
& 650 &  -43 &  3.29 $ \cdot 10^{-2}$  & 6.30 $ \cdot 10^{-5}$  & 4.14 $ \cdot 10^{-8}$  & 5.75 $ \cdot 10^{-10}$  & 2.62 $ \cdot 10^{-12}$  \\ 

\hline \hline
\multirow{6}{*}{$10^{-5}$} & 
400 &  194 &  6.96 $ \cdot 10^{-20}$  & 4.40 $ \cdot 10^{-13}$  & 6.97 $ \cdot 10^{-7}$  & 5.01 $ \cdot 10^{-4}$  & 8.98 $ \cdot 10^{-2}$  \\ 
& 450 &  146 &  3.33 $ \cdot 10^{-19}$  & 2.81 $ \cdot 10^{-12}$  & 5.93 $ \cdot 10^{-6}$  & 1.97 $ \cdot 10^{-3}$  & 1.64 $ \cdot 10^{-1}$  \\ 
& 500 &  99 &  1.32 $ \cdot 10^{-18}$  & 1.20 $ \cdot 10^{-11}$  & 2.71 $ \cdot 10^{-5}$  & 4.13 $ \cdot 10^{-3}$  & 1.57 $ \cdot 10^{-1}$  \\ 
& 550 &  52 &  1.51 $ \cdot 10^{-17}$  & 9.73 $ \cdot 10^{-11}$  & 1.57 $ \cdot 10^{-4}$  & 9.63 $ \cdot 10^{-3}$  & 1.48 $ \cdot 10^{-1}$  \\ 
& 600 &  5 &  1.22 $ \cdot 10^{-12}$  & 9.65 $ \cdot 10^{-8}$  & 1.91 $ \cdot 10^{-3}$  & 2.98 $ \cdot 10^{-2}$  & 1.16 $ \cdot 10^{-1}$  \\ 
& 650 &  -43 &  3.29 $ \cdot 10^{-2}$  & 6.26 $ \cdot 10^{-5}$  & 3.97 $ \cdot 10^{-8}$  & 5.54 $ \cdot 10^{-10}$  & 2.62 $ \cdot 10^{-12}$  \\ 

\hline \hline
\multirow{6}{*}{$10^{-6}$} & 
400 &  194 &  6.96 $ \cdot 10^{-24}$  & 4.42 $ \cdot 10^{-17}$  & 7.03 $ \cdot 10^{-11}$  & 5.04 $ \cdot 10^{-6}$  & 9.04 $ \cdot 10^{-2}$  \\ 
& 450 &  146 &  3.33 $ \cdot 10^{-23}$  & 2.83 $ \cdot 10^{-16}$  & 6.01 $ \cdot 10^{-10}$  & 2.00 $ \cdot 10^{-5}$  & 1.66 $ \cdot 10^{-1}$  \\ 
& 500 &  99 &  1.32 $ \cdot 10^{-22}$  & 1.22 $ \cdot 10^{-15}$  & 2.80 $ \cdot 10^{-9}$  & 4.26 $ \cdot 10^{-5}$  & 1.63 $ \cdot 10^{-1}$  \\ 
& 550 &  52 &  1.51 $ \cdot 10^{-21}$  & 1.01 $ \cdot 10^{-14}$  & 1.70 $ \cdot 10^{-8}$  & 1.04 $ \cdot 10^{-4}$  & 1.59 $ \cdot 10^{-1}$  \\ 
& 600 &  5 &  1.22 $ \cdot 10^{-16}$  & 1.11 $ \cdot 10^{-11}$  & 2.53 $ \cdot 10^{-7}$  & 3.96 $ \cdot 10^{-4}$  & 1.54 $ \cdot 10^{-1}$  \\ 
& 650 &  -43 &  3.29 $ \cdot 10^{-2}$  & 6.26 $ \cdot 10^{-5}$  & 3.97 $ \cdot 10^{-8}$  & 5.53 $ \cdot 10^{-10}$  & 2.62 $ \cdot 10^{-12}$  \\ 

\hline \hline
\multirow{6}{*}{$10^{-7}$} & 
400 &  194 &  6.96 $ \cdot 10^{-28}$  & 4.43 $ \cdot 10^{-21}$  & 7.01 $ \cdot 10^{-15}$  & 5.04 $ \cdot 10^{-8}$  & 9.05 $ \cdot 10^{-2}$  \\ 
& 450 &  146 &  3.33 $ \cdot 10^{-27}$  & 2.83 $ \cdot 10^{-20}$  & 6.01 $ \cdot 10^{-14}$  & 2.00 $ \cdot 10^{-7}$  & 1.66 $ \cdot 10^{-1}$  \\ 
& 500 &  99 &  1.32 $ \cdot 10^{-26}$  & 1.21 $ \cdot 10^{-19}$  & 2.80 $ \cdot 10^{-13}$  & 4.26 $ \cdot 10^{-7}$  & 1.63 $ \cdot 10^{-1}$  \\ 
& 550 &  52 &  1.51 $ \cdot 10^{-25}$  & 1.01 $ \cdot 10^{-18}$  & 1.70 $ \cdot 10^{-12}$  & 1.04 $ \cdot 10^{-6}$  & 1.59 $ \cdot 10^{-1}$  \\ 
& 600 &  5 &  1.22 $ \cdot 10^{-20}$  & 1.11 $ \cdot 10^{-15}$  & 2.54 $ \cdot 10^{-11}$  & 3.96 $ \cdot 10^{-6}$  & 1.54 $ \cdot 10^{-1}$  \\ 
& 650 &  -43 &  3.37 $ \cdot 10^{-2}$  & 6.31 $ \cdot 10^{-5}$  & 3.99 $ \cdot 10^{-8}$  & 5.54 $ \cdot 10^{-10}$  & 2.61 $ \cdot 10^{-12}$  \\ 
\hline \hline
\end{tabular}}
\caption{ Benchmark 1: production cross-section for $\sigma ( p p \rightarrow \tilde{t} \bar{\tilde{t}} 
\rightarrow X)$ at $\sqrt{s}=14$ \TeV, where $X= 2b2j$, $4b2j$, $6b2j$, $1t5b2j$ and $2t4b2j$, as a function of $\lambda_{33i}''$ and for different values of $m_{\tilde{t}} - m_{\chi^+}$. See Tabs.~\ref{tab:input_parameters} and \ref{tab:parameters} for the low-energy values of the MSSM parameters.} 
\label{tab:bench1-7}
\end{center}
\end{table}

 \begin{table}[h]
\begin{center}
\tiny{
\begin{tabular}{|c|c|c|c|c|c|c|c|}
\hline \hline
 $\lambda_{33i}''$ & $\mu$ [GeV] & $ m_{\tilde{t}} - m_{\chi^{+}} [GeV] $ & $ \sigma(2b2j) $ [pb] &  $ \sigma(4b2j) $ [pb] & $\sigma(6b2j)$ [pb] & $\sigma(1t5b2j)$ [pb] & $\sigma(2t4b2j)$ [pb]\\ 
\hline \hline
\multirow{6}{*}{$10^{-1}$} & 
750 &  243 &  3.52 $ \cdot 10^{-5}$  & 4.57 $ \cdot 10^{-4}$  & 1.45 $ \cdot 10^{-3}$  & 1.62 $ \cdot 10^{-6}$  & 4.88 $ \cdot 10^{-10}$  \\ 
& 800 &  195 &  8.53 $ \cdot 10^{-5}$  & 7.56 $ \cdot 10^{-4}$  & 1.64 $ \cdot 10^{-3}$  & 8.09 $ \cdot 10^{-7}$  & 1.00 $ \cdot 10^{-10}$  \\ 
& 850 &  147 &  2.63 $ \cdot 10^{-4}$  & 1.42 $ \cdot 10^{-3}$  & 1.87 $ \cdot 10^{-3}$  & 2.63 $ \cdot 10^{-7}$  & 9.83 $ \cdot 10^{-12}$  \\ 
& 900 &  100 &  5.30 $ \cdot 10^{-4}$  & 1.38 $ \cdot 10^{-3}$  & 8.93 $ \cdot 10^{-4}$  & 5.12 $ \cdot 10^{-8}$  & 1.09 $ \cdot 10^{-12}$  \\ 
& 950 &  52 &  1.02 $ \cdot 10^{-3}$  & 7.80 $ \cdot 10^{-4}$  & 1.48 $ \cdot 10^{-4}$  & 1.42 $ \cdot 10^{-8}$  & 5.78 $ \cdot 10^{-13}$  \\ 
& 1000 &  5 &  1.43 $ \cdot 10^{-3}$  & 2.02 $ \cdot 10^{-5}$  & 7.30 $ \cdot 10^{-8}$  & 2.66 $ \cdot 10^{-10}$  & 3.51 $ \cdot 10^{-13}$  \\ 
\hline \hline
\multirow{6}{*}{$10^{-2}$} & 
750 &  239 &  5.34 $ \cdot 10^{-9}$  & 6.72 $ \cdot 10^{-6}$  & 1.89 $ \cdot 10^{-3}$  & 2.53 $ \cdot 10^{-6}$  & 7.96 $ \cdot 10^{-10}$  \\ 
& 800 &  191 &  1.70 $ \cdot 10^{-8}$  & 1.45 $ \cdot 10^{-5}$  & 2.79 $ \cdot 10^{-3}$  & 1.57 $ \cdot 10^{-6}$  & 2.19 $ \cdot 10^{-10}$  \\ 
& 850 &  143 &  8.74 $ \cdot 10^{-8}$  & 4.39 $ \cdot 10^{-5}$  & 5.35 $ \cdot 10^{-3}$  & 9.25 $ \cdot 10^{-7}$  & 4.05 $ \cdot 10^{-11}$  \\ 
& 900 &  96 &  3.85 $ \cdot 10^{-7}$  & 9.09 $ \cdot 10^{-5}$  & 5.08 $ \cdot 10^{-3}$  & 3.46 $ \cdot 10^{-7}$  & 6.14 $ \cdot 10^{-12}$  \\ 
& 950 &  48 &  4.94 $ \cdot 10^{-6}$  & 3.13 $ \cdot 10^{-4}$  & 4.64 $ \cdot 10^{-3}$  & 1.95 $ \cdot 10^{-7}$  & 2.18 $ \cdot 10^{-12}$  \\ 
& 1000 &  1 &  1.47 $ \cdot 10^{-3}$  & 1.22 $ \cdot 10^{-5}$  & 2.63 $ \cdot 10^{-8}$  & 1.67 $ \cdot 10^{-10}$  & 3.47 $ \cdot 10^{-13}$  \\ 
\hline \hline
\multirow{6}{*}{$10^{-3}$} & 
750 &  239 &  5.37 $ \cdot 10^{-13}$  & 6.38 $ \cdot 10^{-8}$  & 1.90 $ \cdot 10^{-3}$  & 3.35 $ \cdot 10^{-5}$  & 1.50 $ \cdot 10^{-7}$  \\ 
& 800 &  191 &  1.71 $ \cdot 10^{-12}$  & 1.38 $ \cdot 10^{-7}$  & 2.77 $ \cdot 10^{-3}$  & 3.15 $ \cdot 10^{-5}$  & 8.94 $ \cdot 10^{-8}$  \\ 
& 850 &  143 &  8.88 $ \cdot 10^{-12}$  & 4.26 $ \cdot 10^{-7}$  & 5.11 $ \cdot 10^{-3}$  & 3.65 $ \cdot 10^{-5}$  & 6.51 $ \cdot 10^{-8}$  \\ 
& 900 &  96 &  3.98 $ \cdot 10^{-11}$  & 9.04 $ \cdot 10^{-7}$  & 5.16 $ \cdot 10^{-3}$  & 2.27 $ \cdot 10^{-5}$  & 2.49 $ \cdot 10^{-8}$  \\ 
& 950 &  48 &  5.57 $ \cdot 10^{-10}$  & 3.41 $ \cdot 10^{-6}$  & 5.17 $ \cdot 10^{-3}$  & 1.25 $ \cdot 10^{-5}$  & 7.57 $ \cdot 10^{-9}$  \\ 
& 1000 &  1 &  1.47 $ \cdot 10^{-3}$  & 1.18 $ \cdot 10^{-5}$  & 2.57 $ \cdot 10^{-8}$  & 1.90 $ \cdot 10^{-10}$  & 4.29 $ \cdot 10^{-13}$  \\ 
\hline \hline
\multirow{6}{*}{$10^{-4}$} & 
750 &  239 &  5.37 $ \cdot 10^{-17}$  & 3.70 $ \cdot 10^{-10}$  & 6.38 $ \cdot 10^{-4}$  & 1.07 $ \cdot 10^{-3}$  & 4.49 $ \cdot 10^{-4}$  \\ 
& 800 &  191 &  1.71 $ \cdot 10^{-16}$  & 9.37 $ \cdot 10^{-10}$  & 1.28 $ \cdot 10^{-3}$  & 1.40 $ \cdot 10^{-3}$  & 3.82 $ \cdot 10^{-4}$  \\ 
& 850 &  143 &  8.89 $ \cdot 10^{-16}$  & 3.26 $ \cdot 10^{-9}$  & 2.99 $ \cdot 10^{-3}$  & 2.11 $ \cdot 10^{-3}$  & 3.71 $ \cdot 10^{-4}$  \\ 
& 900 &  96 &  3.98 $ \cdot 10^{-15}$  & 7.58 $ \cdot 10^{-9}$  & 3.61 $ \cdot 10^{-3}$  & 1.58 $ \cdot 10^{-3}$  & 1.73 $ \cdot 10^{-4}$  \\ 
& 950 &  48 &  5.58 $ \cdot 10^{-14}$  & 3.06 $ \cdot 10^{-8}$  & 4.21 $ \cdot 10^{-3}$  & 1.02 $ \cdot 10^{-3}$  & 6.12 $ \cdot 10^{-5}$  \\ 
& 1000 &  1 &  1.11 $ \cdot 10^{-3}$  & 9.79 $ \cdot 10^{-6}$  & 2.36 $ \cdot 10^{-8}$  & 2.38 $ \cdot 10^{-9}$  & 6.02 $ \cdot 10^{-11}$  \\ 
\hline \hline
\multirow{6}{*}{$10^{-5}$} & 
750 &  239 &  5.37 $ \cdot 10^{-21}$  & 8.67 $ \cdot 10^{-14}$  & 3.50 $ \cdot 10^{-7}$  & 5.88 $ \cdot 10^{-5}$  & 2.47 $ \cdot 10^{-3}$  \\ 
& 800 &  191 &  1.71 $ \cdot 10^{-20}$  & 2.84 $ \cdot 10^{-13}$  & 1.18 $ \cdot 10^{-6}$  & 1.29 $ \cdot 10^{-4}$  & 3.51 $ \cdot 10^{-3}$  \\ 
& 850 &  143 &  8.89 $ \cdot 10^{-20}$  & 1.34 $ \cdot 10^{-12}$  & 5.02 $ \cdot 10^{-6}$  & 3.54 $ \cdot 10^{-4}$  & 6.22 $ \cdot 10^{-3}$  \\ 
& 900 &  96 &  3.98 $ \cdot 10^{-19}$  & 4.44 $ \cdot 10^{-12}$  & 1.24 $ \cdot 10^{-5}$  & 5.41 $ \cdot 10^{-4}$  & 5.92 $ \cdot 10^{-3}$  \\ 
& 950 &  48 &  5.58 $ \cdot 10^{-18}$  & 2.89 $ \cdot 10^{-11}$  & 3.75 $ \cdot 10^{-5}$  & 9.01 $ \cdot 10^{-4}$  & 5.43 $ \cdot 10^{-3}$  \\ 
& 1000 &  1 &  5.75 $ \cdot 10^{-6}$  & 1.40 $ \cdot 10^{-7}$  & 9.36 $ \cdot 10^{-10}$  & 8.84 $ \cdot 10^{-9}$  & 2.09 $ \cdot 10^{-8}$  \\ 
\hline \hline
\multirow{6}{*}{$10^{-6}$} & 
750 &  239 &  5.37 $ \cdot 10^{-25}$  & 8.79 $ \cdot 10^{-18}$  & 3.60 $ \cdot 10^{-11}$  & 6.04 $ \cdot 10^{-7}$  & 2.54 $ \cdot 10^{-3}$  \\ 
& 800 &  191 &  1.71 $ \cdot 10^{-24}$  & 2.90 $ \cdot 10^{-17}$  & 1.23 $ \cdot 10^{-10}$  & 1.34 $ \cdot 10^{-6}$  & 3.67 $ \cdot 10^{-3}$  \\ 
& 850 &  143 &  8.89 $ \cdot 10^{-24}$  & 1.38 $ \cdot 10^{-16}$  & 5.34 $ \cdot 10^{-10}$  & 3.77 $ \cdot 10^{-6}$  & 6.62 $ \cdot 10^{-3}$  \\ 
& 900 &  96 &  3.98 $ \cdot 10^{-23}$  & 4.66 $ \cdot 10^{-16}$  & 1.36 $ \cdot 10^{-9}$  & 5.98 $ \cdot 10^{-6}$  & 6.55 $ \cdot 10^{-3}$  \\ 
& 950 &  48 &  5.58 $ \cdot 10^{-22}$  & 3.15 $ \cdot 10^{-15}$  & 4.46 $ \cdot 10^{-9}$  & 1.07 $ \cdot 10^{-5}$  & 6.47 $ \cdot 10^{-3}$  \\ 
& 1000 &  1 &  6.53 $ \cdot 10^{-10}$  & 1.84 $ \cdot 10^{-11}$  & 1.42 $ \cdot 10^{-13}$  & 1.34 $ \cdot 10^{-10}$  & 3.17 $ \cdot 10^{-8}$  \\ 
\hline \hline
\multirow{6}{*}{$10^{-7}$} & 
750 &  239 &  5.37 $ \cdot 10^{-29}$  & 8.80 $ \cdot 10^{-22}$  & 3.60 $ \cdot 10^{-15}$  & 6.04 $ \cdot 10^{-9}$  & 2.54 $ \cdot 10^{-3}$  \\ 
& 800 &  191 &  1.71 $ \cdot 10^{-28}$  & 2.90 $ \cdot 10^{-21}$  & 1.23 $ \cdot 10^{-14}$  & 1.34 $ \cdot 10^{-8}$  & 3.67 $ \cdot 10^{-3}$  \\ 
& 850 &  143 &  8.89 $ \cdot 10^{-28}$  & 1.38 $ \cdot 10^{-20}$  & 5.35 $ \cdot 10^{-14}$  & 3.78 $ \cdot 10^{-8}$  & 6.59 $ \cdot 10^{-3}$  \\ 
& 900 &  96 &  3.98 $ \cdot 10^{-27}$  & 4.66 $ \cdot 10^{-20}$  & 1.37 $ \cdot 10^{-13}$  & 5.98 $ \cdot 10^{-8}$  & 6.54 $ \cdot 10^{-3}$  \\ 
& 950 &  48 &  5.58 $ \cdot 10^{-26}$  & 3.16 $ \cdot 10^{-19}$  & 4.47 $ \cdot 10^{-13}$  & 1.08 $ \cdot 10^{-7}$  & 6.49 $ \cdot 10^{-3}$  \\ 
& 1000 &  1 &  6.54 $ \cdot 10^{-14}$  & 1.85 $ \cdot 10^{-15}$  & 1.43 $ \cdot 10^{-17}$  & 1.35 $ \cdot 10^{-12}$  & 3.19 $ \cdot 10^{-8}$  \\ 
\hline \hline
\end{tabular}}
\caption{ Benchmark 2: production cross-section for $\sigma ( p p \rightarrow \tilde{t} \bar{\tilde{t}} 
\rightarrow X)$ at $\sqrt{s}=14$ \TeV, where $X= 2b2j$, $4b2j$, $6b2j$, $1t5b2j$ and $2t4b2j$, as a function of $\lambda_{33i}''$ and for different values of $m_{\tilde{t}} - m_{\chi^+}$. See Tabs.~\ref{tab:input_parameters} and \ref{tab:parameters} for the low-energy values of the MSSM parameters.} 
\label{tab:bench2-7}
\end{center}
\end{table}




\end{document}